\documentclass[useAMS,usenatbib,times]{mn2e}
\usepackage{epsfig,times,amsmath,supertabular,colortbl,amsfonts,amssymb,multirow,array,fleqn,longtable}

\title[STEP2: High precision weak lensing analyses]
  {\vspace{-12mm}The Shear TEsting Programme 2: \\
  Factors affecting high precision weak lensing analyses} 
\author[Massey \etal]
{Richard Massey$^{1}$\thanks{rjm@astro.caltech.edu}, 
Catherine Heymans$^{2}$,
Joel Berg\'e$^{3}$,
Gary Bernstein$^{4}$,  
Sarah Bridle$^{5}$, \newauthor 
Douglas Clowe$^{6}$,
H{\aa}kon Dahle$^{7}$, 
Richard Ellis$^{1}$,
Thomas Erben$^{8}$,
Marco Hetterscheidt$^{8}$,  \newauthor  
F. William High$^{9,1}$, 
Christopher Hirata$^{10}$,
Henk Hoekstra$^{11}$,
Patrick Hudelot$^{12}$, 
Mike Jarvis$^{4}$,   \newauthor
David Johnston$^{13}$,
Konrad Kuijken$^{14}$,
Vera Margoniner$^{15}$,
Rachel Mandelbaum$^{16}$,  \newauthor
Yannick Mellier$^{17,18}$, 
Reiko Nakajima$^{4}$, 
Stephane Paulin-Henriksson$^{19}$,
Molly Peeples$^{20,1}$,  \newauthor
Chris Roat$^{15}$,
Alexandre Refregier$^{3}$,
Jason Rhodes$^{13,1}$,
Tim Schrabback$^{8}$,
Mischa Schirmer$^{21}$, \newauthor
Uro\v{s} Seljak$^{16}$,
Elisabetta Semboloni$^{17,2}$
\& Ludovic Van Waerbeke$^{2}$ \\
$^{1}$California Institute of Technology, 1200 E.\ California Blvd., Pasadena, CA 91125, USA.\\
$^{2}$University of British Columbia, 6224 Agricultural Rd., Vancouver, BC, V6T 1Z1, Canada.\\  
$^{3}$Service d'Astrophysique, CEA Saclay, F-91191 Gif sur Yvette, France.\\
$^{4}$Department of Physics and Astronomy, University of Pennsylvania, Philadelphia, PA 19104, USA.\\ 
$^{5}$Department of Physics and Astronomy, University College London, Gower Street, London, WC1E 6BT, UK.\\
$^{6}$Steward Observatory, University of Arizona, 933 N.\ Cherry Ave., Tuscon, AZ 85721, USA.\\
$^{7}$Institute of Theoretical Astrophysics, University of Oslo, P.O.\ Box 1029, Blindern, N-0315 Oslo, Norway.\\ 
$^{8}$Argelander-Institut f\"ur Astronomie, Universit\"at Bonn, Auf dem H\"ugel 71, 53121 Bonn, Germany.\\
$^{9}$Department of Physics, Harvard University, 17 Oxford St., Cambridge, MA 01238, USA.\\
$^{10}$Institute for Advanced Study, Einstein Drive, Princeton, NJ 08540, USA.\\
$^{11}$University of Victoria, Elliott Building, 3800 Finnerty Rd, Victoria, BC, V8P 5C2, Canada.\\
$^{12}$Observatoire Midi-Pyr\'en\'ees, UMR5572, 14 Avenue Edouard Belin, 31000 Toulouse, France.\\
$^{13}$Jet Propulsion Laboratory, 4800 Oak Grove Drive, Pasadena, CA 91109, USA.\\ 
$^{14}$Leiden Observatory, P.O. Box 9513, NL-2300 RA, Leiden, The Netherlands. \\  
$^{15}$Department of Physics, University of California at Davis, One Shields Avenue, Davis, CA 95616, USA.\\
$^{16}$Department of Physics, Jadwin Hall, Princeton University, Princeton, NJ 08544, USA.\\
$^{17}$Institut d'Astrophysique de Paris, UMR7095 CNRS, Universit\'e Pierre~\&~Marie Curie - Paris, 98 bis bd Arago, 75014 Paris, France.\\
$^{18}$Observatoire de Paris - LERMA, 61 avenue de l'Observatoire, 75014 Paris, France.\\
$^{19}$INAF / Catania Astrophysical Observatory, via S.\ Sofia 78, 95123 Catania, Italy.\\
$^{20}$Department of Astronomy, Ohio State University, 140 W.\ 18th Avenue, Columbus, OH 43210, USA.\\
$^{21}$Isaac Newton Group of Telescopes, Calle Alvarez Abreu 70, 38700 Santa Cruz de la Palma, Spain.
\vspace{-9mm}}

\newcommand{\ie}{\mbox{\it i.e.}}
\newcommand{\eg}{\mbox{\it e.g.}}
\newcommand{\cf}{\mbox{\it c.f.}}
\newcommand{\etal}{\mbox{\it et~al.}}

\def\gs{\mathrel{\raise1.16pt\hbox{$>$}\kern-7.0pt 
\lower3.06pt\hbox{{$\scriptstyle \sim$}}}}         
\def\ls{\mathrel{\raise1.16pt\hbox{$<$}\kern-7.0pt 
\lower3.06pt\hbox{{$\scriptstyle \sim$}}}}         

\begin{document}

\pagerange{\pageref{firstpage}--\pageref{lastpage}} \pubyear{2006} \maketitle
\label{firstpage} \begin{abstract}  \indent The Shear TEsting Programme (STEP)
is a collaborative project to improve the accuracy and reliability of weak
lensing measurement, in preparation for the next generation of wide-field
surveys. We review sixteen current and emerging shear measurement methods in a
common language, and assess their performance by running them (blindly) on
simulated images that contain a known shear signal. We determine the common
features of algorithms that most successfully recover the input parameters. A
desirable goal would be the combination of their best elements into one ultimate
shear measurement method. In this analysis, we achieve previously unattained
discriminatory precision via a combination of more extensive simulations and
pairs of galaxy images that have been rotated with respect to each other. That
removes the otherwise overwhelming noise from their intrinsic ellipticities.
Finally, the robustness of our simulation approach is confirmed by testing the
relative calibration of methods on real data.


Weak lensing measurement has improved since the first STEP paper. Several
methods now consistently achieve better than $2\%$ precision, and are still
being developed. However, we can now distinguish all methods from {\it
perfect} performance. Our main concern continues to be the potential for a
multiplicative shear calibration bias: not least because this can not be
internally calibrated with real data. We determine which galaxy populations are
responsible and, by adjusting the simulated observing conditions, we also
investigate the effects of instrumental and atmospheric parameters. We have isolated several
previously unrecognised aspects of galaxy shape measurement, in which focussed
development could provide further progress towards the sub-percent level of
precision desired for future surveys. These areas include the suitable treatment
of image pixellisation and galaxy morphology evolution. Ignoring the former
effect affects the measurement of shear in different directions, leading to an
overall underestimation of shear and hence the amplitude of the matter power
spectrum. Ignoring the second effect could affect the calibration of shear
estimators as a function of galaxy redshift, and the evolution of the lensing
signal, which will be vital to measure parameters including the dark energy
equation of state.

\end{abstract}




\section{Introduction}

The observed shapes of distant galaxies become slightly distorted by the
(differential) gravitational deflection of a light bundle as it passes near
foreground mass structures. Such ``cosmic shear'' happens regardless of the
nature and state of the foreground mass.  It is therefore a uniquely powerful
probe of the cosmic mass distribution, dominated by dark matter. Observations of
gravitational lensing are directly and simply linked to theories of structure
formation that are otherwise ill-equipped to predict the distribution of light
\citep[for reviews, see][]{Bible,witrev,refrev}. Measurements are {\it not}
limited by astrophysical bias
\citep[\eg][]{oferbias,meg901,hoebias,smithbias,weinbias}, which affects optical
surveys, nor by unknown physics of distant supernov\ae\ 
\citep[\eg][]{sne_review,sne_variety,sne_environ,sne_metals}, nor by the uncertain relations between the mass
of galaxy clusters and their observable X-ray luminosity or temperature
\citep[\eg][]{hutw,pier,vlnew}. Gravitational lensing is a purely geometric effect,
requiring knowledge of only deflection angles and distances. By directly
observing the growth of the mass structures over cosmic time, and by
investigating the large-scale geometry of the universe, it is also an effective
probe of dark energy \citep{cfhtls_deep,cfhtls_wide,jarvis06,quintessence} and
can test alternative theories of gravity that move beyond general relativity
\citep{lensinggravity}.

The practical use of weak lensing in cosmology effectively began with the simultaneous
detection of a coherent cosmic shear signal by four independent groups
\citep{BRE,Kaiser_orig,vW_orig,Wittman_nature}. Since then, the field of weak lensing has
advanced dramatically. Large, dedicated surveys with ground- and space-based telescopes have
recently measured the projected 2D power spectrum of the large-scale mass distribution and
drawn competitive constraints on the matter density parameter $\Omega_m$ and the amplitude
of the matter power spectrum $\sigma_8$
\citep{Maoli,RRG01,vWb01,HYG02,BMRE,RRG02,Jarvis,MLB02,Hamana,Massey,RhodesSTIS,vWb04,HymzGEMS,Jarvis05,
cfhtls_wide,cfhtls_deep,Hettcs,gemscs2,dahleclusters}.
The results from these efforts are found to be in broad agreement and are rapidly becoming
more credible, with the most recent publications presenting several different diagnostic
tests to determine the levels of systematic error. Ambitious plans are being laid for
dedicated telescopes both on the ground (\eg\  VST-KIDS, DES, VISTA darkCAM, Pan-STARRS,
LSST) and in space (\eg\ DUNE, SNAP, JDEM). Indeed, future weak lensing surveys were
recently identified as the most promising route to understanding the nature of dark energy
by the joint NSF-NASA-DOE Astronomy and Astrophysics Advisory Committee (AAAC) and NSF-DOE
High Energy Physics Advisory Panel (HEPAP) Dark Energy Task Force\footnote{\tt
http://www.nsf.gov/mps/ast/detf.jsp}. The importance of weak lensing in future
cosmological and astrophysical contexts seems assured.

However, the detection and measurement of weak gravitational lensing presents a technical
challenge. The $\sim1\%$ distortion induced in the observed shapes of galaxies is an order
of magnitude smaller than their typical intrinsic ellipticities, and a similar factor
smaller than the spurious shape distortions created by convolution with the telescope's
point spread function (PSF). Correction for these effects is crucial and complex. To test
the reliability of weak lensing measurements, it has therefore been necessary since the
first detections to manufacture simulated images that closely resemble real data but contain
a known shear signal. \citet{Baconsims}, \citet{erben} and \citet{HYGBHI} ran their shear
measurement methods on such images. By comparing the input and mean measured shears, they
determined the calibration error inherent to each technique, and in some cases discovered
(and hence corrected) a multiplicative calibration bias. This is most important because it
cannot be self-calibrated from a survey itself. Other systematics can be checked for in real
data via correlation of the galaxies and the PSF, or via an $E$-$B$ decomposition
\citep{SchvWM02,CNPT02,EBring}. These early tests determined that the first successful shear
measurement methods were accurate to $\le10\%$ of the signal.

To maximise progress in this technical field, and to foster the exchange of data and
theoretical knowledge within the weak lensing community, we launched the Shear TEsting
Programme (STEP). In the first STEP paper, \citep[STEP1]{step1}, we parametrized the
performance of methods in terms of their multiplicative shear calibration bias $m$, an
additive residual shear offset $c$ and, in some cases, a nonlinear responsivity to shear
$q$. That analysis confirmed that the main difficulty in weak lensing lies in the
calibration of the shear signal, but encouragingly showed that all of the methods used on
existing weak lensing surveys achieve better than $\sim7\%$ accuracy. Shear measurement
error is therefore not currently a dominant source of error.

Unfortunately, this accuracy will not be sufficient to realise the potential of the
ambitious and much larger future surveys. STEP1 found that the most accurate shear
measurement methods were successfully calibrated to within a few percent, but the limited
size and precision of the first STEP simulations forbade any finer analysis than this. The
morphologies of galaxies in the first simulated images were also overly simplistic, in a way
that did not fully test the assumptions of some shear measurement methods that galaxies lack
substructure and complex shapes. 

In this second STEP paper, we include complex galaxy morphologies and conduct a
more precise test of current and developing shear measurement algorithms to the
$\le0.5\%$ level. We achieve this precision through the combination of a more
extensive set of simulated images and an ingenious use of galaxy pairs rotated
with respect to each other \citep{reiko}. This removes
the otherwise dominant noise from galaxies' intrinsic ellipticities. The new set
of simulated images has also been designed  to span a wide range of realistic
observing conditions and isolate several potentially challenging aspects of
shear calibration in which the accuracy of shear recovery may begin to
deteriorate. The data set is sufficiently large for it to be divided into
different simulated observing conditions and for independent tests to be carried
out within each. We thereby test the effects of the following parameters on
shear measurement precision: 

\newpage

\begin{itemize}
\item Complex galaxy morphology
\item Galaxy size
\item Galaxy magnitude
\item Selection effects related to galaxy ellipticity
\item Direction of the shear signal relative to the pixel grid
\item PSF size
\item PSF ellipticity
\end{itemize}

Sixteen different shear measurement codes have been run on the simulated images.
These can be categorised into four distinct categories. We provide a brief
description of each algorithm, and outline the relative successes of each
method. The STEP programme has dramatically sped the development of new shear
measurement methods \citep[\eg][Bridle \etal\ in
preparation]{shapelets2,bj02,shapelets3,shapeletskk,reiko}, and we particularly
focus on these. However, these methods necessarily remain experimental, and
development continues. The results from such methods should therefore be taken
as an indication of progress rather than a judgement on their ultimate
potential.

This paper is organised as follows.  In \S\ref{sec:data}, we describe the simulated images.
In \S\ref{sec:methods}, we review the different shear measurement methods used by each
author, translating them into a common language for ease of comparison, and categorising
them into four distinct groups. In \S\ref{sec:results}, we compare each author's measured
shear with the input signal, and split the simulations in various ways to isolate areas of
potential difficulty in shear measurement. Because of the number of different methods used,
this is a rather daunting process. In \S\ref{sec:discussion}, we provide some perspective on
the results, assessing the relative performance of the different methods, and the categories
of methods. In \S\ref{sec:conc}, we derive some general conclusions and outline suggestions
for future development.

\section{Simulated images}

\label{sec:data}

We have used the \citet{simage} simulation package to manufacture artificial
images that closely resemble deep $r$-band data taken in good conditions with
the {\it Suprime-Cam} camera on the {\it Subaru} telescope. We specifically
mimic the weak lensing survey data of \citet{subaruobs}. The {\it Subaru}
telescope was built with careful consideration of weak lensing requirements, and
has reliably obtained the highest quality weak lensing data to date
\citep[][Kasliwal \etal\ in preparation]{suprimecam,Wittman05}. It therefore
represents the current state-of-the-art, and will most closely match future
dedicated survey instruments. The simulated images are publicly available for
download from the STEP website\footnote{\tt
http://www.physics.ubc.ca/$\sim$heymans/step.html \label{url:step}}.

To aid the interpretation of our results, the simulated images incorporate
several ``unrealistic'' simplifications: neither the noise level, the input shear
signal nor the PSF vary as a function of position. This does not adversely
affect the validity of the results, as any combination of PSF size, PSF
ellipticity, and shear signal can usually be found in one of the images.
However, it does let us simply average the measured shear for the large number
of galaxies in each image, without explicitly keeping track of either the shear
or PSF applied to each object. As in STEP1, the main figure of merit throughout
our analysis will be the mean shear measured within each image,
$\langle\tilde\gamma\rangle$, and deviations of that from the known input shear
$\gamma^{\rm input}$. If the mean shear can be determined without bias for any input shear
(and for any PSF), all of the commonly-used statistics typical in cosmic shear
analysis should also be unbiased (but the distribution of the shear estimates
will affect their noise level).


\begin{table}
\begin{center}
\begin{tabular}{ccc}
Image set & PSF description & Galaxy type \\ 
\hline\hline 
 A & Typical Subaru PSF ($\sim0.6\arcsec$)	& shapelets \\
 B & Typical Subaru PSF ($\sim0.6\arcsec$)	& pure exponential \\
 C & Enlarged Subaru PSF ($\sim0.8\arcsec$)	& shapelets \\
 D & Elliptical PSF aligned along $x$-axis  & shapelets \\
 E & Elliptical PSF aligned at $45^\circ$   & shapelets \\
 F & Circularly symmetric Subaru PSF        & shapelets \\
\hline\hline 
\end{tabular}
\end{center}
\caption{The six different sets of images used in the STEP2 analysis are
carefully chosen to isolate and test particular aspects of weak shear
measurement. Either the PSF shape, or the form of galaxies' intrinsic
morphologies varies in a prescribed way between sets.} 
\label{tab:psfs}
\end{table}

To address the specific topics outlined in the introduction, we manufactured six
sets of simulated images. These span a range of realistic observing conditions,
in a carefully orchestrated way that will isolate various effects. The
differences between the images are described in table~\ref{tab:psfs}. Each set
contains 128\ $7\arcmin\times7\arcmin$ images, with a pixel scale of
$0.2\arcsec$. In the first simulated image of each set, the galaxies are not
sheared. For the next 63 images, which all feature the same patch of sky in
order to maximise sensitivity to shear calibration, the galaxies are sheared by
a random amount. This amount is chosen with a flat PDF within $|\gamma^{\rm
input}|<6\%$. To concentrate on cosmic shear measurement rather than cluster
mass reconstruction, this limit is smaller than the maximum shears used in
STEP1. However, the shears are now crucially chosen from a continuous
distribution and are allowed to be in any direction relative to the pixel grid.
Note that we are really attempting to measure ``reduced shear''
\citep{SeitzSch97} throughout this analysis, although there is explicitly zero
convergence in the simulations. The input signals were not disclosed to any of
the groups analysing the data.

We can predict the signal to noise ratio in the shear measurement from these images. We
first define a complex ellipticity for each galaxy

\begin{equation}
e = e_1 + i e_2 \equiv \frac{a-b}{a+b}\big(\cos{(2\theta)}+i\sin{(2\theta)}\big) ~,
\label{eqn:edefinition}
\end{equation}

\noindent where $a$ and $b$ are the major and minor axes, and $\theta$ is the orientation of
the major axis from the $x$-axis. This definition is widely used because it is more
convenient than a two-component parametrization involving $\theta$. Both the real and
imaginary parts are well-defined (zero) for a circular object or, on average, for an
unsheared population of objects. In the absence of PSF smearing and shear measurement
errors, the observed galaxy ellipticity $e^{\rm obs}$ is related to its intrinsic
ellipticity $e^{\rm int}$ by

\begin{equation}
e^{\rm obs} = \frac{e^{\rm int} + \gamma}{1 + \gamma^* e^{\rm int}} ~
\label{eqn:gstare}
\end{equation}

\noindent \citep{SeitzSch97}, where $\gamma\equiv\gamma_1+i\gamma_2$ is the complex shear
applied to each image. With only a finite number $N$ of galaxies, all with nonzero intrinsic
ellipticity, measurement of the mean shear $\langle\tilde\gamma\rangle = \langle e^{\rm
obs}\rangle$ is limited by an intrinsic shot noise

\begin{equation}
{\rm SN\ error\ } \approx \langle e^{\rm int}\rangle = 
0 \pm \sqrt{\frac{\langle (e_i^{\rm int})^2 \rangle}{N}} ~.
\label{eqn:gammasnorig}
\end{equation} 

\noindent In the STEP2 simulations, $\sqrt{\langle e_i^2 \rangle}\sim 0.1$, about an order
of magnitude larger than the shear signal.

Since the morphologies of the simulated galaxies are uncorrelated, this noise
can be slowly beaten down by increasing the size of the simulations. But to
dramatically improve the efficiency of the simulations, and circumvent the
meagre $1/\sqrt{N}$ behaviour, we introduce an innovation in the remaining 64
images. Following a suggestion in \citet{reiko}, the entire sky, including the
galaxies, was artificially rotated by $90^\circ$ before being sheared by the
same signals and being convolved with the same PSF as before. This rotation
flips the sign of galaxies' intrinsic ellipticites. To measure biases in shear
measurement methods, we can then consider matched pairs of shear estimators from
the unrotated and rotated version of each galaxy. Averaging these estimators
explicitly cancels the intrinsic shape noise, leaving only measurement noise and
any imperfections in shear measurement. We thus form a shear estimator for each
galaxy pair

\begin{equation}
\tilde\gamma = (e^{\rm obs,unrot} + e^{\rm obs,rot})/2 ~.
\end{equation}

\noindent Since $e^{\rm int,unrot}=e^{\rm int}=-e^{\rm int,rot}$, we can use
equation~(\ref{eqn:gstare}) to find

\begin{eqnarray}
\tilde\gamma &=&  
\left(\frac{ e^{\rm int}  + \gamma}{1 + \gamma^* e^{\rm int}} + 
      \frac{-e^{\rm int}  + \gamma}{1 - \gamma^* e^{\rm int}} \right)/2\nonumber\\
&=& 
\frac{\gamma - \gamma^*(e^{\rm int})^2}{1 - (\gamma^*e^{\rm int})^2} ~.
\end{eqnarray}

\noindent Averaging this shear estimator over $N/2$ galaxy pairs now gives a
shot noise error in $\langle\tilde\gamma\rangle$ of

\begin{equation}
{\rm SN\ error\ } \approx {\gamma \langle (e_i^{\rm int})^2 \rangle} = 
0 \pm \gamma \sqrt{\frac{\langle (e_i^{\rm int})^4 \rangle}{2N}} ~,
\label{eqn:gammasnnew}
\end{equation}

\noindent which has been significantly reduced from equation~(\ref{eqn:gammasnorig}). In the
STEP2 simulations $\sqrt{\langle (e_i^{\rm int})^4 \rangle} \sim 0.05$ and $|\gamma|<0.06$.
Nothing is lost by this approach. All 128 images can still be analysed independently -- and
we do pursue this approach in order to measure the total shape measurement noise in an
ordinary population of galaxies.\\


The \citet{simage} image simulation pipeline required extensive development from previously
published versions to mimic ground-based data. We shall therefore now describe its three
main ingredients: stars ({\it i.e.} PSF), galaxies and noise.

\begin{center}
\begin{figure}
\epsfig{file=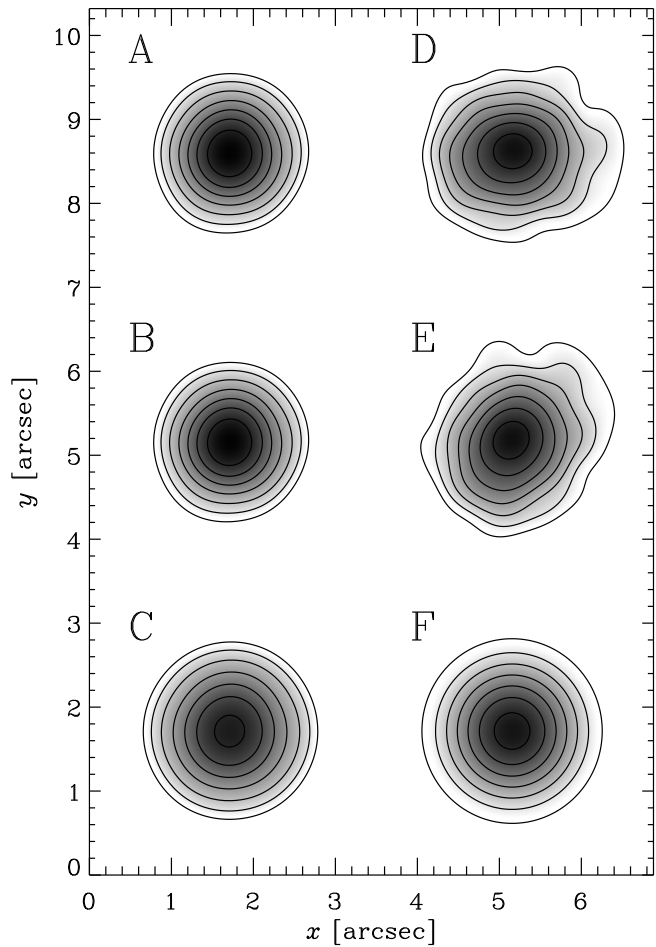,width=8.55cm,angle=0}
\caption{The point spread functions (PSFs) used to generate the six different
sets of simulated images. The colour scale is logarithmic, and the contours,
which are overlaid at the the same absolute value on each PSF, are spaced
logarithmically by factors of two.
They are designed to target specific aspects of weak lensing
measurement that could potentially prove difficult to control. See
table~\ref{tab:psfs} and the text for a description of each PSF.}
\label{fig:psfs}
\end{figure}
\end{center}

\subsection{Stars}

The simulated images are observed after convolution with a various point-spread
functions (PSFs). The PSF shapes are modelled on real stars observed in {\it
Suprime-Cam} images, and are shown in figure~\ref{fig:psfs}. They are modelled
using shapelets \citep{shapelets1,shapelets2,bj02,shapelets3}, a (complete) set
of orthogonal basis functions that can be used to describe the shape any
isolated object. The decomposition of an image into shapelet space acts rather
like a localised Fourier transform, with images $f({\mathbf x})$ being expressed
in shapelet space as a set of indexed coefficients $f_{n,m}$ that weight the
corresponding basis function

\begin{equation} \label{eqn:sseriesp}
f({\mathbf x}) = \sum_{n=0}^\infty \sum_{m=-n}^{n} f_{n,m}
\chi_{n,m}(r,\theta;\beta) ~,
\label{eqn:shapelets1}
\end{equation}

\noindent with $m\le n$, and where the Gauss-Laguerre basis functions are

\begin{eqnarray}
\chi_{n,m}(r,\theta;\beta) = 
  \frac{C_{n,m}}{\beta}
  \left(\frac{r}{\beta}\right)^{|m|}
  L_{\frac{n-|m|}{2}}^{|m|} \left(\frac{r^2}{\beta^2}\right)
  e^\frac{-r^{2}}{2\beta^2}
  e^{-im\theta} ~,
\label{eqn:shapelets2}
\end{eqnarray}

\noindent with a normalising constant $C_{n,m}$ and scale size $\beta$.

The PSFs can therefore take a complex form. They contain substructure, skewness
and chirality. In general, the ellipticity of their isophotes varies as a
function of radius. For computational efficiency, the shapelet series is
truncated at order $n_{\rm max}=12$. The limited wings and the rapid convergence
of the PSFs to zero at large radii compared to those used in STEP1 is {\it not}
a consequence of this truncation, but a confirmation of the excellent optical
qualities of {\it Suprime-Cam}.

PSF A is modelled from a fairly typical star towards the centre of a 40 minute
long {\it Suprime-Cam} exposure (which, in practice is likely to be assembled
from four 10 minute exposures). It has a full-width at half-max (FWHM) of
$0.6\arcsec$. PSF B is identical to PSF A. PSF C is the same star, but enlarged
to model slightly worse seeing, and has a FWHM of $0.8\arcsec$. This is the
worst that might be expected in future weak lensing surveys, with nights during
poorer conditions typically used to obtain data in additional colours. PSF D is
modelled on a star at the edge of the same {\it Suprime-Cam} exposure. The
phases of all of its $m=2$ shapelet coefficients were adjusted to the same value
so that at all radii (and therefore with any radial weight function), its
ellipticity derived from quadrupole moments points in exactly the same
direction. Substructure and skewness apparent in the real {\it Subaru} PSF is
otherwise untouched. As PSF D, the ellipticity is directed parallel to the
$x$-axis of the pixel grid. The star is rotated by $45^\circ$ to make PSF E. It
is an example of extreme ellipticity, which highlights ellipticity-dependent
effects. However, it might be possible to limit such ellipticity in weak lensing
surveys by improving the optical design of future telescopes or optimising
survey tiling and scheduling strategies. PSF F is a circularised version of that
star, obtained by setting all of its $m\ne 0$ shapelet coefficients to zero,
which is equivalent to averaging the PSF over all possible orientations.


\begin{center}
\begin{figure}
\epsfig{file=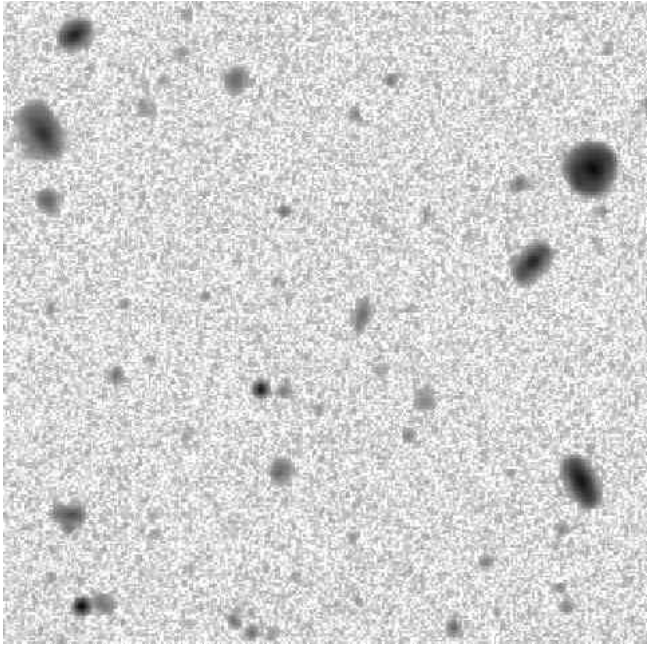,width=8.55cm,angle=0,clip=}
\caption{A $1\arcmin\times 1\arcmin$ section of a simulated image from 
set A, containing shapelet galaxies with complex morphologies. The colour scale
is logarithmic, and the same as that in figure~\ref{fig:idealised_image}.}
\label{fig:shapelet_image}
\end{figure}
\end{center}


\subsection{Shapelet galaxies}

\begin{center}
\begin{figure}
\epsfig{file=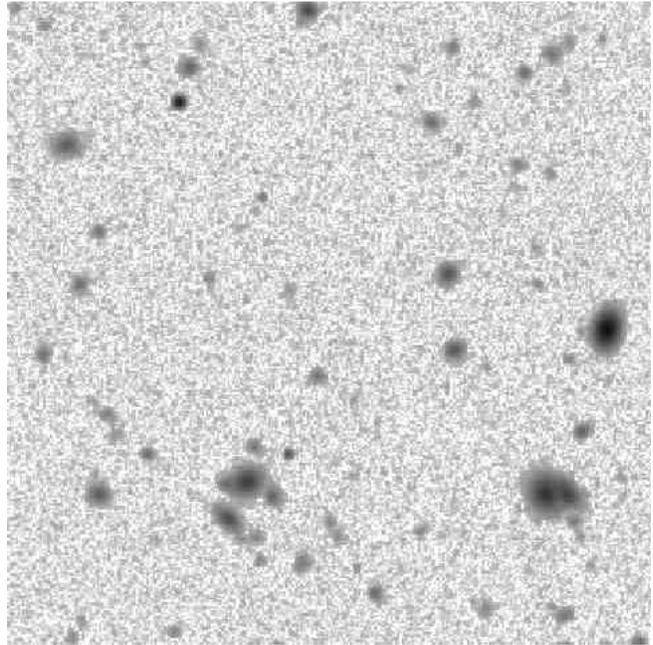,width=8.55cm,angle=0,clip=}
\caption{A $1\arcmin\times 1\arcmin$ section of a simulated image from 
set B, containing idealised galaxies with exponential radial profiles and
simple morphologies. The colour scale is logarithmic, and the same as that in 
figure~\ref{fig:shapelet_image}.}
\label{fig:idealised_image}
\end{figure}
\end{center}

Most of the simulated images contain galaxy shapes also constructed from weighted
combinations of the shapelet basis functions, using a version of the 
\citet{simage} image simulation pipeline similar modified to imitate ground-based
data. The complex and irregular galaxy morphologies that are possible using this
method represent an important advance from the STEP1 analysis using the {\tt
SkyMaker} image simulation package \citep{erben}. The measurement of weak
lensing in STEP1 was considerably simplified by the galaxies' smooth and
unperturbed isophotes. Several shear measurement methods are based on the
assumption that galaxy shapes and the PSF are concentric, elliptical, and in some
cases Gaussian. In addition, the {\tt SkyMaker} galaxies have reflection symmetry
about the centroid which could feasibly cause any symmetrical errors to vanish. By
contrast, PSF correction and galaxy shape measurement are rendered more challenging
in STEP2 by the realistic morphologies that include spiral arms, dust lanes and
small-scale substructure. Our analysis is thus designed to test the robustness of
weak lensing measurement methods. 

The joint size-magnitude-morphology distribution of galaxies was copied from the
{\it Hubble Space Telescope} COSMOS survey (Scoville \etal\ in preparation). This
is a uniform, two square degree set of images taken with the $F814W$ filter on the
{\it Advanced Camera for Surveys} (ACS), to a depth of $28.7$ for a point source at
$5\sigma$. It is deeper than our intended simulations, and with a much finer
resolution, so provides an ideal source population. The extent of the COSMOS survey
also provided sufficient real galaxies to avoid duplication in the simulations
without needing to perturb shapelet coefficients, as in section~4 of
\citet{simage}. We simply used the shapelet models of COSMOS galaxies, randomly
rotated, inverted and repositioned. The positions of galaxies in the simulations
were chosen at random, without attempting to reproduce higher-order clustering.

Since the galaxy models are inevitably truncated at some level in shapelet space, and since we
did not deconvolve the galaxies from the ACS PSF, the smallest simulated galaxies
are intrinsically slightly rounder than those in real {\it Subaru} data.
However, this convolution occurs before shearing and does not alter the necessary
steps for shear measurement. As in real data, the simulated galaxy ellipticity and
morphology distributions do vary with galaxy magnitude and size. We adopt an
alternative definition of ellipticity
\begin{equation}
\big(\varepsilon_1,~\varepsilon_2\big) \equiv \frac{a^2-b^2}{a^2+b^2}
\big(\cos{(2\theta)},~\sin{(2\theta)}\big)~,
\label{eqn:epsilondefinition}
\end{equation}
\noindent where $a$ and $b$ are the major and minor axes, and $\theta$ is the
orientation of the major axis from the $x$-axis. Note the difference from
equation~(\ref{eqn:edefinition}); this version is closer to the notation used by
most shear estimators. Before PSF convolution, the width of this ellipticity
distribution
\begin{equation}
\sigma^{\rm int}_\varepsilon \equiv \big((\sigma^{\rm int}_{\varepsilon_1})^2+(\sigma^{\rm int}_{\varepsilon_2})^2\big)^{1/2}
\end{equation}
\noindent as measured by {\sc SExtractor} \citep{SExt} is
$\sigma^{\rm int}_\varepsilon=0.35\pm0.03$ at $r=22$ and $\sigma^{\rm int}_\varepsilon=0.20\pm0.02$
at $r=26$. Note that this $\varepsilon$ is a different quantity than the $e$ used
in equation~(\ref{eqn:gammasnorig}).

The galaxies were then sheared analytically in shapelet space, using equation~(41) of
\citet{shapelets3}. This operation is to first order in $\gamma$. Terms of order $\gamma^2$
are ignored, but, for typical galaxy shapes, the coefficients by which these are multiplied
are also smaller than those multiplying the first order terms. This therefore introduces
only a very small error. The galaxies were then convolved with the PSF, also in shapelet
space, using equation~(52) of \citet{shapelets1}. They were pixellated by analytically
integrating the shapelet models within adjoining squares, using equation~(34) of
\citet{shapelets3}.

\subsection{Idealised galaxies} 

We have also manufactured one set (B) of simulated images with the same observing conditions
but in which the galaxies have simple, exponential profiles and concentric, elliptical
isophotes. These idealised galaxies provide a contrast to the morphological sophistication
of the shapelet galaxies, and an independent test of the shapelet-based shear measurement
methods. We intentionally chose a very simple form for the idealised galaxy shapes, with a
sharp cusp and extended wings, to most effectively pronounce any difference to the results
from galaxies with realistically complex morphologies. As before, the size-magnitude
distribution of unsheared galaxies was modelled on that observed in the ACS COSMOS images.
Galaxy ellipticities were assigned randomly from a Gaussian distribution. Like STEP1, we
used a constant distribution of intrinsic ellipticity. This had width $\sigma^{\rm
int}_\varepsilon=0.3$ for galaxies at all magnitudes. 


To add a shear signal, the random ellipticities are then perturbed at the
catalogue level. Under a small shear $\gamma_{i}$, the ellipticity $\varepsilon$
defined in equation~(\ref{eqn:epsilondefinition}) transforms as 

\begin{equation}
\varepsilon_{i}^{\rm obs} = \varepsilon_{i}^{\rm int}+
2(\delta_{ij}-\varepsilon_{i}^{\rm int}\varepsilon_{j}^{\rm int})
\gamma_{j} + {\cal O}(\gamma^3),
\label{eqn:epsilontransformation}
\end{equation}

\noindent (\eg\ \citet{RRG00}) where $\delta_{ij}$ is the Kroneker-delta symbol,
and the summation convention was assumed. Similarly, the mean square radius
$d\equiv a^2+b^2$ becomes

\begin{equation}
d^{\prime 2} = d^2 ( 1 + 2 \varepsilon_{i}^{\rm int} \gamma_{i} ) + {\cal O}(\gamma^2) ~.
\end{equation}

\noindent These two expressions are valid up to first order in the shear. Note
that, to this order, the flux $F$ is unaffected by a pure shear. These results are
valid for any galaxy with self-similar isophotes (as long as the moments
converge).

To create a simulated galaxy image $f({\mathbf x})$ with a desired ellipticity,
we first specify the desired size $r_0$ and mean radial profile $p(r^2)$, where
$r^2=x_{1}^2+x_{2}^2$ is the square radius and ${\mathbf x}=(x_{1},x_{2})$ are
Cartesian coordinates on the sky, centered on the centroid of the galaxy. For
convenience, we choose the normalisation and angular scale of the generic
profile such that

\begin{equation}
\label{eq:p_norm}
\iint p(r^{2}) ~{\mathrm d}^{2}{\mathbf x} =
\iint r^{2} p(r^{2})~{\mathrm d}^{2}{\mathbf x} = 1 ~.
\end{equation}

\noindent The exponential profile used in these simulations is given by

\begin{equation}
p(r^2)=\frac{\sqrt{6}}{2\pi r_0} e^{-\sqrt{6(r/r_0)^2}}
\end{equation}

\noindent (\cf\ Refregier 2000 for the alternative case of a Gaussian profile).
Using the conventions of equation~(\ref{eq:p_norm}) and a coordinate
transformation 

\begin{equation}
{\mathbf J} = 
{\mathbf R(\theta)}^{T}
\left( 
\begin{tabular}{cc}
$a^{2}$ & $0$ \\
$0$ & $b^{2}$ \\
\end{tabular}
\right)
{\mathbf R(\theta)}
= d^{2} \left( 
\begin{tabular}{c@{}c}
$1+\varepsilon_{1}$ & $\varepsilon_{2}$ \\
$\varepsilon_{2}$ &  $1-\varepsilon_{1}$ \\
\end{tabular}
\right) ~,
\end{equation}

\noindent where $^{T}$ denotes transpose and the rotation matrix

\begin{equation}
{\mathbf R(\theta)} \equiv \left( 
\begin{tabular}{cc}
$\cos \theta$ & $\sin \theta$ \\
$- \sin \theta$ & $\cos \theta$ \\
\end{tabular}
\right) ~,
\end{equation}

\noindent it is then easy to show that the elliptical galaxy image should have
surface brightness

\begin{equation}
f({\mathbf x})=
F|{\mathbf J}|^{-\frac{1}{2}} p({\mathbf x}^{T}{\mathbf J}^{-1}{\mathbf x}) ~,
\end{equation}

\noindent where the vertical bars denote the matrix determinant. The tails of their
exponential profiles were artificially truncated at elliptical isophotes $5\times r_0$ from
the centre. To pixellate the galaxies, the value of the analytic function was computed at
the centre of each pixel. The PSF was similarly pixellated, and convolution was then
performed in real space to produce the final image $I({\mathbf x})$. Strictly, these
operations should be reversed, and they do not commute. However, the pixels are small and
the PSFs are Nyquist sampled, so the error introduced should be minimal.

\subsection{Noise}

A two-component noise model is then superimposed onto the images. Instrumental
performance mimics that attained with a stack of four ten-minute exposures with
{\it Suprime-Cam} on the 8m {\it Subaru} telescope \citep{subaruobs}. They are
complete to $r=25.5$, and the galaxies selected for lensing analysis are likely
to have a median redshift $z_m\approx 0.9$. This is slightly deeper than most
existing weak lensing surveys, and is towards the deep end of ground-based
surveys planned for the future. The number density of useable galaxies found in
these simulated images is therefore unlikely to be greatly surpassed.

The first component of ``photon counting'' shot noise is first added to the true
flux in every pixel. This is drawn from a Gaussian distribution with a width
equal to the square root of the photon count. The images are then renormalised
to units of counts per second. In the renormalised images, the rms of the
Gaussian is 0.033 times the intensity in a pixel.

A second component of sky background is then added throughout each image, with
an rms of 4.43 counts per second. The DC background level is assumed to be
perfectly subtracted. The model {\it Subaru} images were combined using {\sc
Drizzle}, and the sky background noise is correlated in adjacent pixels. To
mimic this effect, we smoothed the sky noise component (but not the flux in
objects) by a Gaussian of FWHM 3.5 pixels. After this process, the rms of the
sky noise is 1.65 counts per second. A simulated image of a completely blank
patch of sky was also available to measure the covariance between pixels. The
correlated noise particularly affects the detection of small, faint objects, and
impedes the calculation of objects' weights from their detection S/N. It will be
instructive in the future to consider which image resampling kernels and
co-addition methods are optimal for shape measurement, or indeed whether we
should stack the data at all. \citet{Jarvis} suggest measuring galaxy
ellipticities on individual frames and combining these at the catalogue level.
Note that faint simulated galaxies are created to the depth of the COSMOS
survey, below the limiting magnitude of the simulated ground-based images, and
these unresolved sources will also add slightly to the overall sky background.

\begin{table}
\begin{center}
\begin{tabular}{l@{}cl}
\bf{Author}        & \bf{Key} & \bf{Method}  \\\hline\hline
Berg\'e            & JB       & Shapelets \citep{shapelets3} \\\hline
Clowe              & C1       & KSB+ (same PSF model used for all galaxies) \\\hline
Clowe              & C2       & KSB+ (PSF weight size matched to galaxies') \\\hline
Hetterscheidt      & MH       & KSB+ \\\hline
Hoekstra           & HH       & KSB+ \\\hline
Jarvis             & MJ~      & \citet{bj02} \\ \hline
Jarvis             & MJ2      & \citet{bj02} (new weighting scheme) \\ \hline
Kuijken            & KK       & Shapelets \citep{shapeletskk} \\\hline
Mandelbaum         & RM       & Reglens \citep{HirataSeljak03} \\\hline
Nakajima           & RN       & \citet{bj02} (deconvolution fitting) \\ \hline
Paulin-Henriksson  & SP       & KSB+ \\\hline
Schirmer           & MS1      & KSB+ (scalar shear susceptibility) \\\hline
Schirmer           & MS2      & KSB+ (tensor shear susceptibility) \\\hline
Schrabback         & TS       & KSB+ \\\hline
Semboloni          & ES1      & KSB+ (shear susceptibility fitted from population) \\\hline
Semboloni          & ES2      & KSB+ (shear susceptibility for individual galaxies) \\\hline\hline
\end{tabular}
\end{center}
\caption{Table of authors and their shear measurement methods. 
The key identifies the authors in all future plots and tables.} 
\label{tab:methods}
\end{table}

\section{Shear measurement methods}
\label{sec:methods}

Sixteen different shear measurement codes have been run on the simulated
images, by the authors listed in table~\ref{tab:methods}. Those that have been
used elsewhere on real data, attempt to preserve as similar a pipeline as
possible. Each method must first find and measure the shape of stars in each
image. It must interpolate the PSF shape across the field, without assuming that
it is constant. It must then find and measure the shapes of galaxies, correcting
them appropriately for the effects of seeing. Note that we still consider object
identification and classification to be part of a shear measurement method, as
shape biases can easily be introduced at this point
\citep[\eg][]{bj02,HirataSeljak03}; however, that task is likely to be separated
in future STEP projects.

All of the methods work by obtaining, for each galaxy, a two-component
polarisation $\varepsilon_i$ that behaves like a generalised ellipticity.
Precise definitions of polarisation vary between methods, but it is important to
note that easily measurable quantities do not usually change linearly with
applied shear, so that $\langle\varepsilon\rangle\ne\gamma^{\rm input}$  for all
values of $\gamma^{\rm input}$. To obtain an unbiased shear estimator, methods
must determine how their polarisations change under an applied shear, and
compute either a shear susceptibility tensor $P^\gamma_{ij}\equiv\delta
\varepsilon_i/\delta\gamma_j$ or a shear responsivity factor ${\cal R}$. These
are essentially interchangeable concepts, but with the word ``susceptibility''
used to imply measurement from the higher order shape moments of each galaxy
(which are then often averaged or fitted across a galaxy population), and the
word ``responsivity'' to mean an average susceptibility for the population,
measured from moments of the galaxy ellipticity distribution. In either case,
this quantity can be inverted, and used to form a shear estimator

\begin{equation}
\tilde\gamma\equiv(P^\gamma)^{-1}~\varepsilon 
\label{eqn:epsilonoverpgamma}
\end{equation}

\noindent or

\begin{equation}
\tilde\gamma\equiv\frac{\varepsilon}{{\cal R}} ~.
\label{eqn:epsilonoverr}
\end{equation}

\noindent When computing the mean shear from a limited subset of galaxies, such as those in one size or
magnitude bin, we shall investigate two approaches to the calculation of ${\cal R}$. We try using the
constant, global value, as has been done in published work, and we also try calculating ${\cal R}$ from
the statistics of the smaller population. The latter is more noisy, but takes into account the evolution
of galaxy morphology between samples (see \S\ref{sec:variable_calibration}).

In table~\ref{tab:method_classification}, the methods are broadly distinguished 
by their solutions to the two most important tasks in shear measurement. Some
methods correct for the PSF at the catalogue level, by essentially subtracting
the ellipticities of the PSF from that of each galaxy; others attempt to
deconvolve each galaxy from the PSF, and measure the ellipticity of a
reconstructed model. To obtain a polarisation, some  (``passive'') methods measure combinations
of galaxies' observed shape moments; other (``active'') methods shear a model of an intrinsically
circular source until it most closely resembles the observed galaxy. We shall
now provide a brief description of each method, starting in the top-left
quadrant of table~\ref{tab:method_classification}.  Since the STEP program has
dramatically sped the development of new shear measurement methods
\citep[][Bridle \etal\ in
preparation]{shapelets2,bj02,shapelets3,shapeletskk,reiko}, we shall
particularly concentrate on the latest developments in those algorithms.

\begin{table}
\begin{center}
\begin{tabular}{ll}
\begin{tabular*}{1mm}{l}
\rotatebox{90}{{\bf PSF correction scheme~~~~~~~~~~~~~~~~~~~~~~~~~}}
\end{tabular*}
&
\begin{tabular}{c|c|c|}
\multicolumn{3}{c}{\bf ~~~~~~~~~~~~~~~~~~~~~~~~~~~~~~~~~~Shear measurement method} 
\\ \cline{2-3} 
\\ \vspace{-6.7mm} ~ & ~ & ~ \\ 
\multicolumn{1}{c|}{\parbox[s]{18mm}{\centering \vspace{13mm} }}
&
\multicolumn{1}{|>{\columncolor[hsb]{0.96,0.35,1.0}}c|}{\bf Passive}
& 
\multicolumn{1}{|>{\columncolor[hsb]{0.59,0.35,1.0}}c|}{\bf Active}
\\ \hline
\multicolumn{1}{|>{\columncolor[gray]{0.9}}c|}{\parbox[t]{18mm}{ {\bf ~~Subtraction} }}
&
\multicolumn{1}{|>{\columncolor[hsb]{1.00,0.9,0.9}}c|}{\parbox[s]{22mm}{\centering ~ \\ KSB+ (various) \\ Reglens (RM) \\ ~RRG$^*$ ~~K2K$^*$ \\ ~Ellipto$^*$ }}
& 
\multicolumn{1}{|>{\columncolor[hsb]{0.63,0.75,0.9}}c|}{BJ02 (MJ, MJ2)}
\\ \hline
\multicolumn{1}{|>{\columncolor[hsb]{0.17,0.6,1.0}}c|}{\parbox[t]{18mm}{~\\ ~\\ {\bf Deconvolution}}}
& 
\multicolumn{1}{|>{\columncolor[hsb]{0.10,1.0,0.8}}c|}{\parbox[t]{22mm}{\centering ~\\ ~\\ Shapelets (JB)}}
& 
\multicolumn{1}{|>{\columncolor[hsb]{0.33,1.0,0.6}}c|}{\parbox[t]{22mm}{\centering ~ \\  Shapelets (KK) \\ BJ02 (RN) \\ ~im2shape$^*$ \\ ~ }}
\\ \hline
\end{tabular} \\
\end{tabular}
\end{center}
\caption{Broad classification scheme to distinguish different types of shear
measurement methods. Asterisks denote methods not tested in this paper.
The top-left quadrant is red; the top-right blue; the bottom-left orange; and
the bottom-right green.} 
\label{tab:method_classification}
\end{table}

\subsection{Red class methods}

\subsubsection{KSB+ (C1, C2, MH, HH, SP, MS1, MS2, TS, ES1 and ES2)}
\label{sec:KSBmeth}

The shear measurement method developed by \citet{KSB}, \citet{LK97} and
\citet{HFKS98} is in widespread use by many current weak lensing surveys. This
has led to a high level of optimisation of the basic method. The base {\sc
imcat} code is publicly available from the world wide web\footnote{\tt
http://www.ifa.hawaii.edu/$\sim$kaiser/imcat}. Many variations have been
developed, and the ten implementations tested in this paper represent a
cross-section of those that have been applied to real data. The details of each
method are compared fully in the appendix of STEP1. The differences that STEP2
results reveal to be particularly significant are summarised again in
table~\ref{tab:methods_details}.

The core of the method requires the measurement of the quadrupole moments of
each observed galaxy image $I({\mathbf x})$ weighted by a Gaussian of size
$r_g$. From these are formed a polarisation


\begin{equation}
\label{eqn:ksbe}
\big(\varepsilon_1,~\varepsilon_2\big)
  \equiv\frac{\iint I({\mathbf x}) ~W({\mathbf x})~r^2\big(\cos{(2\theta)},~\sin{(2\theta)}\big)~{\mathrm d}^2{\mathbf x}}
             {\iint I({\mathbf x}) ~W({\mathbf x})~r^2~{\mathrm d}^2{\mathbf x}} ~,
\end{equation}

\noindent where

\begin{equation}
W({\mathbf x})=e^{-r^2/2r_g^2} ~.
\end{equation}

The polarisation is corrected for smoothing of the PSF via the smear
susceptibility tensor $P^{\rm sm}$ and calibrated as shears via the shear
polarisability tensor $P^{\rm sh}$: both of which involve higher order shape
moments. Using stars to denote measurements from stars (for which a smaller
weight function is sometimes used) instead of galaxies, these form a shear
estimator

\begin{equation}
\tilde{\gamma} = 
 \left(P^\gamma \right)^{-1}
 \left[\varepsilon - P^{\rm sm}
 \left(P^{\rm sm\star} \right)^{-1}\,
         \varepsilon^{\star}\right] ~,
\label{eqn:ksbshearest}
\end{equation}

\noindent where

\begin{equation}
P^\gamma = P^{\rm sh} - P^{\rm sm} \left(P^{\rm sm \star} \right)^{-1} P^{\rm sh \star} ~.
\label{eqn:ksbPgamma}
\end{equation}

\noindent The tensor inversions can be performed in full, but these measurements
of faint objects are particularly noisy. In practice, since the diagonal
elements of $P^\gamma$ are similar, and its off-diagonal elements are about an
order of magnitude smaller, it can be approximated as a scalar quantity. Many
implementations of KSB+ therefore simply divide by a shear susceptibility
factor. The noise in $P^\gamma$ is also sometimes reduced by fitting it from the
entire population as a function of other observable quantities like galaxy size
and magnitude. Reducing noise in any nonlinear aspect of shear measurement is
vital, because the lensing signal is so much smaller than both the intrinsic
ellipticity and photon shot noise, and must be obtained by linearly averaging
away those sources of noise over a large population of galaxies.

Unfortunately, fundamental limitations in the mathematical formalism of KSB+ introduce
further decisions that must also be resolved to approximate an ideal scenario in practical
implementations. The KSB+ method makes no provision for the effects of pixellisation;
assumes that the PSF isophotes are concentric; and is mathematically ill-defined for
non-Gaussian or non-concentric PSF and galaxy profiles. The various implementations
developed by groups participating in the STEP2 analysis represent a cross-section of those
choices.

Since STEP1, the TS method has incorporated a shear calibration factor of
$0.91^{-1}$, determined from the STEP1 results, but without knowledge of the
STEP2 data. STEP2 therefore tests the robustness of this sort of calibration. As
in STEP1, the C1 and C2 methods incorporate a calibration factor of $0.95^{-1}$
to eliminate the effect of close galaxy pairs. The C1 method uses a constant
model of the PSF for all galaxies; the C2 method lets the size of the weight
function $r_g^\star=r_g$ change to match each galaxy. The new SP method
numerically integrates weight functions within pixels, uses the trace of
$P^\gamma$ from individual galaxies, and similar galaxy weights to the HH
method. The ES1 method is based upon the LV method from STEP1 but, rather than fitting
the shear susceptibility from the galaxy population as a function of size and
magnitude, it finds the twenty most similar galaxies in terms of those
parameters, and uses their average value. This same procedure was used in the
\citet{cfhtls_deep} analysis of the CFHTLS deep survey. Subsequent tests on
STEP1 images suggested that better results could be obtained by using individual
measurements of $P^\gamma$ from each galaxy, and ignoring the galaxy weights.
These improvements have been incorporated into the new ES2 method.

One final finesse is required for methods that use weights $w_i$ on each galaxy $i$ that
could vary between the rotated and unrotated images. For all $N$ pairs of galaxies, we
determine normalised weights

\begin{equation}
w_i^\prime = \frac{N\,w_i}{\sum_{j=1}^{N} w_j} 
\end{equation}

\noindent and then calculate three estimates of the mean shear in each image

\begin{eqnarray}
\label{eqn:gammaunrot}
\langle\tilde\gamma^{\rm unrot}\rangle &=& \frac{1}{N}\sum (w^{{\rm unrot} \prime} e^{\rm obs,unrot}) \\
\label{eqn:gammarot}
\langle\tilde\gamma^{\rm rot}\rangle   &=& \frac{1}{N}\sum (w^{{\rm rot}   \prime} e^{\rm obs,rot})   \\
\label{eqn:gammamatch}
\langle\tilde\gamma\rangle             &=& \frac{1}{2N}\sum (w^{{\rm unrot} \prime} e^{\rm obs,unrot} \,+\, 
                                              w^{{\rm rot}   \prime} e^{\rm obs,rot}) ~.
\end{eqnarray}

\noindent Errors on these are estimated using a bootstrap technique.


\begin{table*}
\begin{center}
\begin{tabular}{cll@{}cl}
\bf{Author} & \bf{Pixellisation} & \bf{Galaxy weighting scheme} & \bf{Cal$^{\rm{\bf n}}$ factor} & \bf{Shear susceptibility}  \\\hline\hline
JB & Analytic integration  & None                                                & ---    & Global mean shear responsivity ${\cal R}=2-\langle\varepsilon^2\rangle$ \\\hline
C1 & Centre of pixel       & $\mathrm{min}(\nu,40)$                              & 1/0.95 & $\frac{1}{2}$Tr[$P^\gamma$], fitted as $f(r_g,\varepsilon_i)$ \\\hline
C2 & Centre of pixel       & $\mathrm{min}(\nu,40)$                              & 1/0.95 & $\frac{1}{2}$Tr[$P^\gamma$], fitted as $f(r_g,\varepsilon_i)$ \\\hline
MH & Numerical integration & $1/(0.15+\sigma_\varepsilon^2 + \sigma(\frac{1}{2}{\rm Tr}[P^\gamma(r_g)])^2 )$ & 1/0.88 & $\frac{1}{2}$Tr[$P^\gamma$], from individual galaxies \\\hline
HH & Numerical integration & $1/(\sigma_\varepsilon^2 + s_\varepsilon^2/((1-\frac{\varepsilon^2}{2})\frac{1}{2}{\rm Tr}[P^\gamma])^2 )$ & ---    & $(1-\frac{\varepsilon^2}{2})\frac{1}{2}$Tr[$P^\gamma$], fitted as $f(r_g)$ \\\hline
MJ~~~& Centre of pixel     & $1/\sqrt{\varepsilon^2+2.25s_\circ^2}$              & ---    & Global mean shear responsivity ${\cal R}$ \\\hline
MJ2& Centre of pixel       & $1/s_\circ^2$                                       & ---    & Global mean shear responsivity ${\cal R}$ \\\hline
KK & Centre of pixel       & $1/(0.1^2+\sigma_{e_1}^2+\sigma_{e_2}^2)$           & ---    & Global mean shear responsivity ${\cal R}=1-\langle\varepsilon^2\rangle$ \\\hline
RM & Centre of pixel       & $f(S/N)$                                            & ---    & Global mean shear responsivity ${\cal R}$ \\\hline
RN & Centre of pixel       & $1/\sqrt{\varepsilon^2+2.25s_\circ^2}$               & ---    & Global mean shear responsivity ${\cal R}$ \\\hline
SP & Numerical integration & $1/(0.15+\sigma_\varepsilon^2 + \sigma(\frac{1}{2}{\rm Tr}[P^\gamma(r_g)])^2 )$ & --- & $\frac{1}{2}$Tr[$P^\gamma$], Individual galaxies \\\hline
MS1& Numerical integration & $1/\sigma_\varepsilon^2(r_g,\mathrm{mag})$          & ---   & $\frac{1}{2}$Tr[$P^\gamma$], fitted as $f(r_g,\mathrm{mag})$ \\\hline
MS2& Numerical integration & $1/\sigma_\varepsilon^2(r_g,\mathrm{mag})$          & ---    & Full $P^\gamma$ tensor, fitted as $f(r_g,\mathrm{mag})$ \\\hline
TS & Numerical integration & None                                                & 1/0.91 & $\frac{1}{2}$Tr[$P^\gamma$], from individual galaxies \\\hline
ES1& Numerical integration & 1/($\sigma_\varepsilon^2(r_g,\mathrm{mag})+0.44^2$) & ---    & $\frac{1}{2}$Tr[$P^\gamma$], smoothed from galaxy population $f(r_g,\mathrm{mag})$ \\\hline
ES2& Numerical integration & None                                                & ---    & $\frac{1}{2}$Tr[$P^\gamma$], from individual galaxies \\\hline
\end{tabular}
\end{center}
\caption{Choices adopted by each of the shear measurement methods that 
significantly affect their performance in this paper. See the appendix in STEP1
for more details about the differences between the various implementations of 
KSB+.} 
\label{tab:methods_details}
\end{table*}

\subsubsection{Reglens (RM)}
\label{sec:rm}

The Reglens (RM) method consists of two parts: the SDSS data processing pipeline {\sc Photo}
\citep{sdssphoto}, followed by the re-Gaussianization pipeline \citep{HirataSeljak03,MandelbaumHirata}.
The magnitude cut was adjusted, and one additional subroutine was required for the STEP2 analysis, to
properly determine the noise variance in the presence of correlated background noise. The STEP2 images are
more crowded than SDSS images, leading to occasional deblending problems. Objects with failed deblending
were automatically eliminated, after visual inspection indicated that nearly all of them were really
several galaxies very close to each other.

PSF correction is performed via a two-step procedure that addresses KSB+'s limitation of being exact only in
the limit of Gaussian PSF and galaxy profile. The PSF is first split into a Gaussian component $G({\mathbf
x})$ plus a small residual $\epsilon({\mathbf x})$, so that the observed image

\begin{equation}
I = (G + \epsilon) \otimes f = G \otimes f + \epsilon \otimes f ~,
\end{equation}

\noindent where $f({\mathbf x})$ is the galaxy image before convolution of the PSF, and
$\otimes$ signifies convolution. Assuming knowledge of $f$, it would be possible 
to find

\begin{equation}
\label{eqn:regauss} 
I^\prime \equiv G \otimes f = I - \epsilon \otimes f ~,
\end{equation}

\noindent the galaxy image as it would appear when convolved with a perfectly
Gaussian PSF.  Although $f$ is not known in practice, it is convolved with a
small correction $\epsilon$ in the final equality, so 
equation~(\ref{eqn:regauss}) is fairly accurate even with an approximation
$f_0$. The SDSS and STEP2 analyses used an elliptical Gaussian as $f_0$, with
its size and ellipticity determined from the difference between the best-fit
Gaussians to the observed image and the full PSF. Possible alternatives to this
approximation are discussed in \citet{HirataSeljak03}.

Correction for the isotropic part of the now Gaussian PSF reqires a subtraction
similar to that in KSB+ equation~(\ref{eqn:ksbshearest}), except that Reglens
directly subtracts moments of the PSF from those of the galaxy (\ie\ the
numerator and denominator of equation~(\ref{eqn:ksbe})) before they are divided
(\ie\ the ratio in equation~(\ref{eqn:ksbe})). Furthermore, the moments are
calculated using weight functions $W_{I^\prime}({\mathbf x})$ and
$W_{G}({\mathbf x})$ that are the best-fitting elliptical Gaussians to the image
and to the PSF respectively. The advantage of these adaptive weight functions is
that they do not bias the shape measurement or require later correction.
Correction for the anisotropic part of the Gaussian PSF is finally performed by
shearing the coordinate system, including $I^\prime$, until $G$ is circular.

In the absence of galaxy weights, a shear estimate for each galaxy would be
computed via equation~(\ref{eqn:epsilonoverr}). The shear responsivity

\begin{equation}
{\cal R} = 2 - \sigma_\varepsilon^2 \equiv
2 - \left\langle\varepsilon_1^2
  + \varepsilon_2^2 
  -  s_{\varepsilon_1}^2
  -  s_{\varepsilon_2}^2\right\rangle ~,
\end{equation}

\noindent is calculated from shape distribution statistics of the entire galaxy
population and the error on each polarisation, $s_{\varepsilon_i}$, is
calculated by propagating measured photon shot noise in the image. During our
analysis, it became apparent that, for the RM, MJ, MJ2 and RN methods, it
is necessary to recalculate ${\cal R}$ in each bin of galaxy size or magnitude
when the catalogue is so split.

To improve the signal to noise, galaxies are each weighted by a factor

\begin{equation}
w = \frac{1}{\sigma_\varepsilon^2 + s_{\varepsilon_1}^2} ~.
\end{equation}

\noindent An estimate of the mean shear in each image is then simply

\begin{equation}
\langle\tilde{\gamma}\rangle = \sum{w\frac{\varepsilon}{{\cal R}}}~\Big/~\sum{w} ~,
\label{eqn:eover2r}
\end{equation}

\noindent with a shear responsivity \citep{bj02}


\begin{equation}
{\cal R} = 
\sum w\left(2-2k_0-k_1|\varepsilon|^2\right)
~\Big/~\sum w ~,
\label{eqn:responsivity}
\end{equation}

\noindent where $k_0=\sigma_\varepsilon^2-w\sigma_\varepsilon^4$ and
$k_1=w^2\sigma_\varepsilon^4$.

Note that this calculation of ${\cal R}$ in the STEP2 images is much more
uncertain than in SDSS data, because the correlated background noise in the
STEP2 images is not as well understood. Consequently, this may introduce some
bias into the STEP2 results that does not exist with the real data.


%

\subsubsection{Other methods not tested in this paper}

\citet[][RRG]{RRG00} is a modification of the KSB+ method for space-based data in which the PSF is small. In
this limit, $\varepsilon^\star$ becomes noisy. Like Reglens, RRG therefore deals directly with moments rather than
polarisations for as long as possible, and performs the subtraction before the division. The moments use a
circular weight function, and therefore require correction for this truncation as well as the PSF. RRG uses a
global shear responsivity ${\cal R}\approx2-\langle\varepsilon^2\rangle$.

\citet[][K2K]{K00} also seeks a resolution of the Gaussian PSF limitation in
KSB+. The galaxy image is first convolved by an additional ``re-circularising
kernel'', which is a modelled version of the observed PSF that has been rotated
by $90^\circ$. PSF correction and shear measurement is thereafter fairly similar
to KSB. However, particular efforts are made to correct biases that arise from
the use of $P^\gamma$ measured after shear rather than before shear.

Ellipto \citep{ellipto} also uses a re-circularising kernel to eliminate the anisotropic component of the
PSF, following \citep{FischerTyson97}. It then repeats object detection to remove PSF-dependent selection
biases. Galaxy polarisations are derived from moments weighted by the best-fit elliptical Gaussian. It is a
partial implementation of BJ02, discussed in the next section, and primarily differs from BJ02 by
using a simpler re-circularising kernel.

\subsection{Blue class methods}

\subsubsection{BJ02 (MJ and MJ2)}
\label{sec:BJ02MJ}

The remaining methods are based upon expansions of the galaxy and PSF shapes
into Gauss-Laguerre (``shapelet'') basis functions. The JB and KK methods use
them with a circular basis function, as defined in
equations~(\ref{eqn:shapelets1}) and (\ref{eqn:shapelets2}), while the MJ, MJ2
and RN methods use more general elliptical versions. Shapelets are a natural
extension of KSB+ to higher order. The first few shapelet basis functions are
precisely the weight functions used in KSB+, with $r_g$ reinterpreted as the
shapelet scale size $\beta$. Generalised versions of the $P^{\rm sh}$ and
$P^{\rm sm}$ matrices are derived in \citet{shapelets2}. Extending the basis set
to higher order than KSB+ allows complex shapes of galaxies and PSFs to be well
described, even when the ellipticity varies as a function of object radius. The
shapelet basis set is mathematically well-suited to shear measurement because of
the simple transformation of shapelet coefficients during typical image
manipulation.

The two Jarvis (MJ, MJ2) methods correct for the anisotropic component of the
PSF by first convolving the image with an additional, spatially-varying kernel
that is effectively $5 \times 5$ pixels. This convolution is designed to null
both the Gaussian-weighted quadrupole of the PSF as well as its next higher
$m=2$ shapelet coefficient (since it is the $m=2$ components of the PSF that
mostly affects the observed shapes of galaxies). For PSF ellipticities of order
$\sim 0.1$ or less, a $5 \times 5$ pixel kernel is sufficient to round a typical
PSF up to approximately 50 pixels in diameter: much larger than the PSFs used in
this study.

The shapelet basis functions are sheared, to make them elliptical, then
pixellated by being evaluated at the centre of each pixel. Shapelet coefficients
$f_{n,m}=0$ are determined for each galaxy in distorted coordinate systems, and
the polarisability $\varepsilon$ is defined as $-1$ times the amount of distortion that makes
each object appear round (\ie\ $f_{2,2}=0$). Some iteration is required to get
this measurement to converge. In the distorted coordinate frame where the galaxy
is round, the weight function for this coefficient is a circular Gaussian of the
same size as the galaxy. Matching the shape of the weight function to that of
the galaxy has the advantage that the polarisability no longer requires
correction for truncation biases introduced by the weight function.

Finally, a correction for the PSF dilution (the circularising effect of the PSF)
is applied by also transforming the PSF into this coordinate system, then using
formul\ae\ proposed by \citet{HirataSeljak03}. 

%
%

The two methods (MJ, MJ2) differ only in the weights applied to each galaxy. The MJ method
is identical to the MJ method used for the STEP1 study. It uses weights

\begin{equation}
w_{\rm MJ} = \frac{1}{\sqrt{e^2 + 2.25 s_\circ^2}} ~, \label{eqn:mjweight}
\end{equation}

\noindent where $s_\circ$ is the uncertainty in the polarisability due to image
shot noise, as measured in the coordinate system where the galaxy is round. 
STEP1 revealed that this optimised weight gave incorrect responsivities as the
input shear became large ($\approx 0.1$).  For this study, method MJ2 was
therefore added, which is identical except that it uses weights that are not a
function of the galaxies' polarisations

\begin{equation}
w_{\rm MJ2} = \frac{1}{s_\circ^2} ~. \label{eqn:mj2weight}
\end{equation}

\noindent These weights should be less biased for larger input shears.
The MJ weight might be more appropriate for cosmic shear
measurements, and the MJ2 weight for cluster lensing.

The shear responsivity ${\cal R}$ for the MJ2 method is the same as that in
equation~(\ref{eqn:responsivity}). For the ellipticity-dependent weight used by
the MJ method, this is generalised to

\begin{equation}
{\cal R} \equiv 
\frac{\sum \Big[ w\left(2-2k_0-k_1|\varepsilon|^2\right)
+ \varepsilon\frac{\partial w}{\partial\varepsilon}(1-k_0-k_1|\varepsilon|^2)
\Big]}{\sum w} ~,
\label{eqn:responsivityfull}
\end{equation}

\noindent where the summations are over the entire galaxy population, or for
each size or magnitude bin. For either method, an estimate of the mean shear in
each image is then

\begin{equation}
\langle\tilde\gamma\rangle = \sum{w\frac{\varepsilon}{{\cal R}}}~\Big/~\sum{w} ~.
\label{eqn:eoverr}
\end{equation}

\noindent Note that, in the absence of shape noise,
equation~(\ref{eqn:responsivityfull}) reproduces the extra $(1-\varepsilon^2/2)$
term multiplying $P^\gamma$ in the HH implementation of KSB+ (see
table~\ref{tab:methods_details}).

\subsection{Orange class methods}

\subsubsection{Shapelets (JB)}
\label{sec:JBmeth}

The Berg\'{e} (JB) shear measurement method uses a parametric shapelet model to
attempt a full deconvolution of each galaxy from the PSF. Deconvolution is an 
ill-defined operation in general, since information is irrevocably lost during
convolution. In shapelet space, however, it is easy to restrict the galaxy model
to include only that range of physical scales in which information is expected
to survive. \citet{shapelets3} describes an iterative algorithm designed to
optimise the scale size of the shapelets and to thus capture the maximum range
of available scales for each individual galaxy. A complete software package to
perform this analysis and shapelet manipulation is publicly available from the
shapelets web site\footnote{\tt
http://www.astro.caltech.edu/$\sim$rjm/shapelets}.

To model a deconvolved galaxy shape, the basis functions are first convolved
with the PSF in shapelet space, then integrated analytically within pixels: thus
undergoing the same processes as real photons incident upon a CCD detector. The
convolved basis functions are then fit to the data, with the shapelet
coefficients as free parameters. Reassembling the model using {\it un}convolved
basis functions produces a deconvolved reconstruction of each galaxy. This
performs better than a Wiener-filtered deconvolution in Fourier space, because
shapelets have a preferred centre. The available basis functions act as a prior
on the reconstruction, localising it in real space (and also allowing a slightly
higher resolution at the central cusp than at large radii). The deconvolved
model can also be rendered free of noise by ensuring that a sufficient range of
scales are modelled to lower the residual $\chi^2_{\rm reduced}$ to exactly
unity. Unfortunately, achieving exactly this target is hindered by the presence
of correlated background noise in the STEP2 simulations. Incorporating the noise
covariance matrix is mathematically trivial but computationally unfeasible, and
a practical implementation has not yet been developed. Proceeding regardless,
the shape of this analytic model can be directly measured 
\citep[see][]{polshapetest,shapelets4}, including its {\it unweighted} moments.
These can not be measured directly from real data because observational noise
prevents the relevant integrals from converging. 

Once a deconvolved model is obtained, extraction of a shear estimator is easy.
It could mimic the KSB method. However, removing the weight function (like the
Gaussian in equation~(\ref{eqn:ksbe})), makes the polarisation itself into an
unbiased shear estimator

\begin{equation}
\tilde\gamma=
  \frac{\iint f({\mathbf x}) ~r^2~\big(\cos{(2\theta)},~\sin{(2\theta)}\big)~{\mathrm d}^2{\mathbf x}}
       {\iint f({\mathbf x}) ~r^2~{\mathrm d}^2{\mathbf x}} ~.
\end{equation}

\noindent The numerator of this expression has a shear susceptibilty equal to
the denominator. But that denominator is a scalar quantity, with explicitly 
zero off-diagonal elements in the susceptibility tensor, which can therefore be
easily inverted. It is also a simple product of a galaxy's flux and size, both
low-order quantities that can be robustly measured. The method is intended to be
completely linear for as long as possible, and to introduce minimal bias for
even faint objects in this final division. Since the denominator also changes
during a shear, a population of galaxies acquires an overall shear responsivity
factor
\begin{equation}
\mathcal{R}=2-\langle\varepsilon^2\rangle ~.
\end{equation}

The method is still under development. The shear responsivity factor has
currently been calculated only from the entire galaxy population. No weighting
scheme has yet been applied to the shear catalogue when calculating mean
shears. Once galaxies have passed crude cuts in size, flux, and flags (which
indicate successful convergence of the shapelet series and of the iteration),
they are all counted equally. These aspets will be improved in the future.


\subsection{Green class methods}

\subsubsection{Shapelets (KK)}

The Kuijken (KK) shear measurement method assumes that each galaxy was intrinsically circular, then shears
it, and smears it by the PSF, until it most closely matches the observed image. The shear required is the
stored as the polarisation $\varepsilon$. As described in \citet{shapeletskk}, this approach is desirable,
because it is understood precisely how a circular object changes under a shear. 

This process could operate in real space; however, the convenient properties of
shapelets make the required image manipulations easier and faster in shapelet
space. The pixellated image need be accessed only once, when each galaxy is
initially decomposed into shapelets (without deconvolution). Models of circular
sources can have arbitrary radial profiles, parametrized by shapelet
coefficients with $m=0$ and $n\le12$. This is sheared in shapelet space to first
order in $\gamma$, although, in principle, this could also be increased to
accommodate more highly elliptical objects. Also in shapelet space, it is
smeared by a model of the PSF. Since there is only one shapelet decomposition
overall, and one forward convolution for each object, the code is much faster
than the Berg\'{e} (JB) method. Furthermore, the decomposition uses completely
orthogonal shapelet basis functions, so the errors on shapelet coefficients are
also uncorrelated at that stage. To avoid iterating the decomposition, the
optimum scale size $\beta$ for each object is approximated from {\sc SExtractor}
parameters, and the range of scales is fixed in advance. In the current
implementation, the basis functions are evaluated at the centre of each pixel.
Since both the PSF and the galaxy are pixellated, its effects ought to drop out.
In terms of the orthogonality of the shapelet basis functions, this approach is
satisfactory as long so the range of scales is small, and oscillations in the
basis functions remain larger than the pixel scale \citep[\cf][]{shapeletsrb}.

To determine the shear required to make a circular source match each real
galaxy, a fit is performed using a numerical recipes Newton-Raphson algorithm,
which is quadratic in shapelet coefficients, the centroid and the shear. Since
the galaxies are not really all circular, in practice the global population does
have a non-trivial shear susceptibility or ``responsivity'' ${\cal R}$. For an
ensemble population of galaxies, this is a scalar quantity. As can be deduced
from equation~(\ref{eqn:epsilontransformation}), it involves the variance of the
intrinsic polarisation distribution

\begin{equation}
{\cal R}\equiv 1-\langle e^2\rangle ~.
\end{equation}

\noindent Unlike other methods that use a shear responsivity correction, this
quantity was calculated only once for the KK method, from the entire galaxy
population. However, the calculation of $\langle e^2\rangle$ properly takes into
account the galaxy weights

\begin{equation}
\langle e^2\rangle = 
   \frac{\sum\left[w(e_1^2+e_2^2-s_{e_1}^2-s_{e_2}^2)\right]}{\sum w}
 - \left(\frac{\sum w(e_1+e_2)}{\sum w}\right)^2 ,
\end{equation}

\noindent where $s_{e_i}$ is the noise on each polarisation calculated
by propagating photon shot noise, and the weight for each galaxy is

\begin{equation}
w=\frac{1}{(\sigma_e^{\rm int})^2 + s_{e_1}^2 + s_{e_2}^2 } ~.
\end{equation}

\noindent Note that the estimates of errors on the polarisations did not take into account
the fact that the background noise was correlated between adjacent pixels, and
are therefore likely to be underestimated.

Shear estimates for individual galaxies are then computed similarly to equation~(\ref{eqn:eoverr}),
but where $\tilde{\gamma}\equiv e/{\mathcal R}$ here.

\subsubsection{BJ02 (RN)}
\label{sec:BJ02RN}

The ``deconvolution fitting method'' by Nakajima (RN) implements nearly the full
formalism proposed by BJ02, which is further elaborated in \citet{reiko}. Like
MJ and MJ2, it shears the shapelet basis functions until they match the
ellipticity of the galaxy. The amount of distortion that makes an object appear round (\ie\ 
$f_{2,2}=0$) defines the negative of its polarisability $\varepsilon$. 

Since no PSF interpolation scheme has yet been developed, the
pipeline deviates from the STEP rules by using prior knowledge that the PSF is
constant across each image (but not between images).
Deconvolution from the PSF is performed in a similar fashion to
the JB method. The Gauss-Laguerre basis functions are convolved with the PSF to
obtain a new basis set. These are evaluated at the centre of each pixel. The new
basis functions are fitted directly to the observed pixel values, and should
fully capture the effect of highly asymmetric PSFs or galaxies, as well as the
effects of finite sampling. The fit iterates until a set of sheared
Gauss-Laguerre basis functions are obtained, in which the coefficients
$f_{2,0}=f_{2,2}=0$ and hence the deconvolved galaxy appears round. All PSF
coefficients were obtained to $n\le12$, and galaxy coefficients to $n\le8$.

The weights applied to each galaxy are optimised for small shears, using the
same prescription as the MJ2 method in equation~\ref{eqn:mj2weight}. The shear
responsivity ${\cal R}$ is similarly calculated using
\ref{eqn:responsivityfull}, averaged over the entire galaxy population or within
size and magnitude bins as necessary.

The evolution of the RN method during the STEP2 analysis highlights the utility
of even one set of STEP simulations. In the first submission, it was noticed
that a few outlying shear estimates in each field were destabilising the result.
These were identified as close galaxy pairs, so an algorithm was introduced to
remove these, and the size and magnitude cuts were also gradually adjusted over
several iterations to improve stability.

\subsubsection{Other methods not tested in this paper}

Im2shape \citep{im2shape} performs a similar PSF
deconvolution, but parametrizes each galaxy and each PSF as a sum of elliptical Gaussians.
The best-fit parameters are obtained via a Markov-Chain Monte-Carlo sampling
technique. Concentric Gaussians are usually used for the galaxies, in which case the
ellipticity is then a direct measure of the shear via equations~(\ref{eqn:edefinition}) 
and (\ref{eqn:gstare}). For alternative galaxy models using non-concentric Gaussians, shear 
estimators like that of the JB method could also be adopted. The ``active'' or ``passive''
classification of this method is somewhat open to interpretation.


\section{Results}
\label{sec:results}

Individual authors downloaded the simulated images and ran their own shear
measurement algorithms, mimicking as closely as possible the procedure they
would have followed with real data. None of the authors knew the input shears at
this stage. Their galaxy catalogues were then compiled by Catherine Heymans and
Richard Massey. Independently of the other authors, the mean shears in each
image were compared to the input values. Galaxies in the measured
catalogues were also matched to their rotated counterparts and to objects in the
input catalogues, with a $1\arcsec$ tolerance. Except for determining false
detections or stellar contamination in the measured catalogues (which were
removed in the matched catalogues), no results using the input shapes are
presented in this paper.

In this section, we present low level data from the analyses, in terms of direct
observables. For further discussion and interpretation of the results in terms
of variables concerning global survey and instrumental performance, see
\S\ref{sec:discussion}. To conserve space, only a representative sample of the
many results are displayed here. The rest is described in the text, in relation
to the illustrative examples, and is also available from the STEP
website$^{\ref{url:step}}$. First,
we shall describe the measurement of stars; then the number density of galaxies
and then shears in each set of images. Finally, we shall split the galaxy
catalogues by objects' observed sizes and magnitudes.

\subsection{PSF modelling}

\begin{table}
\begin{center}
\begin{tabular}{ccr@{$\pm$}lr@{$\pm$}l}
\multirow{2}{*}{\bf{Image set}} & 
\multicolumn{5}{c}{\bf{PSF model from TS implementation of KSB+}} \\
 & FLUX\_RADIUS &
\multicolumn{2}{c}{$\varepsilon_1$} &
\multicolumn{2}{c}{$\varepsilon_2$} \\
\hline\hline 
 A & $0.334\arcsec$ & -(0.68 & $0.10)\%$ &  (1.21 & $0.07)\%$ \\
 B & $0.334\arcsec$ & -(0.66 & $0.07)\%$ &  (1.28 & $0.05)\%$ \\
 C & $0.406\arcsec$ & -(0.47 & $0.07)\%$ &  (0.97 & $0.06)\%$ \\
 D & $0.390\arcsec$ & (11.49 & $0.11)\%$ &  (2.20 & $0.14)\%$ \\
 E & $0.390\arcsec$ & -(2.21 & $0.14)\%$ & (11.29 & $0.16)\%$ \\
 F & $0.392\arcsec$ & -(0.01 & $0.12)\%$ &  (0.01 & $0.01)\%$ \\
\hline\hline 
\end{tabular}
\end{center}
\caption{PSF models for the six sets of images used in the 
STEP2 analysis by the TS implementation of KSB+, averaged over stars in 
the simulated images. These quantities may be more familiar to some 
readers. FLUX\_RADIUS is directly from SExttractor, and the ellipticities
are all measured using a Gaussian weight function of rms size 
$r_g=0.6\arcsec=3$pixels.}
\label{tab:psfmeasurements}
\end{table}

The first task for all shear measurement methods is to identify stars and measure
the shape of the PSF. Table~\ref{tab:psfmeasurements} lists parameters of the PSF
model generated by the TS implementation of KSB+. These quantities are more
familiar than those derived analytically from the shapelet models, and also
demonstrate the differences between measured PSF ellipticities and inputs described
in table~\ref{tab:psfs}. The few percent polarisations measured for components of
PSFs D and E that should be zero are typical of several other methods. These may
explain the peculiar residual shear offsets described in
\S\ref{sec:psfellipticity}.

\subsection{Galaxy number counts and the false detection rate}

\begin{table*}
\begin{center}
\begin{tabular}{c@{~}cc@{~}cccc@{~}c}
\multirow{2}{*}{\bf{Author}} & 
\bf{Image} & 
\multicolumn{2}{c}{\bf{$n_{\rm gals}$}} &
\bf{mean mag} & 
\bf{$\%$ mag} & 
\multicolumn{2}{c}{\bf{$\sigma\gamma$}} \\
 & 
\bf{set} & 
\bf{original /} & 
\bf{matched} & 
\bf{(original)} & 
\bf{decrease} & 
\bf{original /} & 
\bf{matched} \\
\hline\hline
\multirow{2}{*}{JB} & A &$  37 (0) $&$  25$&$  24.04$&$ 1.2$&$     0.012$&$   0.007$\\ 
                    & C &$  28 (1) $&$  21$&$  23.50$&$ 1.0$&$     0.014$&$   0.008$\\ \hline
\multirow{2}{*}{C1} & A &$  51 (2) $&$  45$&$  23.70$&$ 0.3$&$     0.008$&$   0.003$\\ 
                    & C &$  46 (2) $&$  40$&$  23.64$&$ 0.4$&$     0.009$&$   0.003$\\ \hline
\multirow{2}{*}{C2} & A &$  50 (2) $&$  45$&$  23.70$&$ 0.3$&$     0.008$&$   0.003$\\ 
                    & C &$  45 (2) $&$  40$&$  23.64$&$ 0.4$&$     0.009$&$   0.003$\\ \hline
\multirow{2}{*}{MH} & A &$  38 (0) $&$  35$&$  23.68$&$ 0.4$&$     0.008$&$   0.003$\\ 
                    & C &$  33 (0) $&$  29$&$  23.56$&$ 0.5$&$     0.009$&$   0.004$\\ \hline
\multirow{2}{*}{HH} & A &$  28 (0) $&$  26$&$  23.05$&$ 0.2$&$     0.010$&$   0.002$\\ 
                    & C &$  24 (0) $&$  21$&$  22.97$&$ 0.3$&$     0.012$&$   0.002$\\ \hline
\multirow{2}{*}{MJ~~} & A &$  27 (1) $&$  24$&$  23.30$&$ 0.3$&$     0.009$&$   0.003$\\ 
                    & C &$  25 (0) $&$  22$&$  23.26$&$ 0.4$&$     0.009$&$   0.003$\\ \hline
\multirow{2}{*}{MJ2}& A &$  27 (1) $&$  24$&$  22.58$&$ 0.1$&$     0.014$&$   0.002$\\ 
                    & C &$  25 (0) $&$  22$&$  22.48$&$ 0.2$&$     0.016$&$   0.002$\\ \hline
\multirow{2}{*}{KK} & A &$  32 (0) $&$  26$&$  23.46$&$ 0.5$&$     0.009$&$   0.003$\\ 
                    & C &$  27 (0) $&$  21$&$  23.35$&$ 0.5$&$     0.010$&$   0.003$\\ \hline
\multirow{2}{*}{RM} & A &$  36 (0) $&$  32$&$  23.41$&$ 0.3$&$     0.009$&$   0.002$\\ 
                    & C &$  27 (0) $&$  23$&$  23.21$&$ 0.4$&$     0.010$&$   0.003$\\ \hline
\multirow{2}{*}{RN} & A &$  22 (1) $&$  19$&$  23.10$&$ 0.3$&$     0.009$&$   0.003$\\ 
                    & C &$  16 (1) $&$  13$&$  23.03$&$ 0.5$&$     0.011$&$   0.004$\\ \hline
\multirow{2}{*}{SP} & A &$  27 (11)$&$  15$&$  23.13$&$ 0.4$&$     0.014$&$   0.003$\\ 
                    & C &$  25 (10)$&$  13$&$  23.10$&$ 0.4$&$     0.016$&$   0.004$\\ \hline
\multirow{2}{*}{MS1}& A &$  43 (1) $&$  39$&$  23.68$&$ 0.3$&$     0.007$&$   0.003$\\ 
                    & C &$  37 (1) $&$  33$&$  23.55$&$ 0.3$&$     0.008$&$   0.003$\\ \hline
\multirow{2}{*}{MS2}& A &$  41 (1) $&$  36$&$  23.46$&$ 0.1$&$     0.010$&$   0.004$\\ 
                    & C &$  35 (1) $&$  30$&$  23.26$&$ 0.1$&$     0.013$&$   0.006$\\ \hline
\multirow{2}{*}{TS} & A &$  40 (0) $&$  36$&$  23.74$&$ 0.5$&$     0.008$&$   0.004$\\ 
                    & C &$  34 (0) $&$  29$&$  23.64$&$ 0.6$&$     0.010$&$   0.005$\\ \hline
\multirow{2}{*}{ES1}& A &$  40 (0) $&$  34$&$  23.81$&$ 0.6$&$     0.008$&$   0.003$\\ 
                    & C &$  35 (0) $&$  30$&$  23.71$&$ 0.7$&$     0.008$&$   0.003$\\ \hline
\multirow{2}{*}{ES2}& A &$  40 (0) $&$  34$&$  23.74$&$ 0.6$&$     0.016$&$   0.009$\\ 
                    & C &$  35 (0) $&$  30$&$  23.69$&$ 0.7$&$     0.017$&$   0.009$\\ \hline\hline

\end{tabular}
\end{center}

\caption{Number density of galaxies used by each method, and the shear 
measurement noise from those galaxies. The number of galaxies per square
arcminute are listed for the unmatched unrotated/rotated catalogues and after
matching. The number in brackets is the percentage of stars or false detections}

\label{tab:numbercounts}
\end{table*}

The methods used a variety of object detection algorithms and catalogue selection criteria.
For each method and each PSF, table~\ref{tab:numbercounts} lists the density of objects per
square arcminute, $n_{\rm gals}$, their mean magnitude, and the percentage of false
detections. Clearly, methods that are able to successfully measure the shapes of more
(fainter) galaxies, while avoiding false detections, will obtain a stronger measurement of
weak lensing, especially because the lensing signal grows cumulatively with galaxy redshift.
The false detection and stellar contamination rate is generally low, and the effective
survey depth is lowered by less than 0.1 magnitudes for all methods after matching rotated
and unrotated catalogues. Nor does matching have a significant effect upon the overall mean
polarisation of galaxies, which is always consistent with zero both before and after
matching -- as might not have been the case in the presence of selection effects
\citep{bj02,HirataSeljak03}.

Table~\ref{tab:numbercounts} also shows the measured dispersion of shear estimators
$\sigma_\gamma$ for each population. This statistic represents a combination of the
intrinsic ellipticity of galaxies and the shape measurement/PSF correction noise introduced
by each method. Lower values will produce stronger measurements of weak lensing. Since shear
measurement is more difficult for smaller or fainter galaxies, and the intrinsic morphology
distribution of galaxies varies as a function of magnitude in images other than set B,
$n_{\rm gals}$ and $\sigma_\gamma$ are likely to be correlated in a complicated fashion.
Galaxy selection effects and weighting schemes are discussed in
\S\ref{sec:selectioneffects} and \S\ref{sec:weightingschemes}.

\subsection{Shear calibration bias and residual shear offset}
\label{sec:mcresults}

\begin{center}
\begin{figure}
\epsfig{file=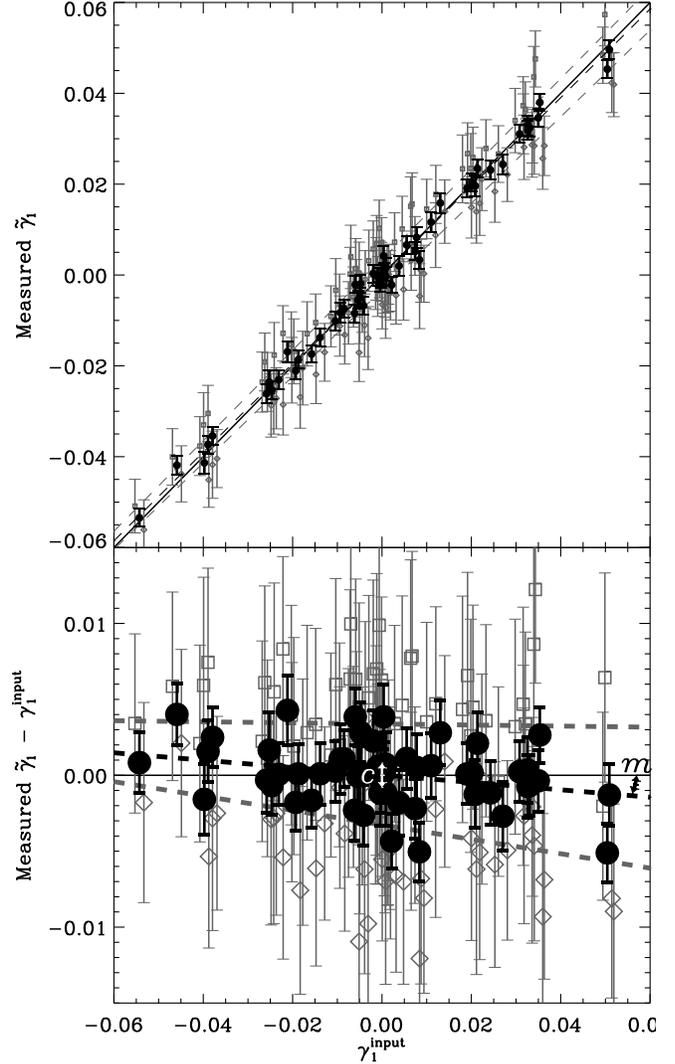,width=8.55cm,angle=0,clip=}
\caption{An example of the input {\it vs} measured shear for one representative
method. This is for the first component of shear measured by the KK method in
image set F. It is neither the best method on this image set, nor the best 
image set for this method, but shows behaviour that is typical of most.
The grey squares and diamonds show results from independent analyses of the
rotated and unrotated images; the black circles show the effect of matching
pairs of otherwise identical galaxies. The bottom panel shows deviations from
perfect shear recovery, which is indicated in both panels by solid lines. Linear fits to the
data are shown as dashed lines. The fitted parameters $m$ (shear calibration 
bias) and $c$ (residual shear offset) are plotted for all methods and all
for all images sets in figure~\ref{fig:mcresults}.}
\label{fig:pinkandblue}
\end{figure}
\end{center}

As with STEP1, we assess the success of each method by comparing the mean shear measured in
each image with the known input shears $\gamma_i^{\rm input}$. We quantify deviations from
perfect shear recovery via a linear fit that incorporates a multiplicative ``calibration
bias'' $m$ and an additive ``residual shear offset'' $c$. With a perfect shear measurement
method, both of these quantities would be zero. Since the input shear is now applied in
random directions, we measure two components each of $m$ and $c$, which correspond to the
two components of shear,
\begin{eqnarray}
\langle\tilde\gamma_1\rangle - \gamma_1^{\rm input} &=& m_1\gamma_1^{\rm input} + c_1 \nonumber\\
\langle\tilde\gamma_2\rangle - \gamma_2^{\rm input} &=& m_2\gamma_2^{\rm input} + c_2 ~.
\label{eqn:m1c1}
\end{eqnarray}

An illustrative example of one typical measurement of the first component of
shear is shown in figure~\ref{fig:pinkandblue}. The grey points correspond to
sets of rotated and unrotated galaxies, and are explained in
\S\ref{sec:rotunrot}. In this example, the negative slope of the black dashed
line in the bottom panel ($m_1$) shows that this method systematically
underestimates shear by $\sim 2.5\%$. However, the negligible $y$-intercept
shows that the PSF was successfully corrected and no residual shear calibration
($c_1$) remained. The measurement of the second component of shear is not shown.
Note that the range of input shear values is smaller than STEP1 and, in this
weak shear r\'egime, none of the methods exhibit the non-linear response to
shear seen with the strong signals in STEP1. We therefore do not attempt to fit
a quadratic function to any of the shear in {\it vs} shear out results.

\begin{figure*}
\begin{center}
\newlength{\figurewidth}
\setlength{\figurewidth}{84mm}
\epsfig{file=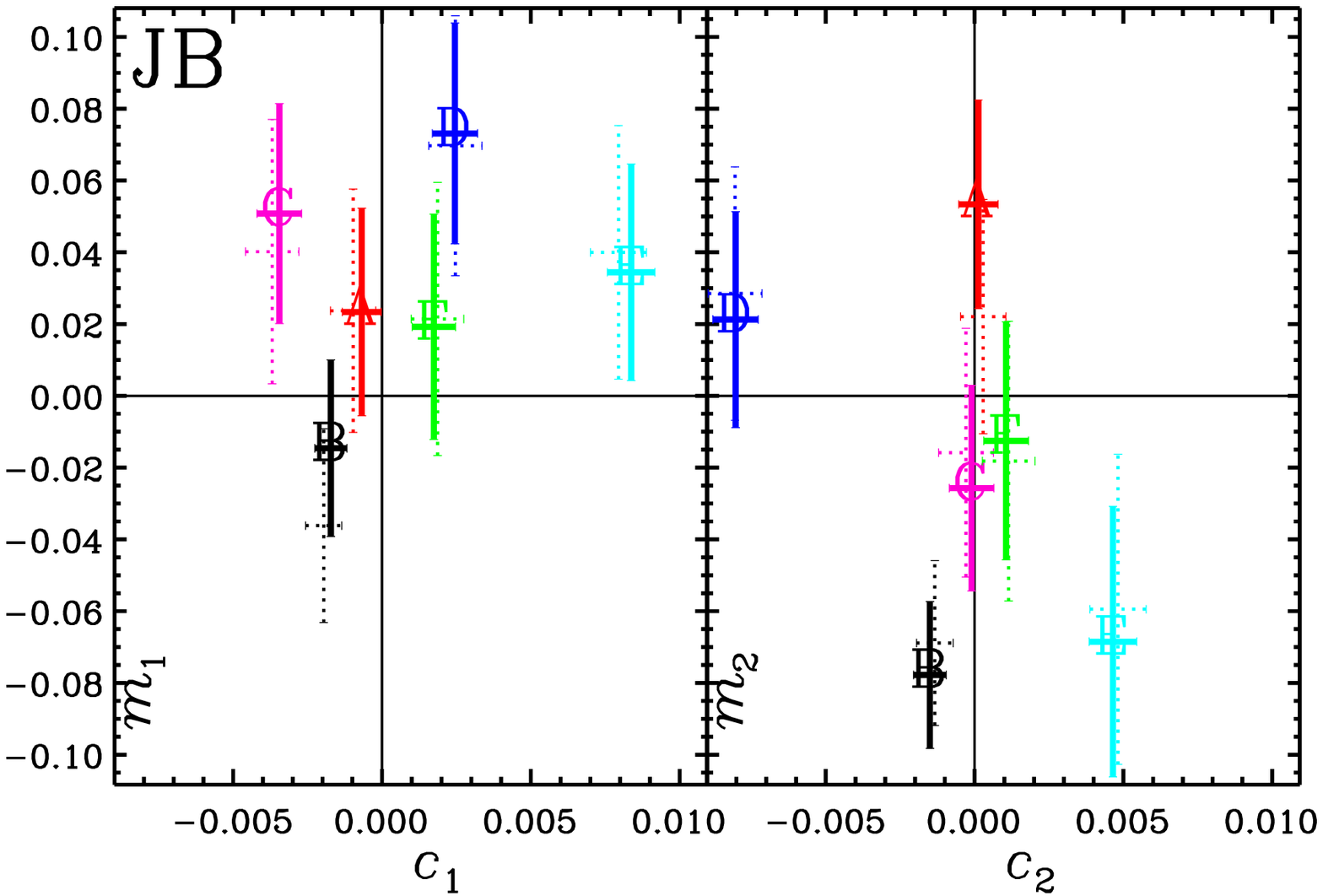,width=\figurewidth}
\epsfig{file=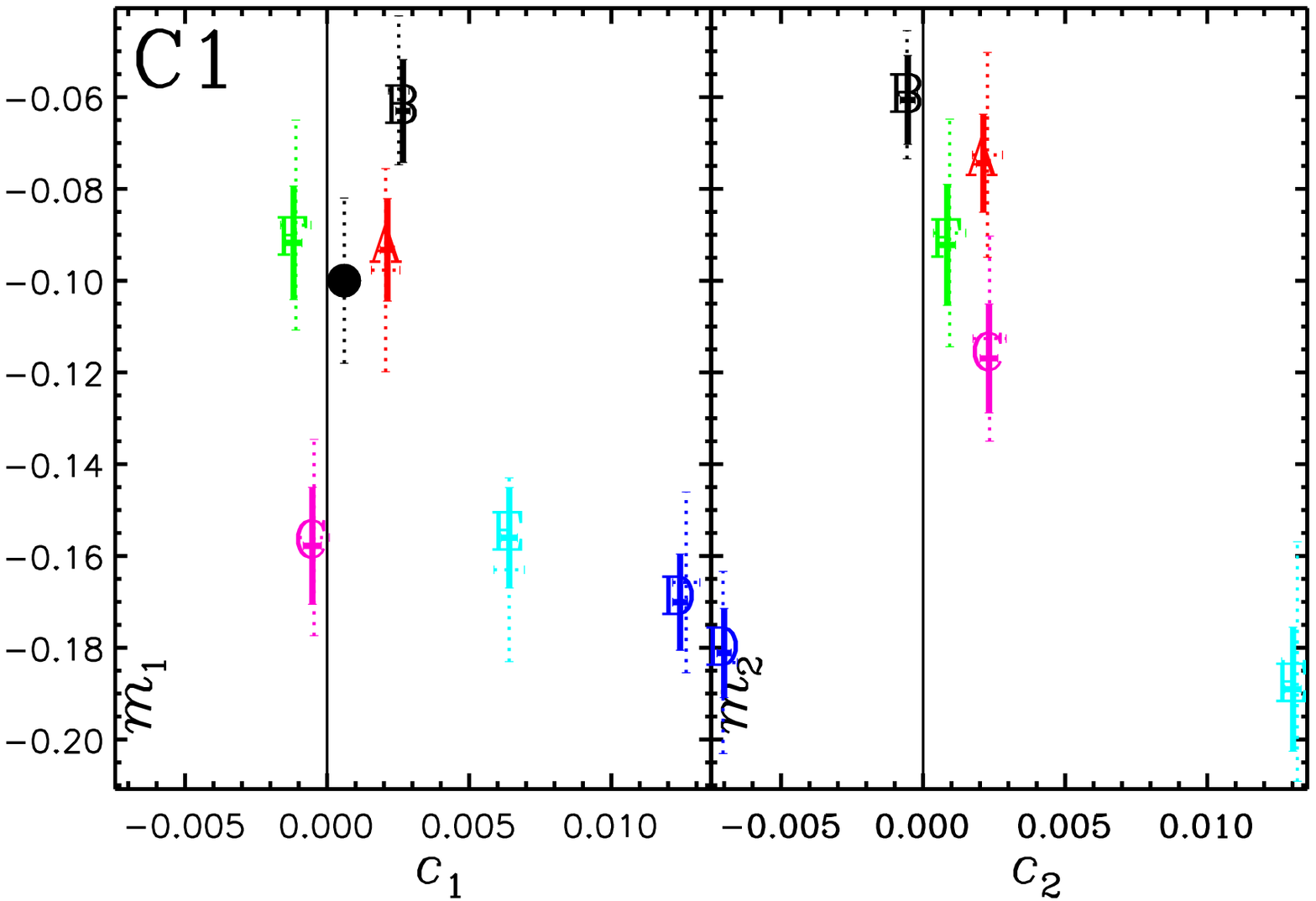,width=\figurewidth}
\epsfig{file=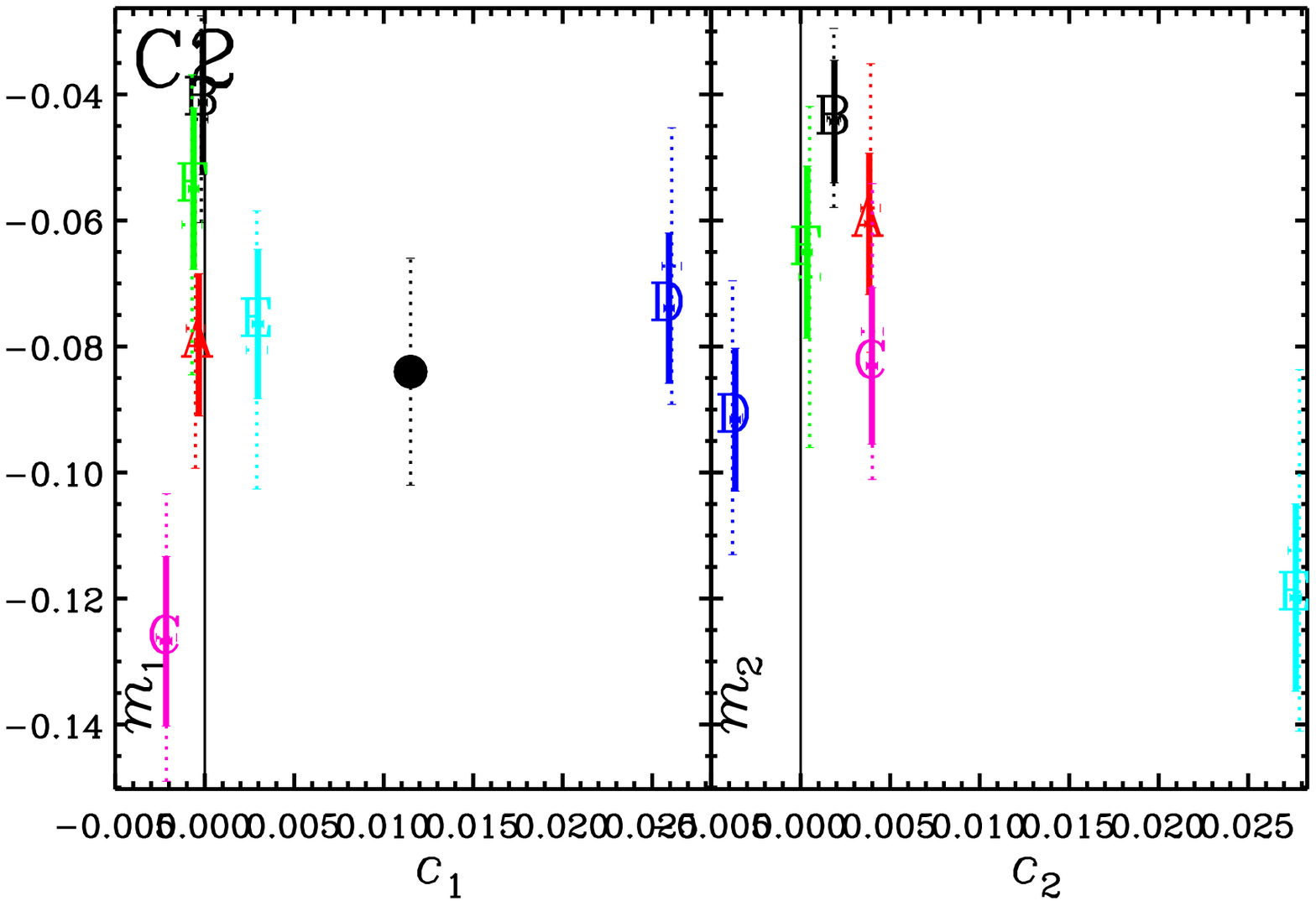,width=\figurewidth}
\epsfig{file=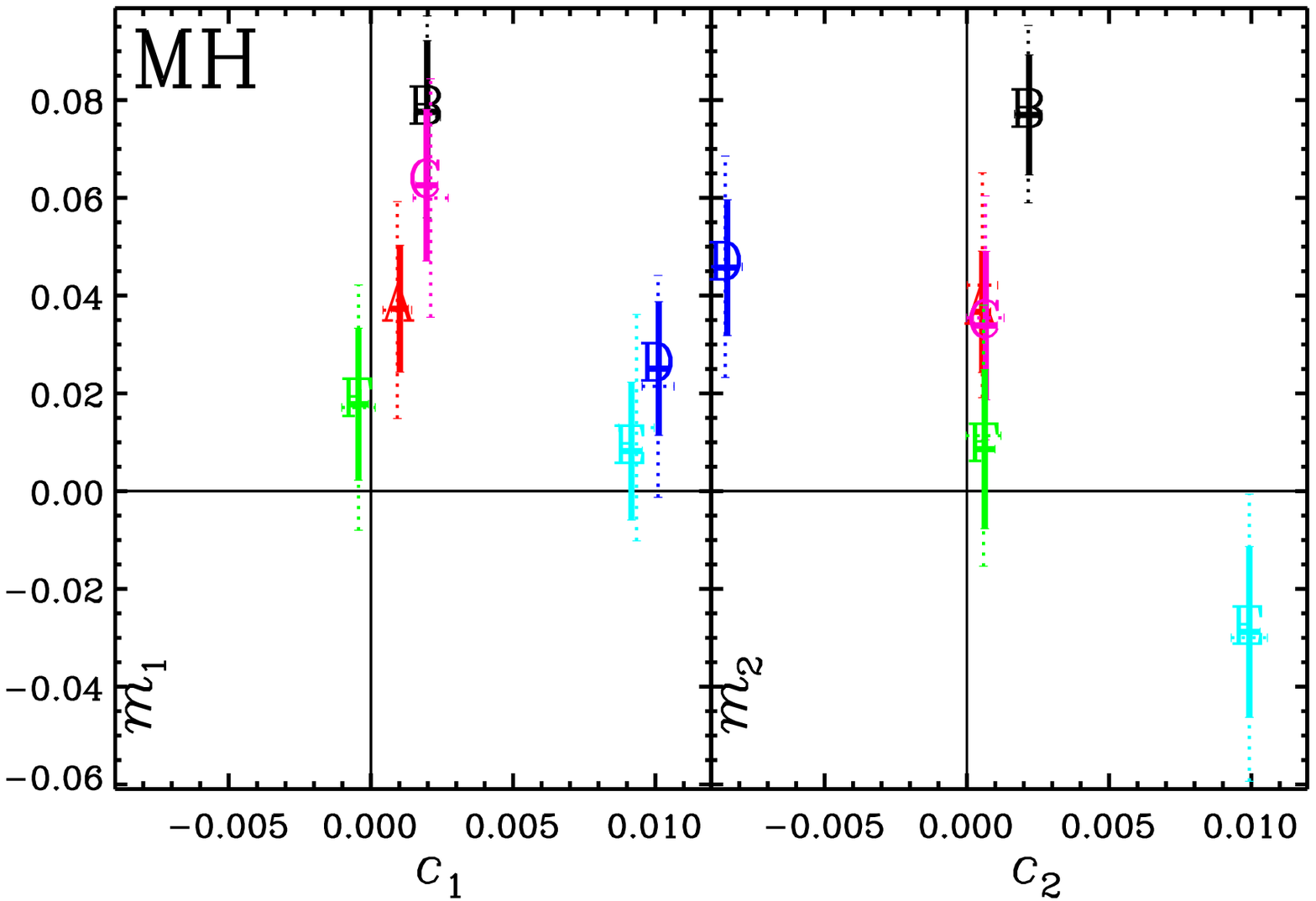,width=\figurewidth}
\epsfig{file=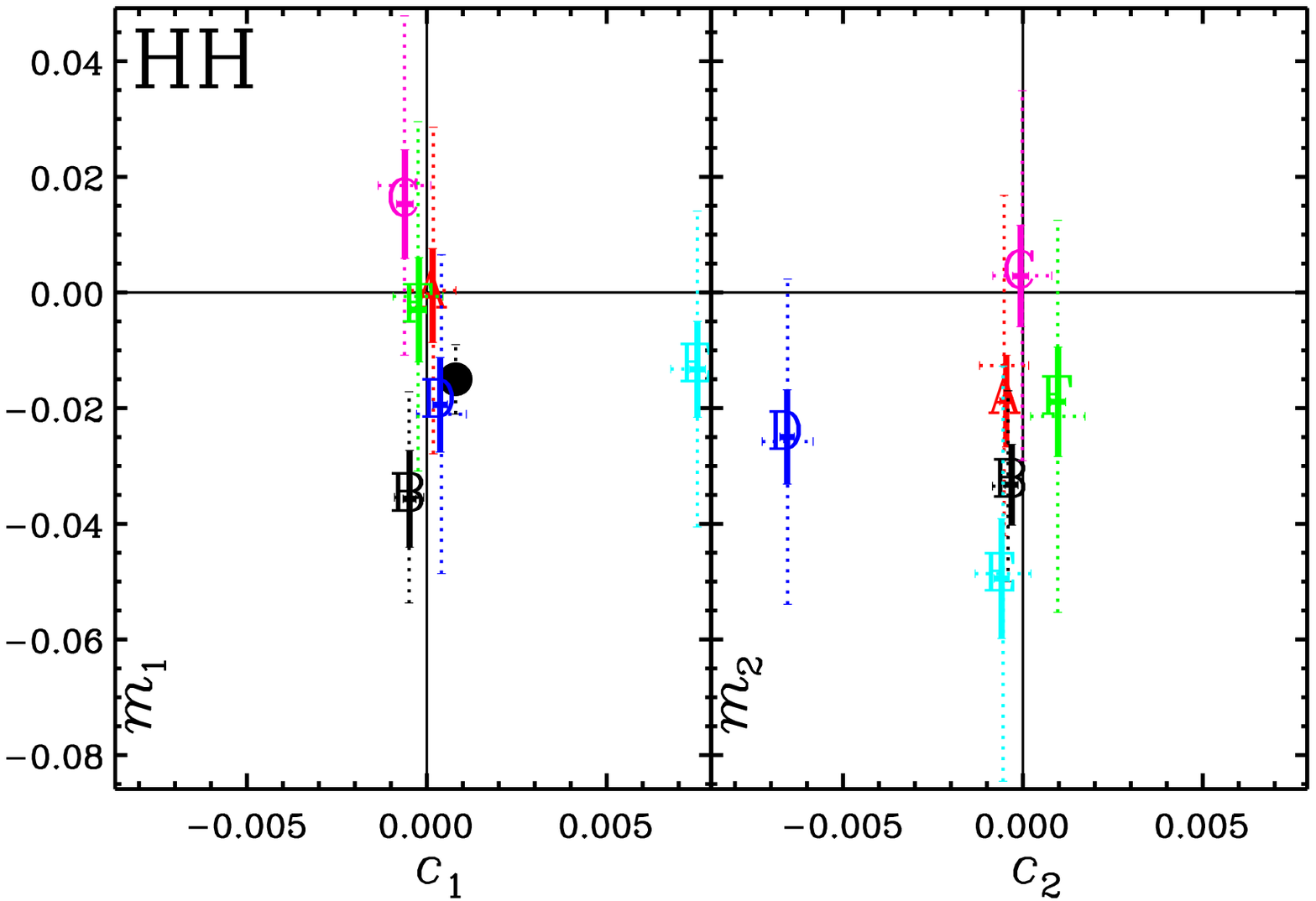,width=\figurewidth}
\epsfig{file=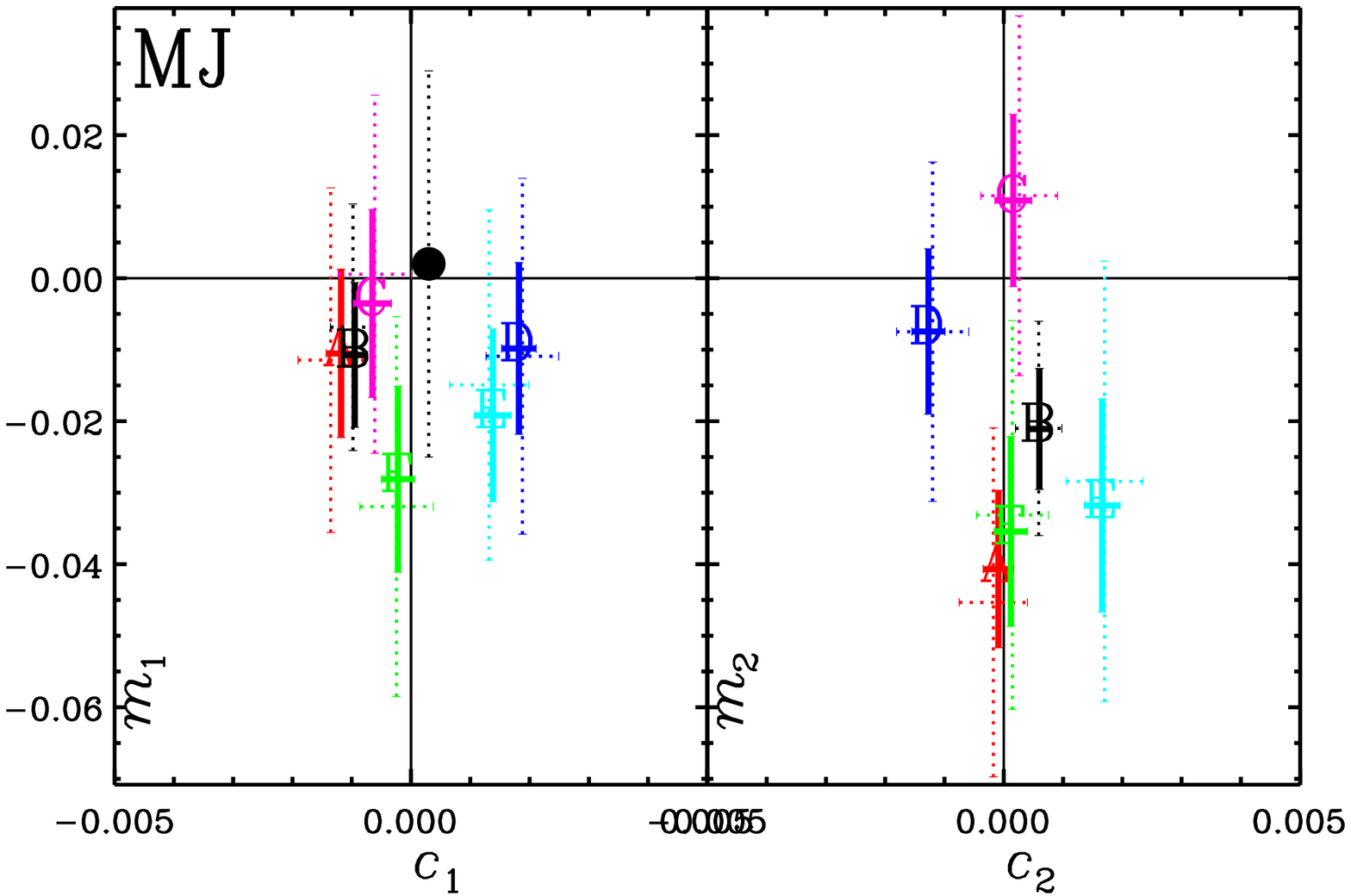,width=\figurewidth}
\epsfig{file=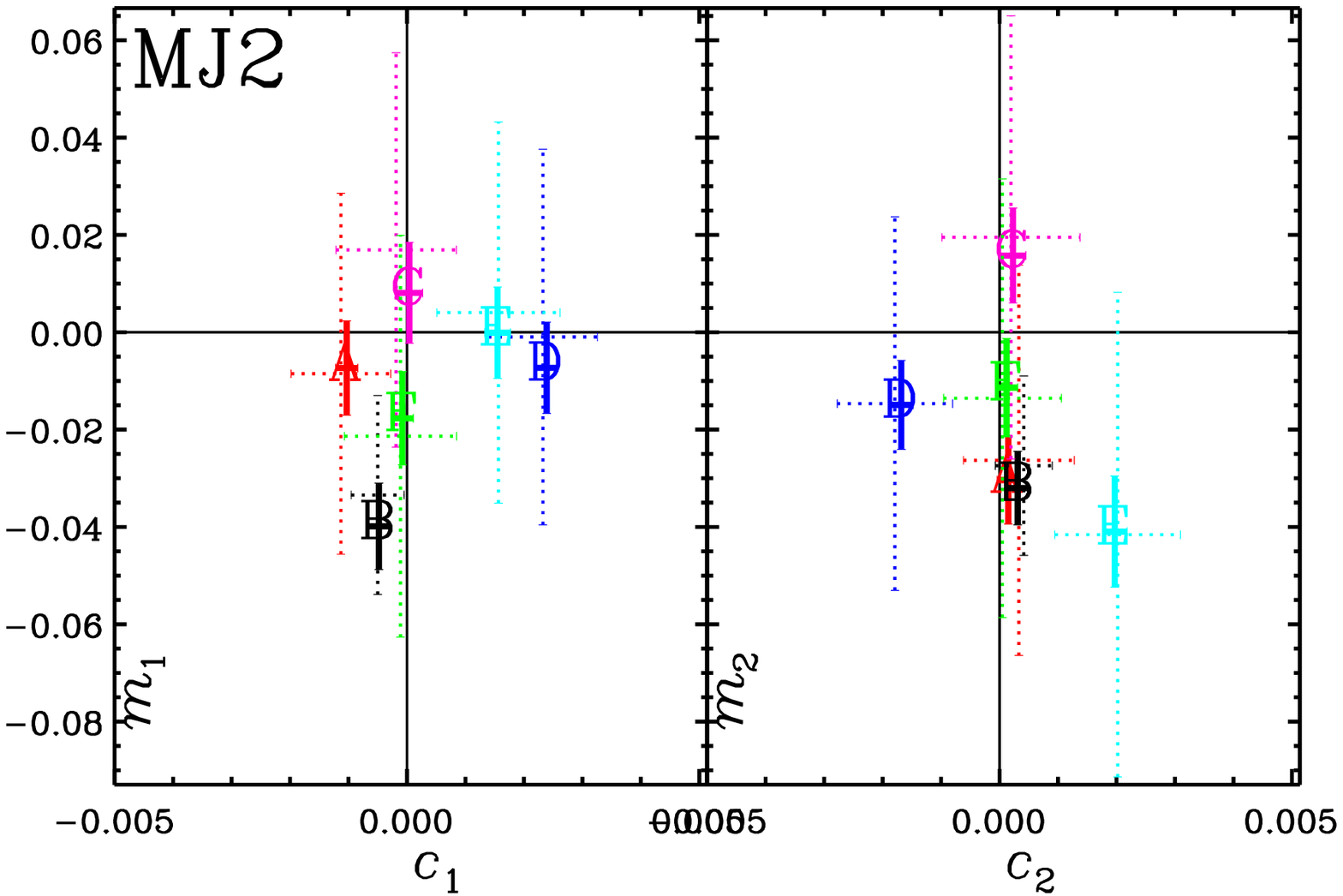,width=\figurewidth}
\epsfig{file=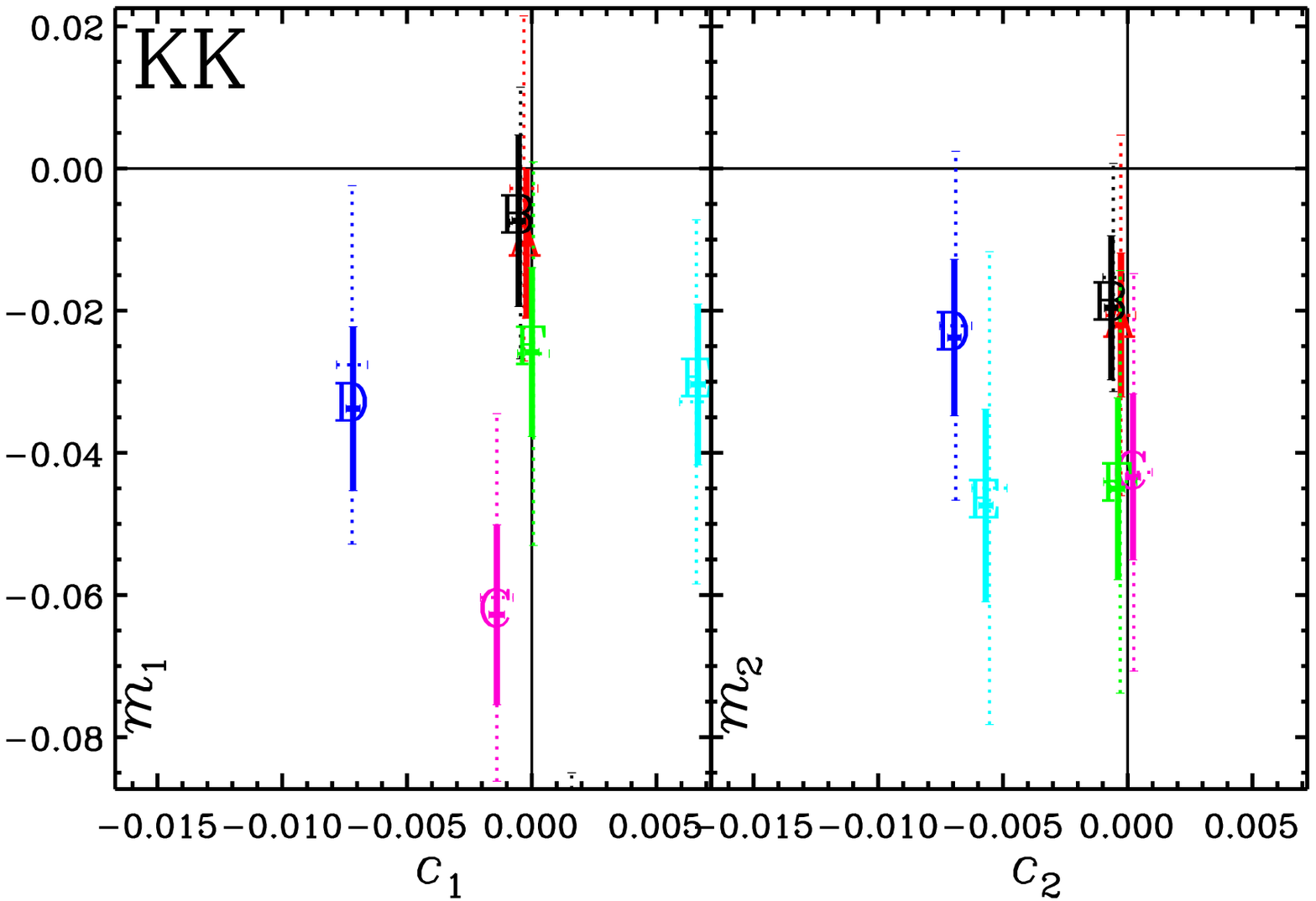,width=\figurewidth}
\caption{
Fitted values of residual shear offset and shear calibration bias for each
method and for each PSF. In all cases, the left hand panel shows results for the
$\gamma_1$ component of shear, and the right hand panel for the $\gamma_2$
component. 
The dotted lines show rms errors after a combined analysis of the
rotated and unrotated galaxies, after the two catalogues have been matched (and
only common detections kept). The solid lines show the reduced errors after
removing...}
\label{fig:mcresults}
\end{center}
\end{figure*}

\begin{figure*}
\begin{center}
\setlength{\figurewidth}{84mm}
\epsfig{file=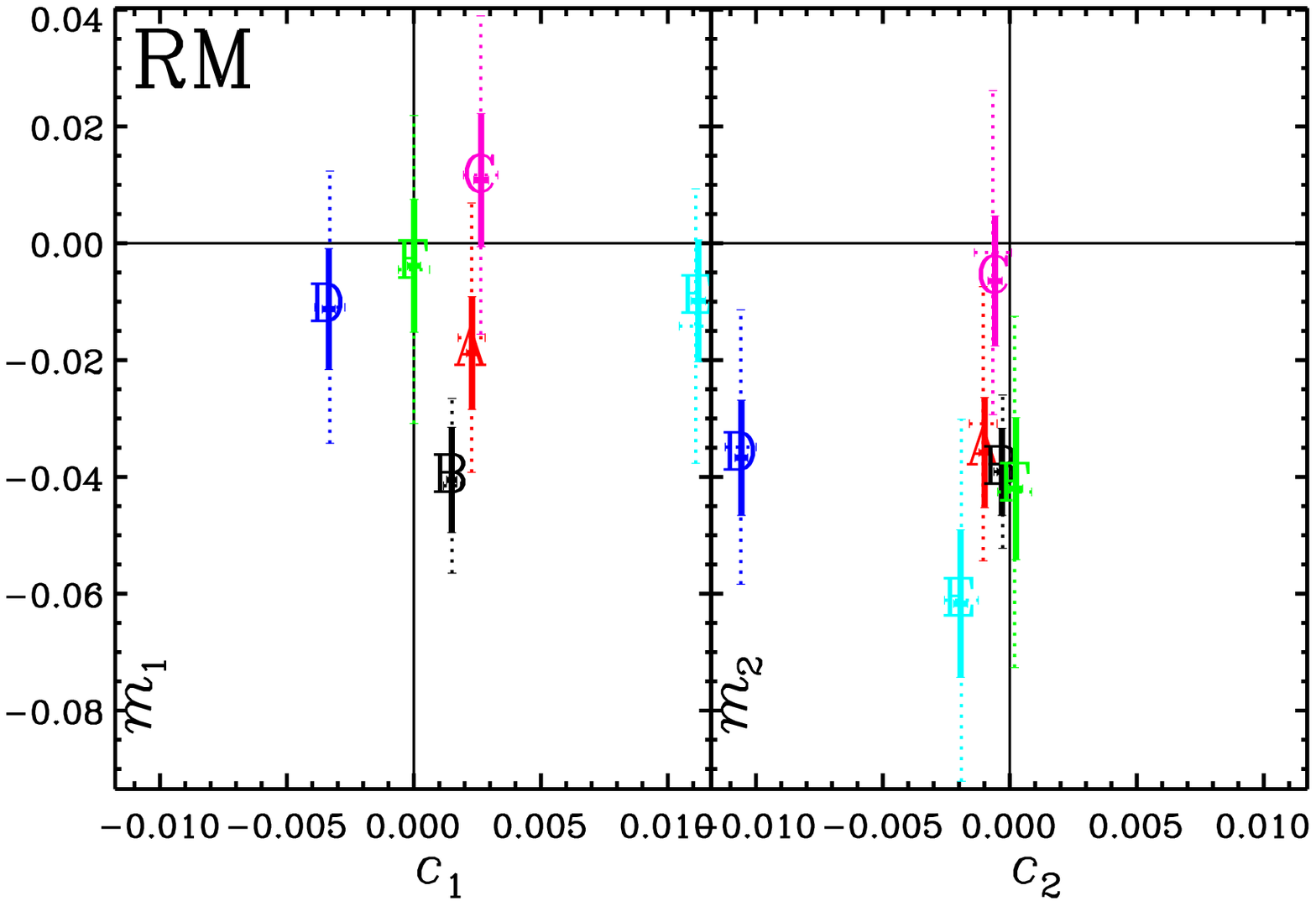,width=\figurewidth}
\epsfig{file=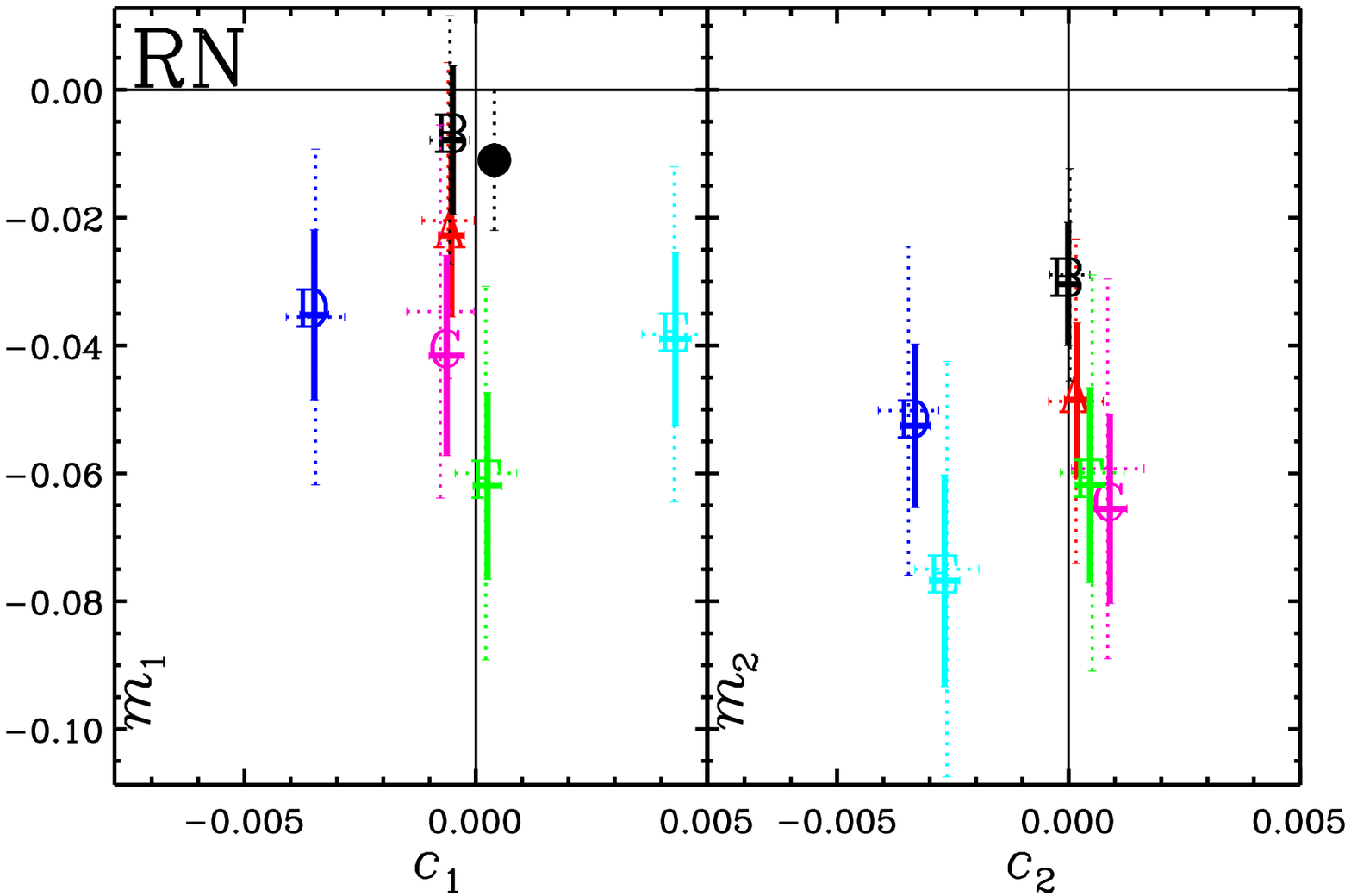,width=\figurewidth}
\epsfig{file=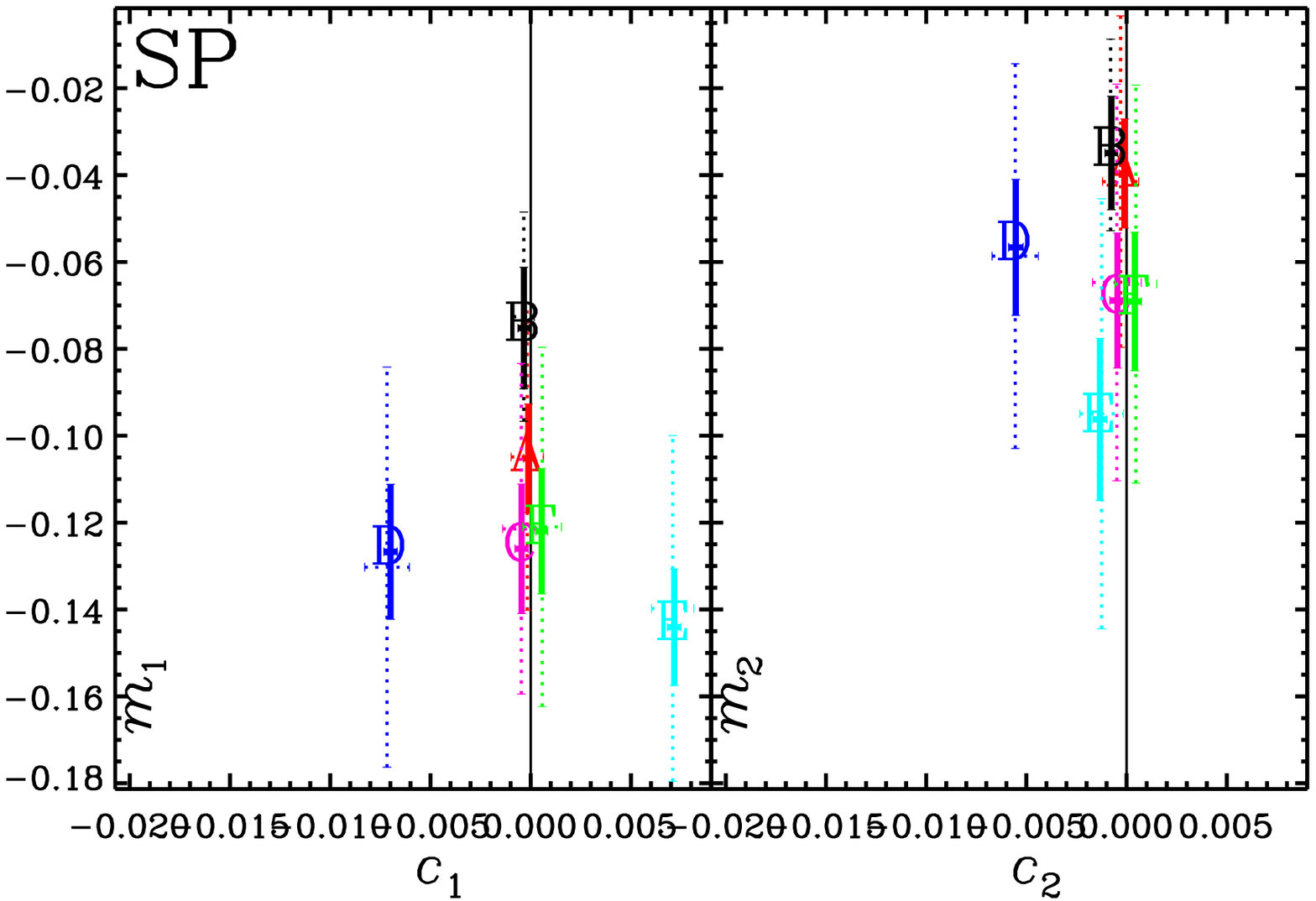,width=\figurewidth}
\epsfig{file=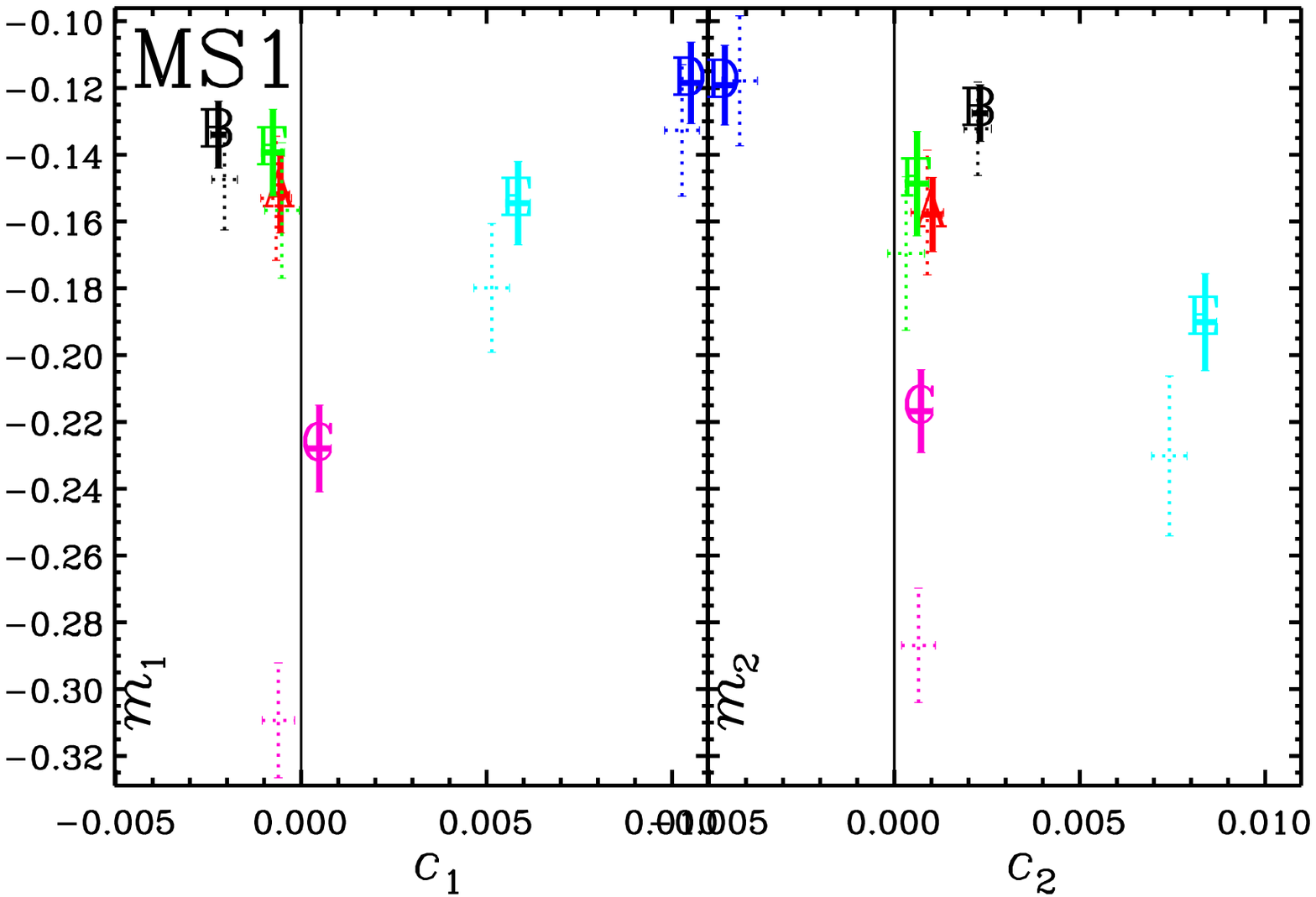,width=\figurewidth}
\epsfig{file=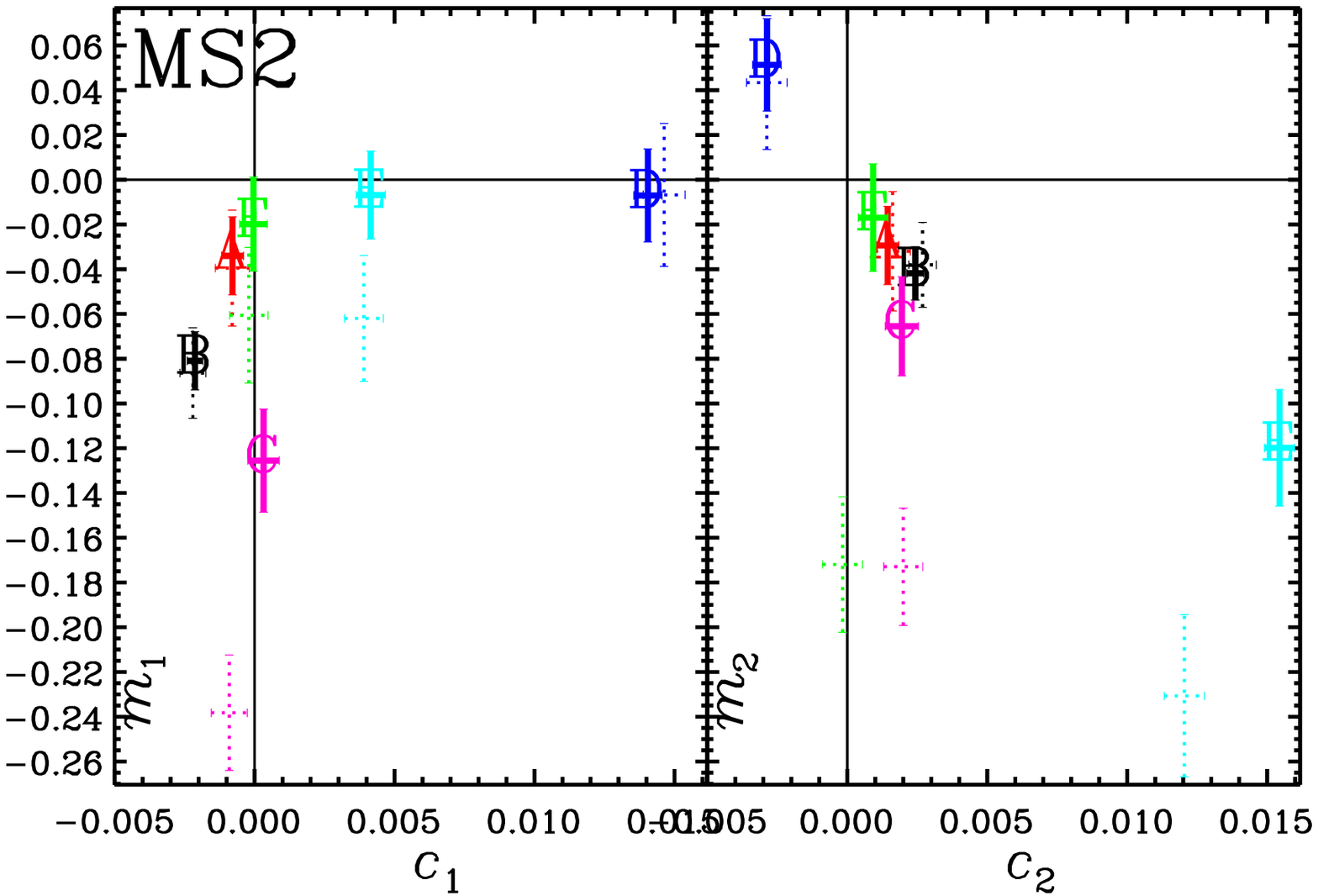,width=\figurewidth}
\epsfig{file=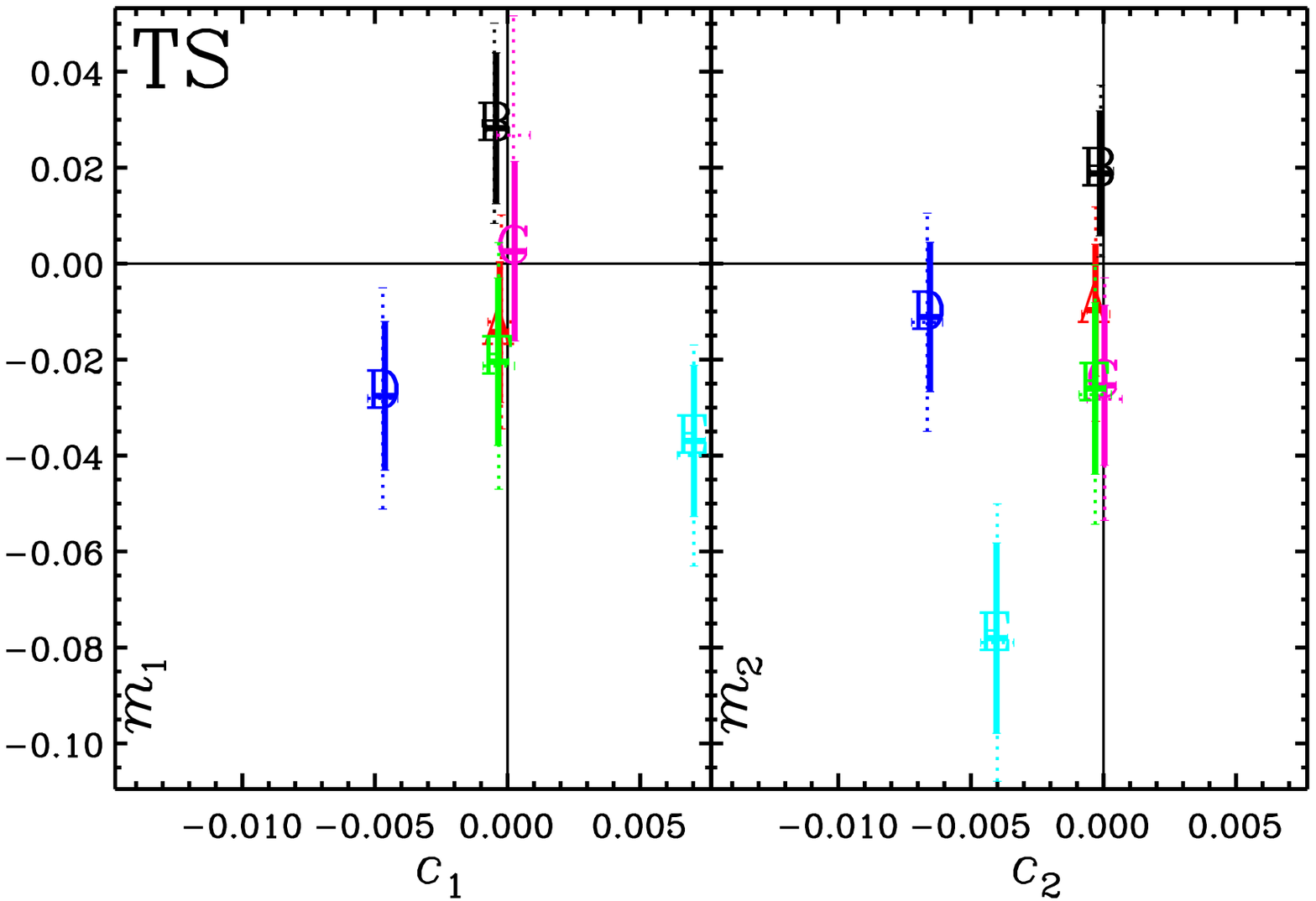,width=\figurewidth}
\epsfig{file=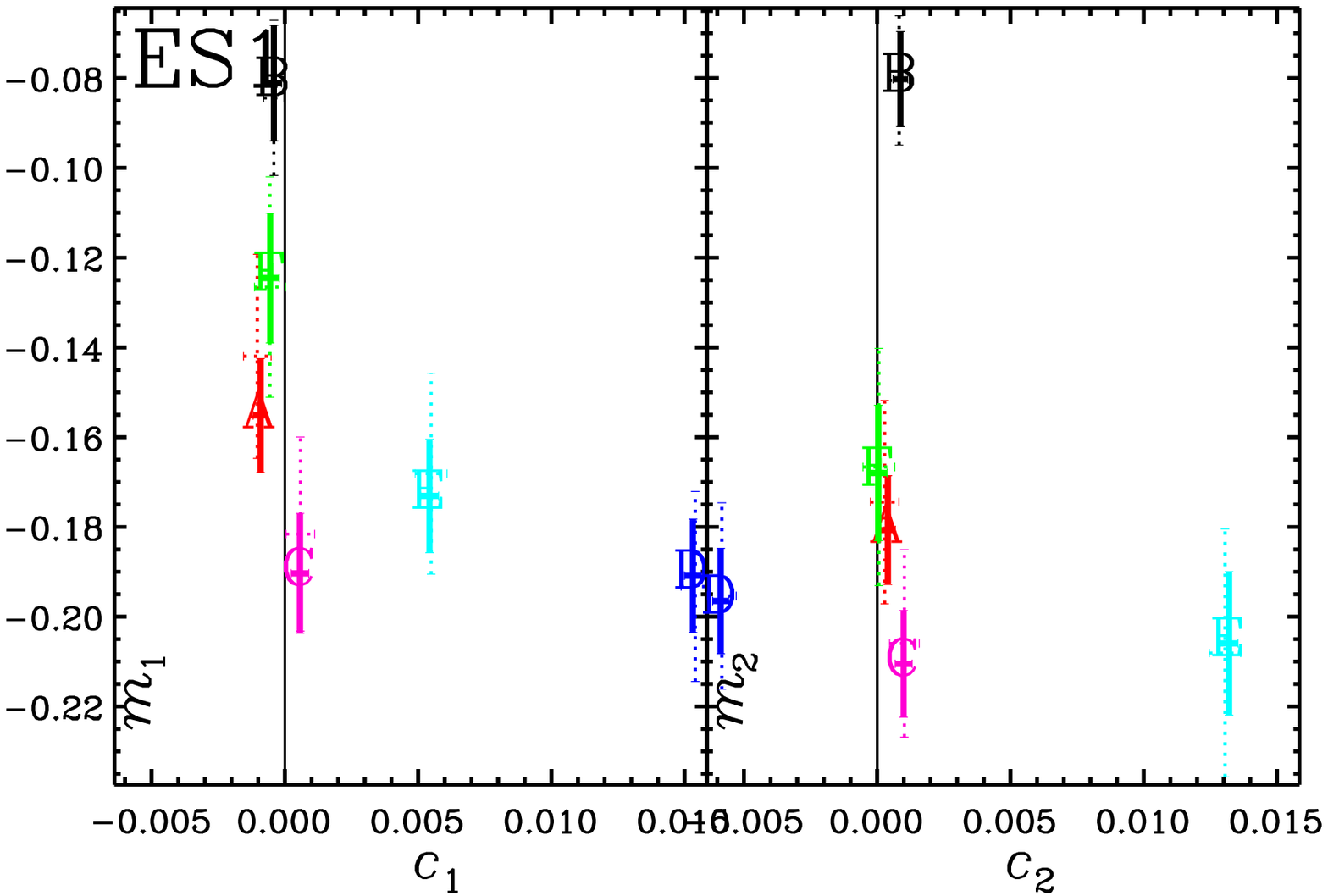,width=\figurewidth}
\epsfig{file=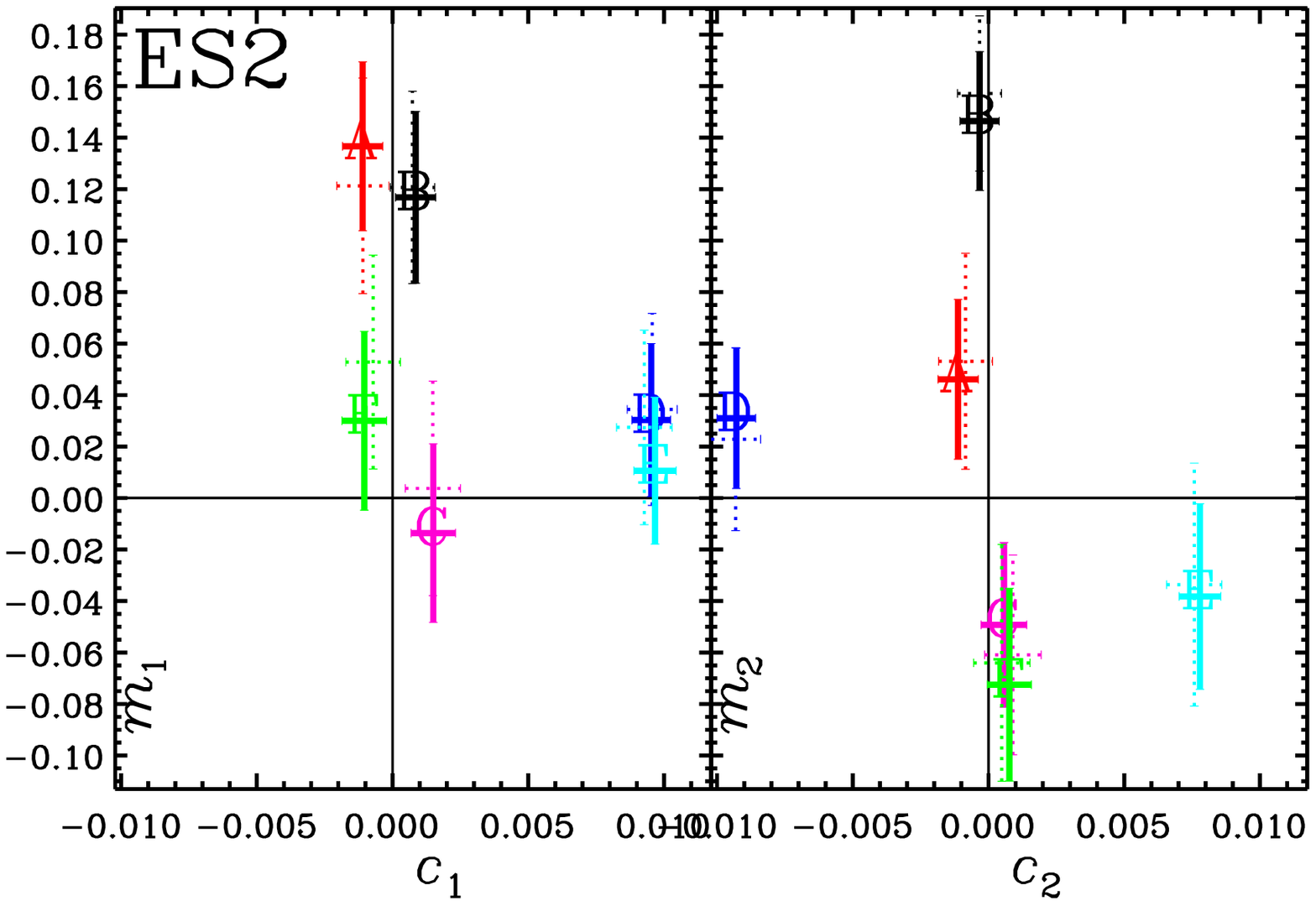,width=\figurewidth}
\end{center}
\begin{flushleft}
{\bf Figure~\ref{fig:mcresults} (continued).} ~~~~~~~~
...intrinsic galaxy shape noise from the matched the pairs of galaxies.
Note that the scales on each panel are different, but the frequency of the axis
labels is preserved. The red points correspond to image set A. The black points 
correspond to image set B, and, where available, the filled black 
circles reproduce results from STEP1. The pink, dark blue, light blue and green
points correspond to image sets C, D, E and F respectively.
\end{flushleft}
\end{figure*}

\subsection{Combining rotated and unrotated galaxies}
\label{sec:rotunrot}

An important advance in this second STEP project is the simultaneous analysis of
galaxies that had been rotated by $90^\circ$ before the application of shear and
convolution with the PSF. This can largely remove noise due to scatter in galaxies'
intrinsic morphology, but complicates the production of a joint shear catalogue,
especially where the galaxies are given different weights in the two catalogues.

Taking the rotated and unrotated sets of images individually, we  obtain two
sets of mean shear estimators $\langle\tilde\gamma^{\rm unrot}\rangle$ and
$\langle\tilde\gamma^{\rm rot}\rangle$, which are defined in
equations~(\ref{eqn:gammaunrot}) and (\ref{eqn:gammarot}). We typically find
that $m_i^{\rm rot} \approx m_i^{\rm unrot}$ and $c_i^{\rm rot} \approx
-c_i^{\rm unrot}$. Such stability to changes in image rotation is to be
expected: cross-talk between ellipticity and shear directions are second order
in $\gamma$ according to equation~(\ref{eqn:gstare}), and the mean ellipticity
is overwhelmingly dominated by the intrinsic ellipticities of a finite number of
galaxies (as demonstrated by the offset between the squares and diamonds in
figure~\ref{fig:pinkandblue}). Intruigingly, for the MS1 and MS2 methods, the
shear calibration bias changes significantly between the rotated and the
unrotated catalogues, and when the two are matched. These methods use smaller
galaxies than most, including some 10--$25\%$ around or below the stellar locus
on a size {\it vs} magnitude plane, and this effect may be caused by
instabilities in the PSF correction of the smallest. As an alternative
explanation, there are also second-order effects inherent in the non-linear
lensing equation that involve the dot product of ellipticity and shear, which
would become significant in the presence of an ellipticity-dependent selection
bias. However, we do not understand why this would affect only this pipeline and
not others. We have not attempted to investigate this isolated effect in more
detail.

\begin{figure}
\begin{center}
\epsfig{file=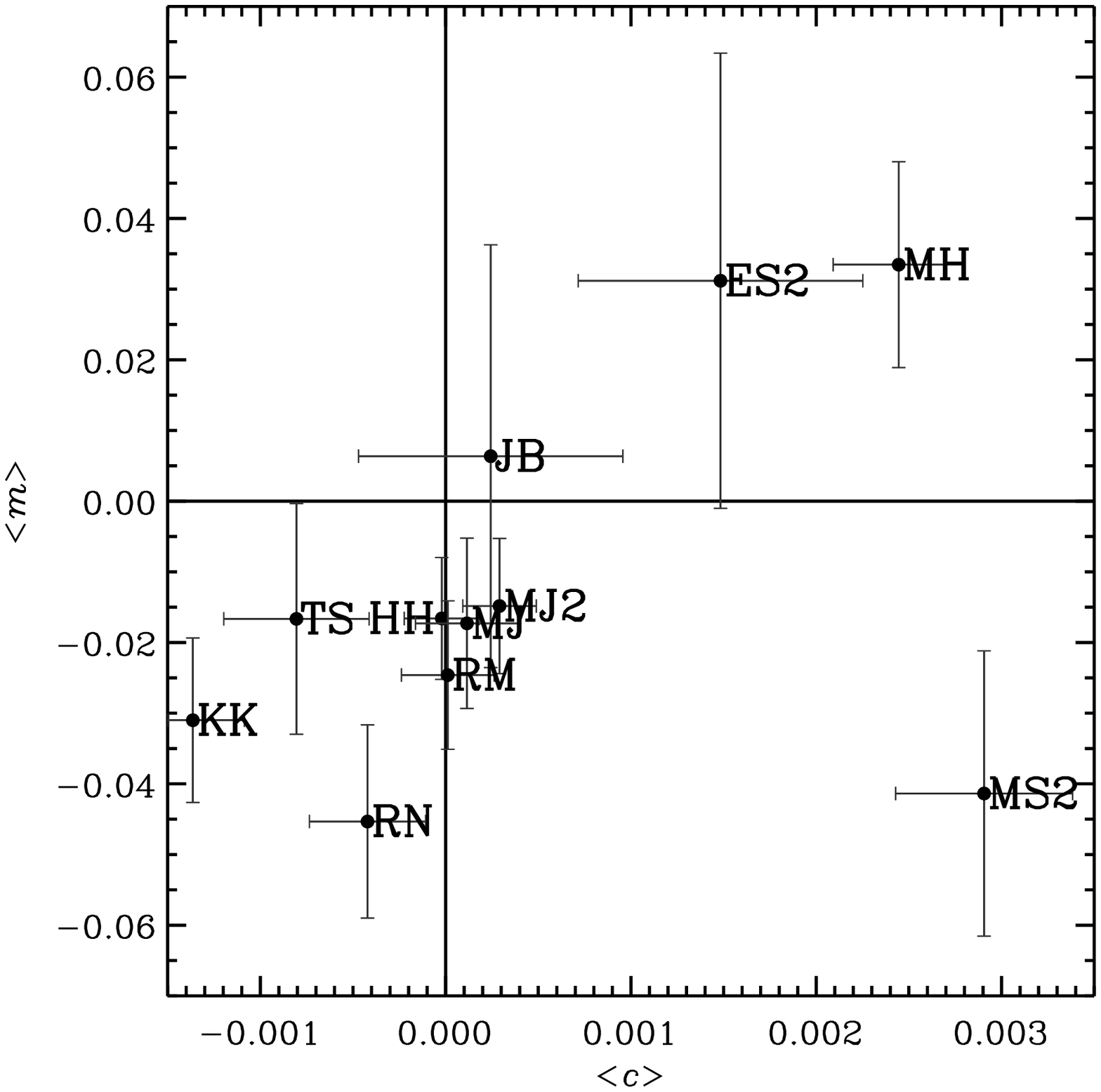,width=8.55cm}
\epsfig{file=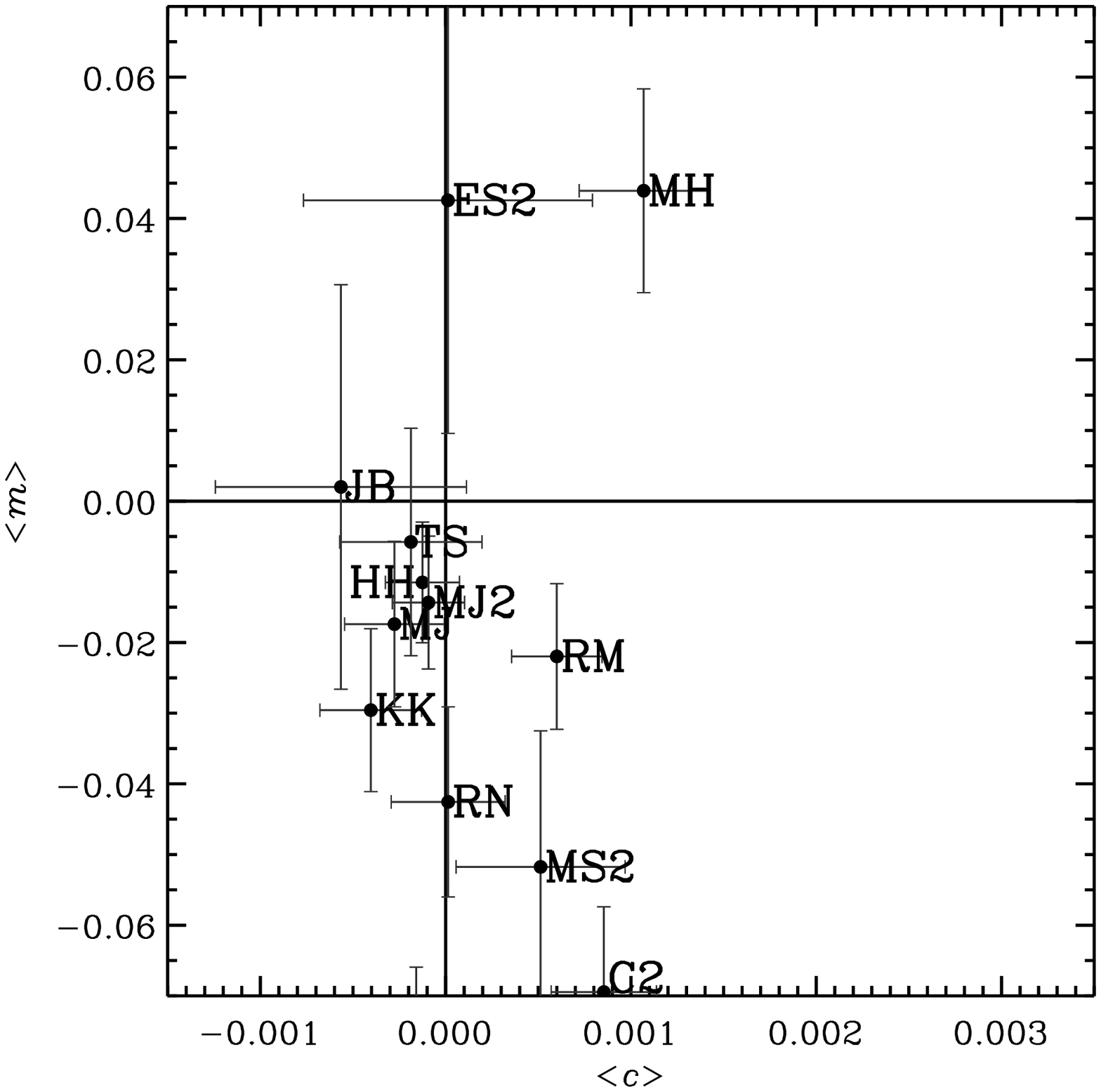,width=8.55cm}
\caption{
Comparison of shear measurement accuracy from different methods, in terms of 
their mean residual shear offset $\langle c\rangle$ and mean shear calibration  bias
$\langle m\rangle$. In the top panel, these parameters have been averaged over
both components of shear and all six sets of images; the bottom panel includes only 
image sets A, B, C and F, to avoid the two highly elliptical PSFs. 
Note that the {\it entire region} of these plots lie inside the grey
band that indicated good performance for methods in figure~3 of STEP1.
The results from methods C1, SP, MS1 and ES1 are not shown here.}
\label{fig:mc_everyone}
\end{center}
\end{figure}

We obtain a third set of parameters $m_i$ and $c_i$ from the matched catalogue with
$\langle\tilde\gamma\rangle$ defined in equation~(\ref{eqn:gammamatch}). In general, we find
that $m_i \simeq (m_i^{\rm unrot} + m_i^{\rm rot})/2$ and $c_i \simeq c_i^{\rm unrot} -
c_i^{\rm rot}$, with significantly smaller errors in this matched analysis. An example of
all three shear estimators for the KK method on image set F are plotted in
figure~\ref{fig:pinkandblue}. The fitted parameters for all of the shear measurement
methods, on all of the PSFs, are shown in figure~\ref{fig:mcresults}. Parameters measured
from the matched pair analysis are also tabulated in the appendix. Results from the most
successful methods are averaged across all of the sets of simulated images and compared
directly in figure~\ref{fig:mc_everyone}.

\subsection{Analysis as a function of galaxy population}
\label{sec:mcmagsizeresults}

\begin{figure*}
\begin{center}
\setlength{\figurewidth}{84mm}
\epsfig{file=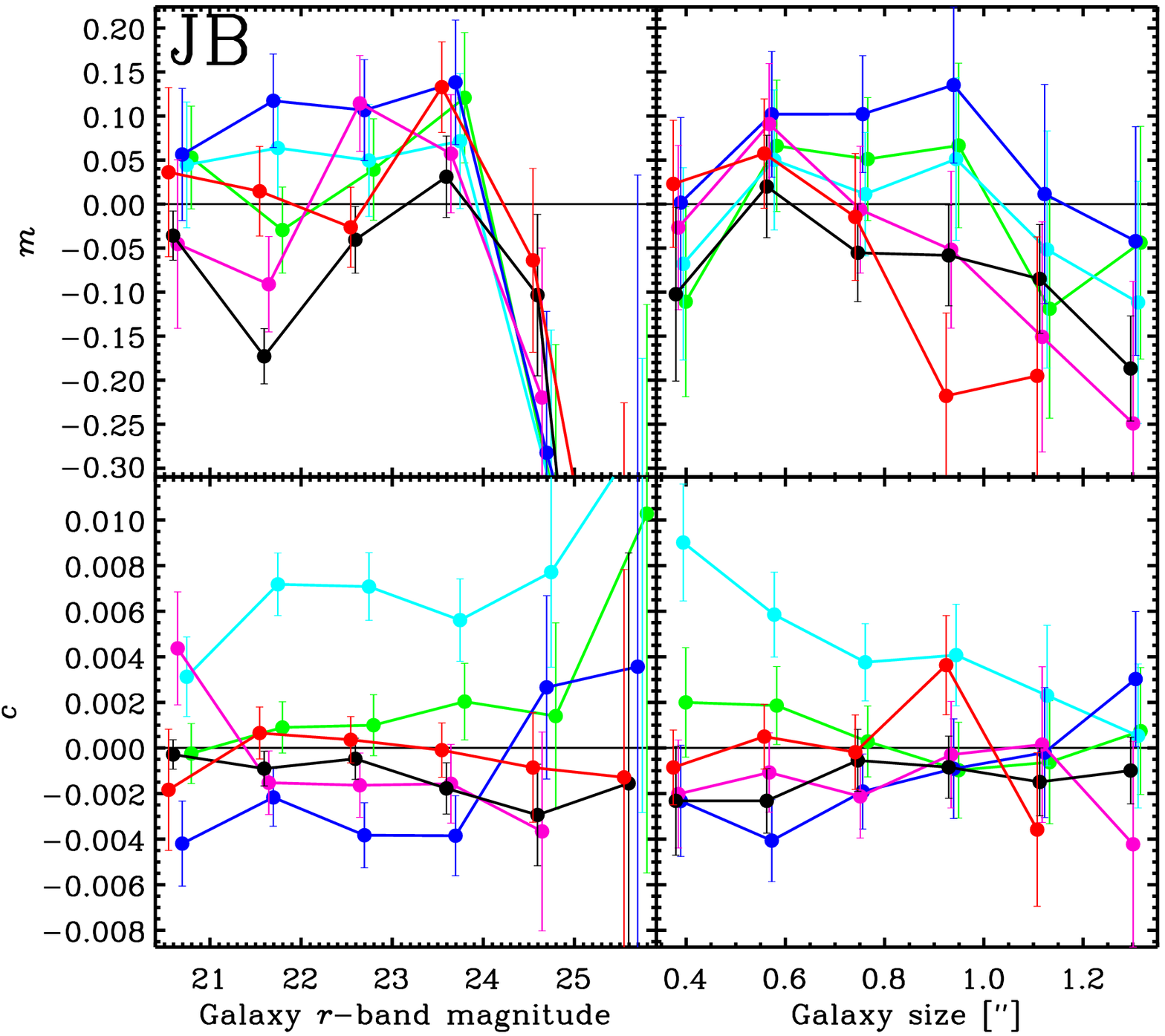,width=\figurewidth}
\epsfig{file=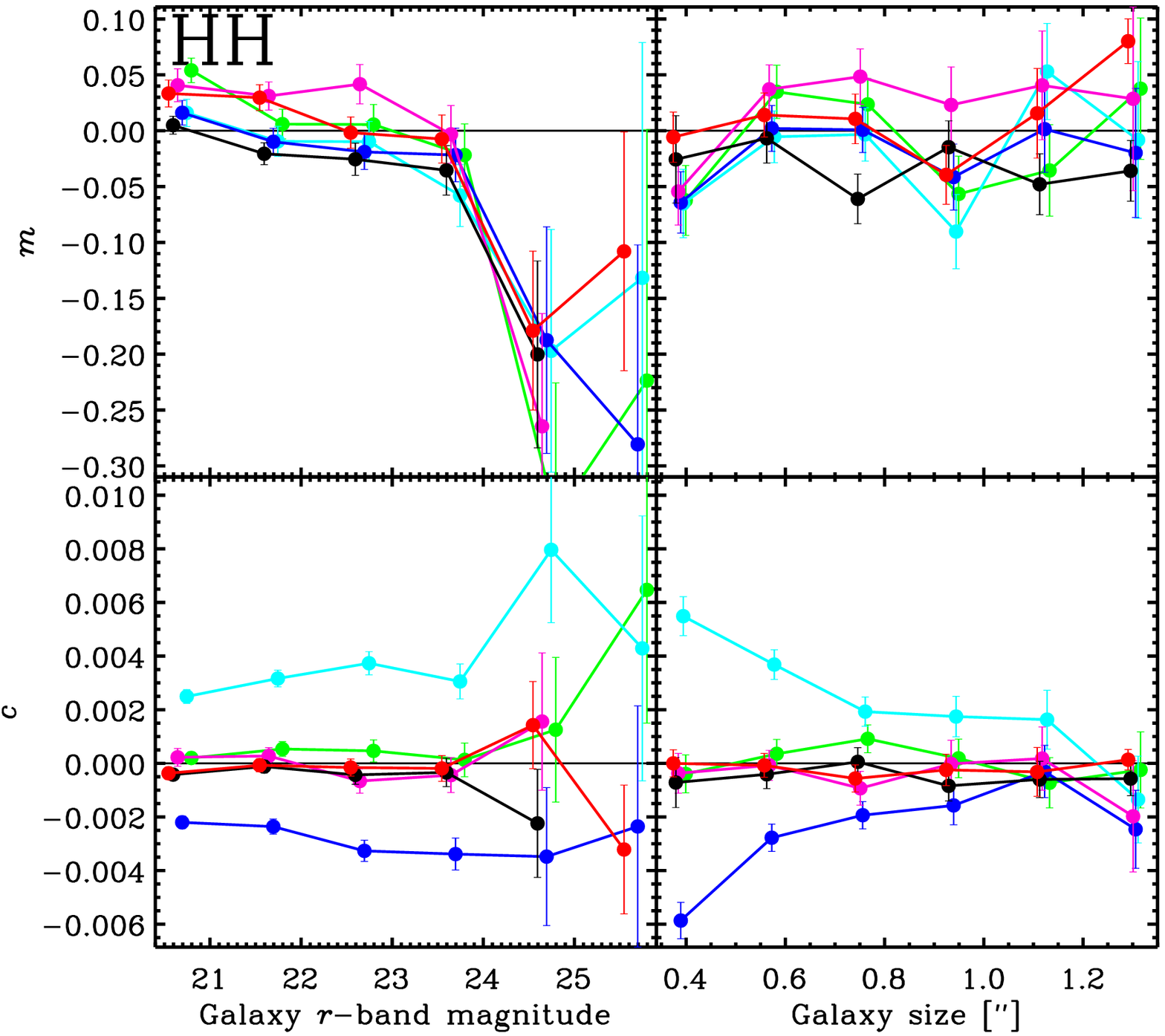,width=\figurewidth}
\epsfig{file=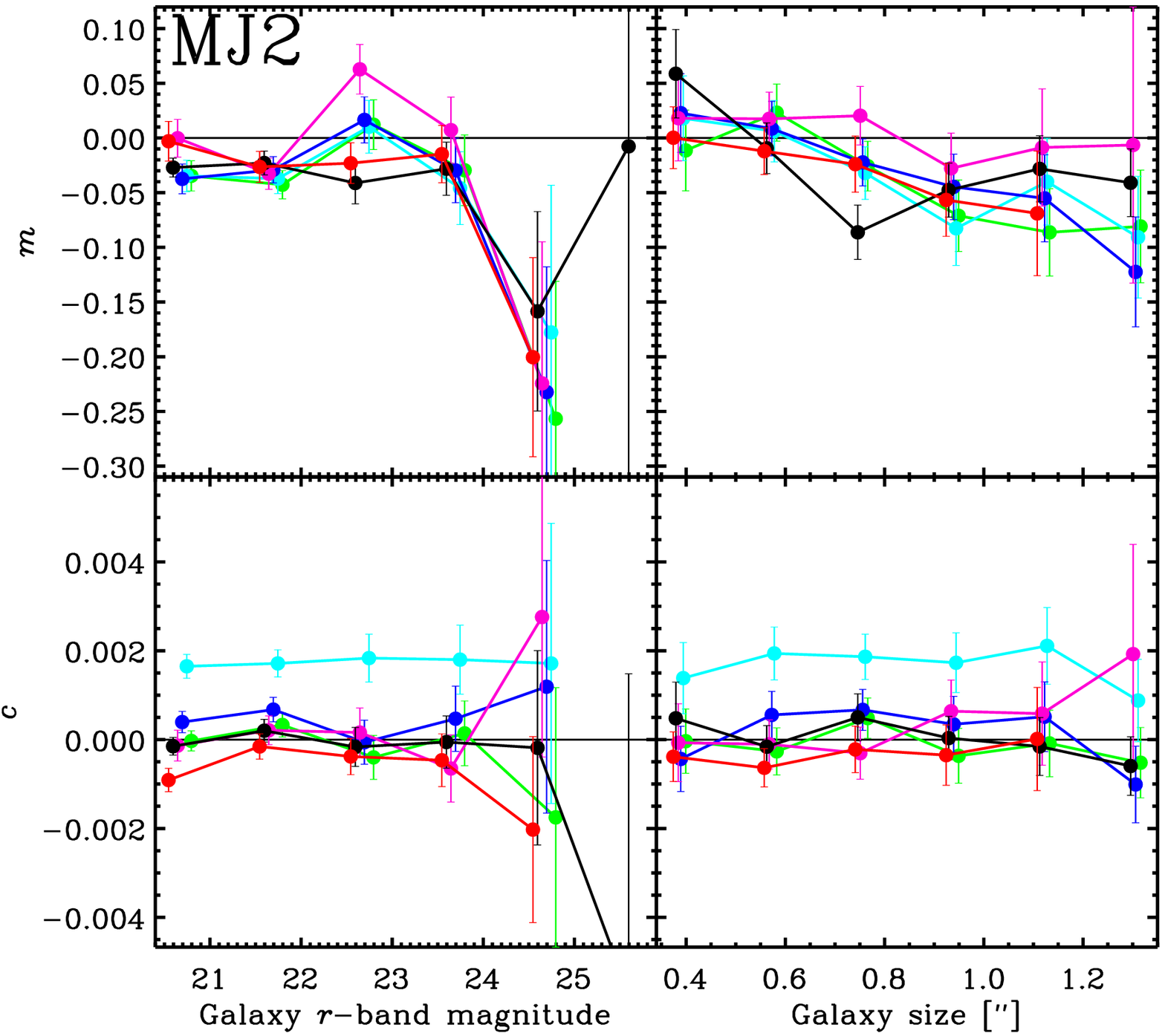,width=\figurewidth}
\epsfig{file=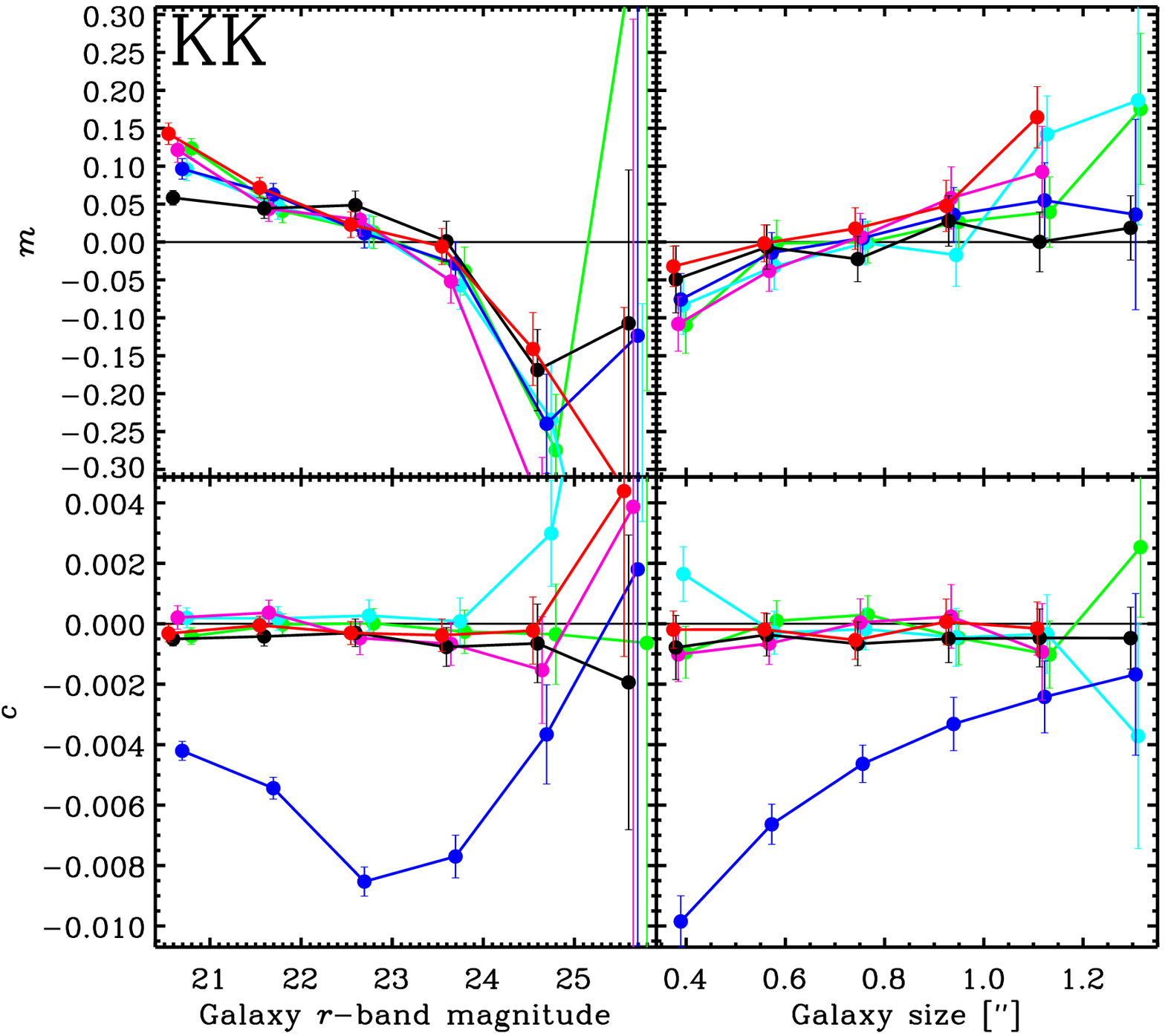,width=\figurewidth}
\epsfig{file=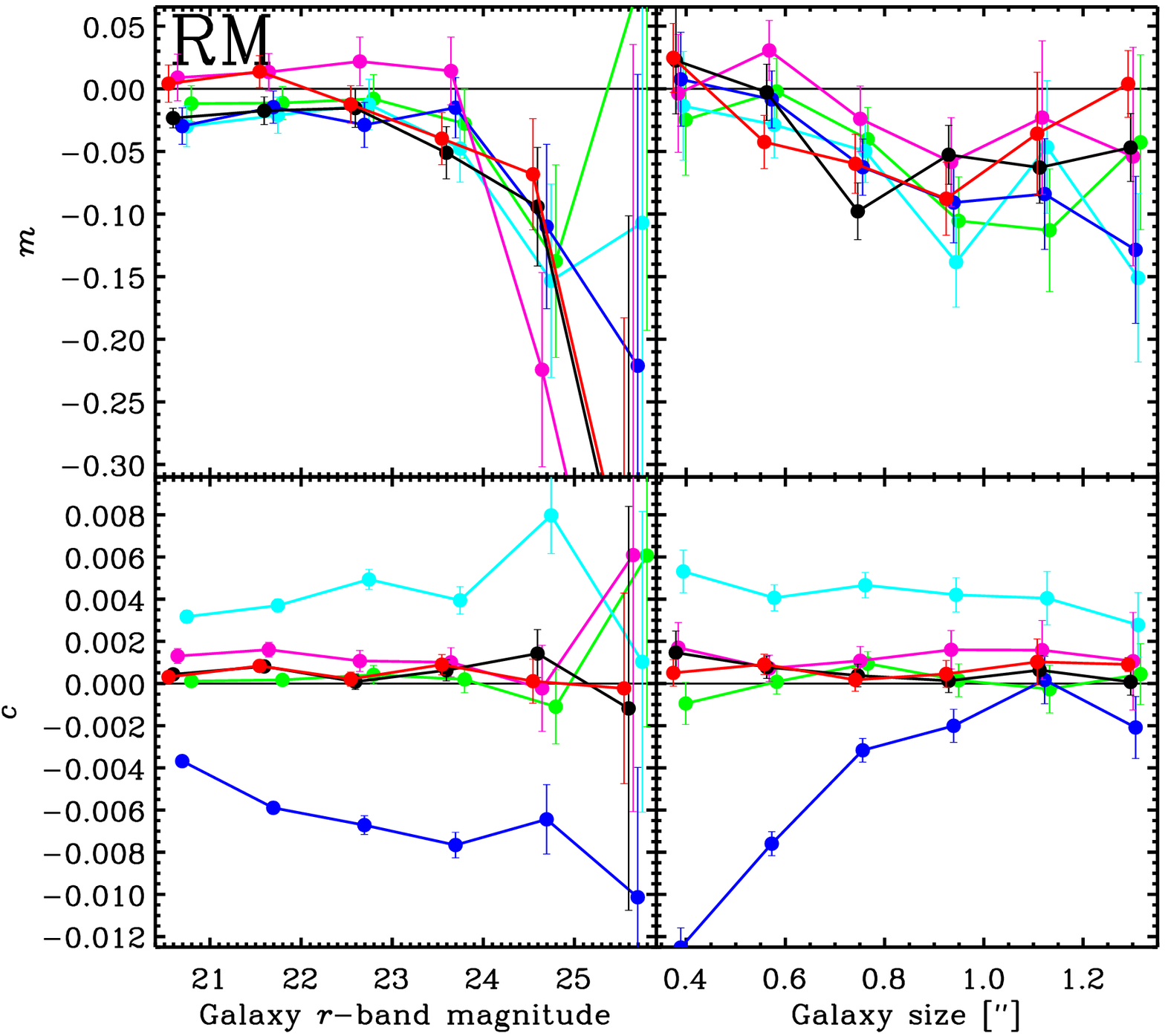,width=\figurewidth}
\epsfig{file=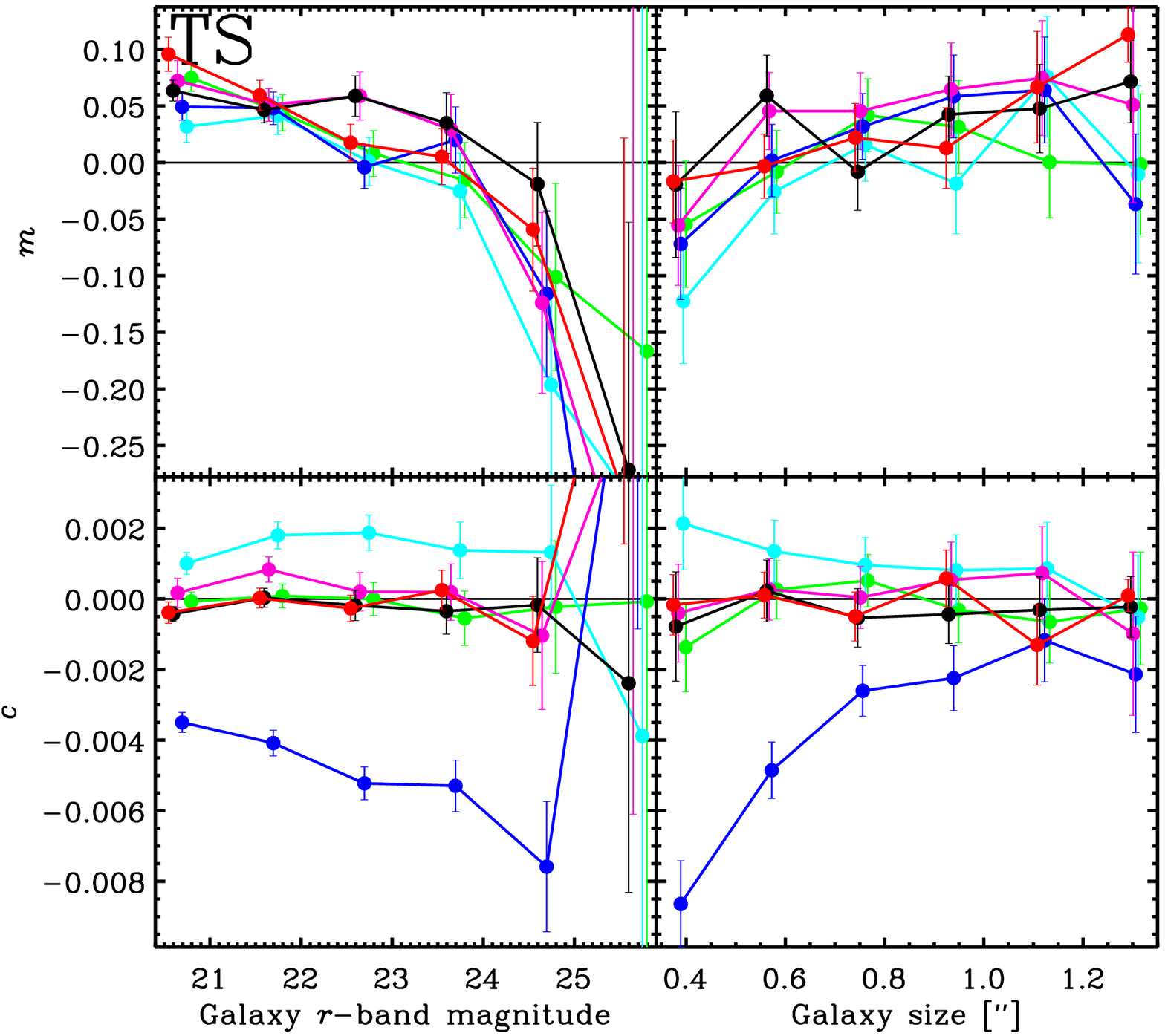,width=\figurewidth}
\caption{
Variation in shear calibration bias and residual shear offset as a function of
galaxy magnitude and size, for a representative sample of methods. The input
values of these are used, which do not have noise. The ``size'' on the 
absciss\ae\ is the unweighted rms size of galaxies from equation~(53) in 
\citet{shapelets3}. The six
coloured lines in each plot correspond to the six sets of images, coloured in
the same way  as in figure~\ref{fig:mcresults}. In all cases, measurements of
the two components of shear have been averaged.}
\label{fig:mcsizemag}
\end{center}
\end{figure*}

It is possible to measure the mean shear correctly from a large population of
galaxies, but to underestimate the shears in some and overestimate it in others.
This was frequently found to be the case in STEP2 data as a function of galaxy
size or magnitude, but correlations could also be present as a function of
galaxy morphological type. Anything that correlates with galaxy redshift is
particularly important, and figure~\ref{fig:mcsizemag} shows the correlation of
shear calibration bias and residual shear offset with galaxy size and magnitude
for an illustrative selection of shear measurement methods. Of course, these
proxies are not absolute: the fundamental parameters of interest are the size of
galaxies relative to the pixel or PSF size, and the flux of galaxies relative to
the image noise level. This must be taken into account before drawing parallel
conclusions on data sets from shallower surveys or those taken in different 
observing conditions.

The results for the TS method are fairly representative of most implementations of KSB+. The
calibration bias changes by 0.2--0.3 between bright and faint galaxies. The mean shear
calibration bias changes between methods by merely raising or lowering this curve. The ES2
curve is least affected, with only a $\sim5\%$ change. The shear calibration bias also
generally changes as a function of galaxy size. The HH method controls this the best, no
doubt due to its fitting of $P^\gamma$ as a function of size only. However, this method
still displays significant variation as a function of magnitude; it is not clear in
figure~\ref{fig:mcsizemag} because the final point expands the $y$-axis scale. The fairly
constant residual shear offset as a function of galaxy magnitude is typical; as is the
dramatic improvements for bigger galaxies in the image sets D and E with highly elliptical
PSFs. That demonstrates that it is a PSF-correction problem. The RM method behaves similarly
to the implementations of KSB+.

Other methods exhibit more idiosyncratic behaviour. The main difference is between
the KK method and the others that use a global shear responsivity ${\cal R}$.
This was calculated only once for the KK method, from the entire galaxy
population. For the other methods, it was recalculated using a subset of
galaxies for each size and magnitude bin. The large trends in the shear
calibration bias as a function of size and magnitude merely reflect the evolving
distribution of intrinsic galaxy ellipticities. The MJ, MJ2, RM and RN methods
also all look like this with a single value of ${\cal R}$, and the KK method
would presumably be improved by this step. The JB results are atypical, but
their additional noise level represents that in all analyses lacking an optimal
galaxy weighting scheme.

\section{Interpretation}
\label{sec:discussion}

We shall now revisit the questions posed in the introduction, concerning the accuracy with
which current methods can measure shear, and in which r\'egimes that accuracy begins to
deteriorate. By noting the variation of results with different PSFs, we shall investigate
the effects of changing atmospheric and observing conditions. We shall also investigate the
effects of image pixellisation, galaxy morphology and morphology evolution, selection biases
and weighting effects. In light of our results, we shall then review the consequences for
previously published measurements of cosmic shear.

The rotated pairs of galaxies provide an unprecedented level of discriminatory power, and we
can now identify high level causes of shear measurement error. Overall, both the shear
calibration (multiplicative) bias and anisotropic PSF correction (additive) errors depend
upon the PSF model. From this information, we can deduce that some aspects of shape
measurement have been suitably controlled. We can deduce that others still provide
difficulty, and it is work in these identified areas that will provide a route to the
desired sub-percent level of precision. This section describes various lessons that we have
learned from our tests, in terms of high level variables.

\subsection{PSF size}

Within the precision accessible by this analysis, all of the methods are
reassuringly tolerant to reasonable changes in observing conditions. Image set A
($0.6\arcsec$ FWHM PSF) represents typical seeing at a good site, and image set 
C ($0.8\arcsec$ FWHM PSF) the worst that might be expected for a weak lensing
survey after appropriate telescope scheduling.

Differences in the residual shear offsets between the two sets of images with
different seeing are generally not significant. The few methods with a
significant difference are JB, MH, KK and ES. In all four cases, the
2--3$\sigma$ offset is in $c_1$ but not $c_2$. The two KSB+ methods have a
positive offset, and the two shapelets methods have a negative one, but no
general conclusion seems manifest.

As expected, most methods demonstrate minimal shear calibration bias with image
set A, and fare slightly worse on image set C. Shear calibration bias for the JB
and RN methods is stable to changes in observing conditions at the $\sim0.5\%$
level. The MH KSB+ method achieves $\sim1\%$ consistency, although its applied
shear calibration factor is apparently a little overzealous.


No global trends emerge that are able to include all of the KSB+ methods. However, for the
generally most successful KSB+ implementations by MH, HH and TS, as well as the BJO2 (MJ,
MJ2) methods, $m$ is higher in image set C than in set A. These methods are all on the top
row of table~\ref{tab:method_classification}, and correct for the PSF by subtracting
combinations of shape moments. The trend is reversed in the KK deconvolution method on the
bottom row, and the calibration bias does not vary in the JB and RN methods. These correct
for the PSF via a full deconvolution. Although all implementations of KSB+ do not
necessarily fit this trend, it does suggest that the isotropic component of the PSF might be
being overcorrected by some moment subtraction schemes. Furthermore, as the PSF moments get
larger, this oversubtraction exaggerates pixellisation effects (see
\S\ref{sec:psfellipticity}). The best PSF correction is generally attained by methods that
model the full PSF and attempt to deconvolve each galaxy -- but this currently works on
slightly fewer galaxies (see \S\ref{sec:selectioneffects}).

\subsection{PSF ellipticity (and skewness)}

Image sets D and E demonstrate the ability of methods to correct for highly elliptical PSFs,
and can be compared to image set F, which has a circularly symmetric PSF. Imperfect
correction for PSF anisotropy will emerge mainly as a residual additive shear offset, $c$.
The method that was most efficient at removing all the different strengths of PSF anisotropy
to better than $0.2\%$ accuracy was MJ/MJ2, and all of the PSF deconvolution methods had
better than $1\%$ accuracy. The most successful KSB+ correction was the HH implementation.
The residual shear offsets are smallest with large galaxies, and deteriorate only as
galaxies get smaller. This behaviour is as expected if the problems are caused by
imperfect PSF correction. 

Many methods have a spurious residual shear offset in both components of shear, while the
PSF is highly elliptical in only the $\varepsilon_1$ or $\varepsilon_2$ direction. This
cross-contamination might come from the ignored off-diagonal elements of the $P^{\rm sm}$
tensor in KSB+, and is indeed slightly better controlled in MS2 (with the full tensor
inversion) than in MS1. However, this can not explain all of the effect; the off-diagonal
elements {\it are} exactly zero for the circular PSF in image set F, and a few methods (JB,
C1, RN, SP, MS1, ES2) have a significantly non-zero residual shear offset for even this set
of images.

A more likely source of the contamination lies in the measurement of stellar
ellipticities. The non-zero residual shear offsets with image set F probably
come from shot noise in the measurement of PSF ellipticity, which is higher than
the shot noise for galaxies because of the smaller number of stars. It will
therefore be worthwhile to make sure that future methods gather the maximum
possible amount of information about the PSF. In particular, small galaxies
provide as much information about the PSF as their own shapes, and this is
currently discarded. Furthermore, PSFs D and E are not only highly elliptical,
but also skewed. The centre of those PSFs therefore depends strongly on the size
of the weight function used. While the main direction of ellipticity is not in
doubt, changing the centre of the PSF also perturbs its apparent ellipticity.
The C1 method, with a fixed stellar weight function and a constant PSF model,
removes stellar ellipticity more consistently that the C2 method, in which the
size of the stellar weight function is altered to match each galaxy (although
matching the galaxy weight function provides a better shear calibration).
Methods that involve deconvolution from a full model of the PSF, or correction
of PSF non-Gaussianity, and which allow the galaxy centroid to iterate during
this process, do indeed seem to be able to better control PSF ellipticity and
centroiding errors. 

We cannot conclusively explain the cross-contamination of both shear components by a PSF
strongly elongated in only one direction, but hypothesise that it is introduced by skewness
and substructure in the PSF. Neither of these are addressed by the formalism of KSB+, and
they are both controlled more reliably by newer methods that explicitly allow such
variation. However, it is also worth noticing the remarkable success of most methods on
other image sets with more typical PSF ellipticities, and remarking that this is still a
small effect that will not dominate shear measurement for the near future.


Our investigation of PSF effects in the STEP2 images is confused by other competing
manifestations of imperfect shear measurement, and the realism of the simulations. The
combination of image pixellisation (see \S\ref{sec:psfellipticity}), correlated galaxy sizes
and magnitudes, and the evolution of intrinsic galaxy size and morphology as a function of
redshift all hinder interpretation. Higher precision tests in the future will
counterintuitively require {\it less} realistic simulated images: for example, ones that are
tailored to compare otherwise identical galaxies at fixed multiples of the PSF size.

\subsection{Pixellisation effects}
\label{sec:psfellipticity}

This is the first STEP project in which the input shear has been applied in many
directions, and in which the two components of shear can be measured
independently. In general, residual shear offsets $c$ are consistent between
components. However, we find that the $\gamma_1$ component, aligned with the
square pixel grid, is typically measured more accurately than the $\gamma_2$
component, along the diagonals.  This is even observed for image set F, in which
the analytic PSF is circularly symmetric. Since there is no other preferred
direction, this phenomenon must therefore be an effect of pixellisation. Image
pixellisation, which is similar (but not identical) to convolution, slightly
circularises galaxies, thereby reducing their ellipticity. Not explicitly
correcting for pixellisation may therefore explain both the general $1-3\%$
underestimation of $\gamma_1$, and the slightly larger underestimation of
$\gamma_2$, in which direction the distance between pixels is exaggerated. For
almost all methods, we consistently find that $m_1>m_2$.

In KSB+, there is no formal mathematical framework to deal with image
pixellisation. Two different approaches have been adopted to approximate the
integrals in equation~(\ref{eqn:ksbe}) with pixellated data. The C1 and C2
implementations calculate the value of the weight functions at the centre of
each pixel and then form a discrete sum; all of the others numerically integrate
the weight functions by subdividing pixels into a number of smaller regions.
Neither approach is ideal. Independent experiments by Tim Schrabback, running
objects with Gaussian radial profiles though his implementation of KSB+, have
shown that pixellisation can cause a systematic underestimation of $\varepsilon$
and $P^{\rm sm}$, and an overestimation of $P^{\rm sh}$. This effect can be up
to $\sim10\%$ for small objects. However, as stars and faint galaxies are
similarly affected, the error on the shear estimate approximately cancels.
Integration using linearly interpolated sub-pixels makes the measurement more
stable to the sub-pixel position of the object centroid, but slightly increases
the individual bias. \citet{Baconsims} tested a variant of the C1 method, and
found a similar $\sim13\%$ overall calibration bias, which was used to correct
subsequent measurements. With hindsight, the different calibration of $\gamma_1$
and $\gamma_2$ are also already visible in that work. 

The MJ2, KK and TS methods are least affected by pixellisation. This might have suggested
that the extraction of a shear estimator by shearing circular objects removes the problem,
were it not for the peculiar behaviour of the RN method. For this method, image sets A and C
follow the usual pattern that $m_1>m_2$, but that bias is reversed when PSF is circular
(image set F and the zero-ellipticity components of PSFs D and E). The SP method is similar.
Strangely, the JB method, which ostensibly tries the hardest to treat pixellisation with
mathematical rigour, displays the most difference between $m_1$ and $m_2$. However, this
method does break a trend by not having an overall negative shear calibration bias. If this
bias is indeed caused by pixellisation, this method appears to have most successfully
eliminated it.

Pixellisation could also hinder shear measurement, and bring about the observed results, via two
additional mechanisms. Firstly, it may exaggerate astrometric errors in the PSF, and produce the
consequences described in the previous section. We would be unable to distinguish these effects. Secondly,
the undersampling of objects may also fundamentally prevent the measurement of their high order shape
moments. All of the STEP2 PSFs (and hence the galaxies) are Nyquist sampled. It would be unfortunate for
lensing if Nyquist sampling were theoretically sufficient to measure astrometry, but not shapes. As it
happens, for methods other than MJ, the pixellisation bias is more pronounced for image set C (with poor
seeing, and therefore better sampled) than on image set A (with good seeing). This suggests that the
pixellisation effects are {\it not} due to undersampling. The STEP1 simulations had the same pixel scale
but worse seeing ($\sim 1\arcsec$ FWHM), so objects were better sampled there.

We therefore hypothesise that the circularising effects of pixellisation explain the general
underestimation of shear and the differential calibration of the $\gamma_1$ and $\gamma_2$ components.
Indeed, a dedicated study of simulated images with varying pixel scales by High \etal\ (in preparation)
supports this view. They find that the shear calibration bias of the RRG method tends to zero  with
infinitely small pixels, grows linearly with pixel scale, and that the bias $m_2\approx\sqrt{2}m_1$.
Because of the isotropy of the Universe, this differential calibration of shear estimators ought not
affect two-point cosmic shear statistics. But it can certainly affect the reconstruction of individual
cluster mass distributions, and is inherently quite disconcerting. The next STEP project will feature sets
of images with varying pixel scales to investigate this effect on a wider scale. In the mean time, dealing
properly with pixellisation will provide a promising direction for further improvement in shear
measurement methods.

\subsection{Galaxy morphology}

The introduction of complex galaxy morphologies tends to hinder shear measurement with KSB+ methods. The shear
calibration bias is more negative with image set A (shapelet galaxies) than with image set B (simple galaxies)
for the C1, C2, MH, SP, MS1, TS and ES1 implementations. Of the implementations of KSB+, only HH and MS2 reverse
this trend. This is perhaps not surprising, given the inherent limitation of KSB+ in assuming that the
ellipticity of a galaxy does not change as a function of radius.

Many of the newer methods deal with complex galaxy morphologies very successfully. Particularly KK, but also the MJ and MJ2
methods, have no significant difference in the shear calibration bias or residual shear offset measured between image sets A
and B. Future ground-based shear surveys are therefore unlikely to be limited at the ~0.5\% level by complex galaxy
morphologies. Indeed, it is apparent in figure~\ref{fig:shapelet_image} that most of the substructure in galaxies that will
be used for lensing analyis is destroyed by the atmospheric seeing. Although complex galaxy morphologies may become important
at the level of a few tenths of a percent, they do not currently pose a dominant source of error or instability in shear
measurement from the ground.


One of the crucial findings of this study, however, concerns the effect of
galaxy morphology {\it evolution}. This could potentially affect the calibration
of shear measurement as a function of galaxy redshift, and is investigated
further in the next section.


In the next STEP project, which will simulate space-based observations, we shall repeat our investigation of
galaxy morphology by comparing three similar sets of image simulations. Galaxy substructure will be better
resolved from space and, because the galaxies observed there are likely to be at a higher redshift, their
intrinsic morphologies may be both more irregular and more rapidly evolving. Both of these effects will amplify
any differences seen from the ground.

\subsection{Shear calibration for different galaxy populations}
\label{sec:variable_calibration}

The STEP2 results reveal that the calibration bias of some shear measurement methods depends
upon the size and magnitude of galaxies. There seem to be two causes. There is often a
sudden $\sim30\%$ deterioration of performance at very faint magnitudes, due to being noise
blown up during the nonlinear process of shear measurement (and exacerbated by
ellipticity-dependent galaxy weighting schemes). This is even observed with many methods that
are otherwise robust (\eg\  HH, MJ2, RN), and may urge more caution in the use of faint
galaxies at the limits of detection. There is also a gradual transition in shear calibration
between bright and faint galaxies that is probably caused by evolution of the intrinsic
morphology distribution as a function of redshift. The observed variation is least
pronounced for image set B, in which the galaxies explicitly do not evolve.

Shear calibration bias that changes gradually as a function of galaxy redshift has important
consequences for any weak lensing measurement. In a 2D survey, it will change the effective
redshift distribution of source galaxies, with all the consequences discussed by
\citet{vWzfudge}. In a 3D analysis, it will affect the perceived redshift evolution of the
matter power spectrum, and the apparent large-scale geometry of the universe. During the
STEP2 analysis, we have developed ways to partially control this, as a function of other
observables like galaxy size and magnitude. To first order, these act as suitable proxies
for redshift, but the underlying causes will need to be well understood, because neither of
these are redshift. Even if the mean shear in size/magnitude bins could be made correct,
this doesn't necessarily imply that the mean shear would be correct in redshift bins. The
techniques could be applied in multicolour surveys as a function of photometric redshift,
but this is not perfect either, not least because of the inevitable presence of catastrophic
photo-$z$ failures.

The obvious place to start looking for shear calibration errors is in the shear
susceptibility and responsivity factors. All of the KSB+ implementations allow
variation in $P^\gamma$ as a function of at least one of galaxy size and galaxy
magnitude. However, the behaviour is neither well understood, nor stable at the
desired level of precision. \citet{Massey} have already observed that $P^\gamma$
fitted from a population ensemble varies for any given object as a function of
the catalogue selection cuts. There is less variation in the shear calibration
bias of the MS1 method ($\Delta m\approx0.1$), which fits only the trace of
$P^\gamma$, than of the MS2 method ($\Delta m\approx0.2$), which models the
entire tensor -- except for image set B, in which there is little variation in
either. Realistic galaxy morphologies therefore do not have shear susceptibility
that is a simple functions of these observables; and trying to model the
variation of all the components of this tensor merely adds noise. The TS
implementation of KSB+, which uses $P^\gamma$ from individual objects, suffers
particularly from this noise, which enters into the denominator
equation~(\ref{eqn:epsilonoverpgamma}), and has at least as much sudden
deterioration at faint magnitudes as other methods. However, this method is
about the least affected by gradual variation in shear calibration bias, with
$\Delta m\approx0.05$. Size galaxy size and magnitude are correlated, the
variation with galaxy magnitude usually carries over to variation with galaxy
size. However, the HH method has notably little variation in $m$ as a function
of galaxy size. This is presumably due to the particularly individual form of
the function used to model $P^\gamma(r_g)$. Unfortunately, $P^\gamma$ is not
fitted as a function of galaxy magnitude, and the HH method still shows strong
($\Delta m\approx0.1$) variation with this. The shear susceptibility in this
implementation is calculated separately in three magnitude bins, and correction
of the faintest galaxies therefore required an extrapolation.

Many of the other shear measurement methods require global calibration via a responsivity
${\cal R}$ factor, which is determined from the distribution of galaxy ellipticities. This
factor is designed to ensure that the mean shear in a population is unbiased. However, it
must be calculated from precisely that population. For the KK method, it was calculated only
once, from the entire catalogue. Although it estimated the overall mean shear correctly, it
then underestimated the shear in small/faint galaxies, and overestimated that in
large/bright galaxies. This bias was addressed for the MJ, MJ2, RM and RN methods by
recalculating ${\cal R}$ within each size and magnitude bin. There is no particular reason
why this should not, in future, be fitted and allowed to vary continuously like the shear
susceptibility in KSB+ methods. The estimates of ${\cal R}$ in bins were more noisy, but
removed the differential shear calibration (in fact, the variation as a function of galaxy
magnitude was slightly overcorrected in the case of the MJ2 and RM methods).\\

\subsection{Galaxy selection effects}
\label{sec:selectioneffects}

There is a marked difference between the depth of the various galaxy catalogues. At one
extreme, the C1/C2 catalogues are deeper, and more ambitious, than all others. At the other,
the RN catalogue (and to some extent the MJ/MJ2 catalogue) is very shallow. The RN method
obtained extremely good results, but only from large and bright galaxies, and it would be
interesting to test whether its PSF deconvolution iteration can converge with a deeper
sample. The JB catalogue of individual rotated and unrotated images is deeper, but not all
of the galaxies at the magnitude limit converged successfully, leading to a relatively
shallow matched catalogue. We could conclude from this that the full deconvolution of every
galaxy is an overly ambitious goal: it is a panacea for many image analysis problems, but
all that we require is one shear estimator. Maximising the number density of useable
galaxies will remain crucial in the near future, to overcome noise from their intrinsic
ellipticities. However, there has been far less time spent developing the deconvolution
methods than the moment subtraction methods, so we reserve judgement for now because of
their promise of robust PSF correction. Furthermore, it is not only the methods that require
complicated iterations that suffer from catalogue shortcomings: the SP catalogue includes a
significant number of spurious detections (10\%) and stars (1\%). Neither of these contain
any shear signal, and their presence partly explains the large, negative calibration bias of
the SP method in the rotated and unrotated images (they are removed during the galaxy
matching).

Most other methods use a fairly standard density of $\sim30$ galaxies per square arcminute
in this simulated data.
This is unlikely to be increased dramatically by any future weak lensing observations. Since
selection effects in the STEP2 analysis must be measured from the individual unrotated and
rotated catalogues, rather than the matched catalogues, the results about catalogue
selection biases are hardly more profound than those of STEP1.\\

\subsection{Galaxy weighting schemes}
\label{sec:weightingschemes}

The weighting schemes applied to galaxies also vary significantly between
methods used in this paper, and these do affect the results in the matched
catalogue. Most of the methods increase the contribution to the estimated mean
shear from those galaxies whose shapes are thought to be most accurately
measured. Such schemes have long been used in the analysis of real 2D data, but
the exact form of the weighting scheme as a function of size, magnitude and
ellipticity varies widely. Even more sophisticated weighting schemes will also
need to be developed for the 3D analyses essential to fully exploit future weak
lensing surveys.

In this analysis, the effectiveness of each weighting scheme can be seen in the
difference between the size of error bars in the analysis of independent
galaxies and of rotated/unrotated pairs of matched galaxies. In the independent
analysis, the scatter includes components from intrinsic galaxy shapes and
measurement noise (\eg\ due to photon shot noise). The former is essentially
removed by matching pairs of galaxies. If a set of error bars shrink
dramatically by matching, the method was dominated by intrinsic galaxy shapes:
this is an ideal situation. If the error bars change little, the measurement
was dominated by measurement noise.

The weighting schemes of MJ2 and KK are very effective in this analysis: their
error bars shrink by up to $75\%$. The weighting schemes of HH, SP and MJ
are similarly effective -- but these methods weight ellipticities using a
function of ellipticity, which may be less accurate in regimes where the mean
shear is large, such as cluster mass reconstruction. Indeed, the aggressive
weighting scheme of MJ was shown in STEP1 to be useful with small input shears,
but introduced a non-linear shear response that became  important if the shear
was high.  A new weighting scheme was developed for MJ2 to address this concern;
however, the range of input shears in STEP2 does not provide sufficient lever
arm to evaluate the potential nonlinear response of any method. 

The value of a successful weighting scheme is demonstrated by the lesser performance of
methods without one. The JB, TS and ES2 methods apply crude weighting schemes that are
merely a step function (cut) in galaxy size and magnitude. Their error bars shrink by only
30--$50\%$ during galaxy matching. Their results are also less stable to the sudden
deterioration of performance seen in several methods with galaxies fainter than or smaller
than a particular limit. This shortfall is easy to correct, and we urge the rapid adoption
of a more sophisticated weighting scheme in those methods.

\begin{figure}
\begin{center}
\epsfig{file=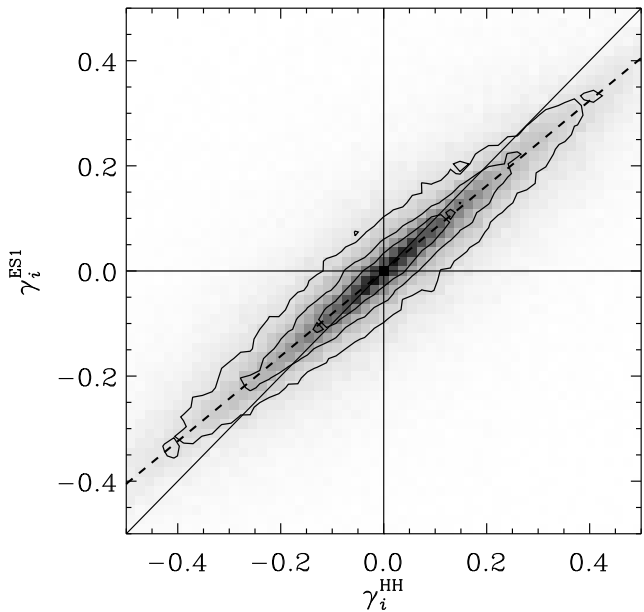,width=8.55cm,angle=0,clip=}
\caption{Comparison of shear measurement in real CFHTLS deep data, from 
a galaxy-by-galaxy comparison of matched
catalogues from the ES1 analysis \citep{cfhtls_deep} and a reanalysis
using the HH method.
The relative calibration of both components of shear are indistinguishable,
and are here included in the same plot.
A slope of unity would imply perfect agreement.
The dashed line indicates the relative calibration of the
two methods in simulated
image set~C, which is the most closely matched to actual observing conditions.
Although this should not be regarded as a strict prediction, since there are
many image parameters that are not matched, its agreement
with the real data is striking.}
\label{fig:cfhtls_wvsd}
\end{center}
\end{figure}

It is important to remember the limitations of the STEP simulations to optimise
a galaxy weighting scheme, because of their inherent simplification that all
galaxies are sheared by the same amount. In real data, the lensing signal
increases cumulatively with redshift, and the distant galaxies therefore contain
the most valuable signal. However, when weighting objects by the accuracy of
their shape measurement, it is the contribution of these small, faint sources
that is usually downweighted. It would instead be better to set weights that
vary as a function of the signal to noise in shear signal -- although the exact
variation of the signal is of course unknown in advance. A statistically
``optimal'' weighting scheme verified from the STEP simulations will therefore
not be optimal in practice. Weighting schemes can also act like calibration
biases as a function of galaxy redshift, exacerbating the problems of
differential shear calibration discussed in the previous section.

\subsection{Consequences for previously published measurements}


\begin{figure}
\begin{center}
\epsfig{file=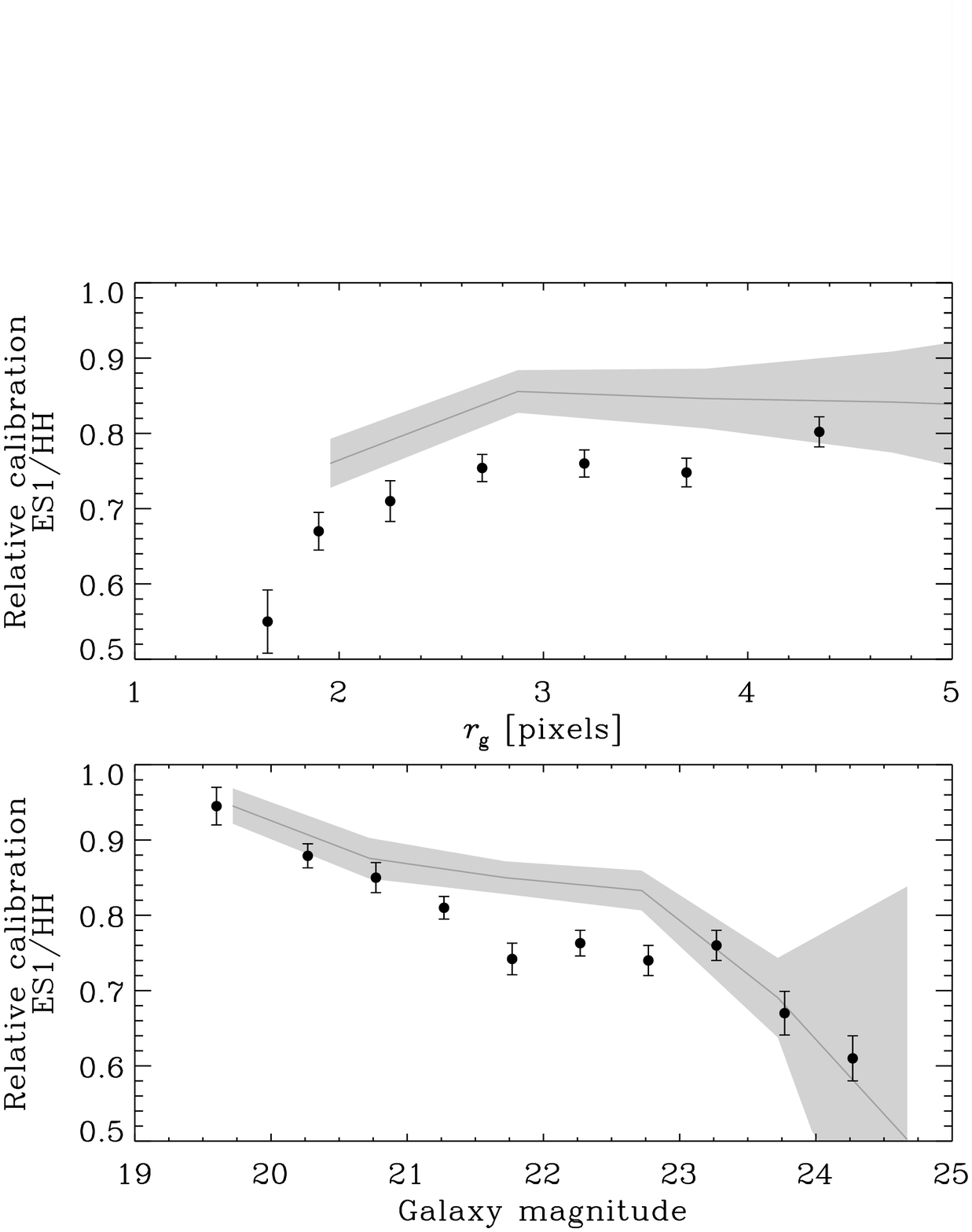,width=8.55cm,angle=0,clip=}
\caption{Comparison of shear measurement in real CFHTLS deep data, as a function
of galaxy size and magnitude. The relative shear calibration of the ES1 and HH
methods is obtained from the ratio of the mean shear calculated
in $3\arcmin\times3\arcmin$ subfields of each CFHTLS deep field.
A value of unity would imply perfect agreement between the catalogues.
Note that we have reconciled the
different definitions of galaxy size in the simulations compared to real data by
approximating $R\approx r_g$. We have dealt with the different relationship
between galaxy magnitude and signal-to-noise (\cf~\S\ref{sec:mcmagsizeresults})
by offseting the magnitudes of objects in the deeper simulated data by -1.
The grey band indicates the relative calibration of the two methods in simulated
image set~C, which is the most closely matched to the CFHTLS data.}
\label{fig:cfhtls_wvsdsizemag}
\end{center}
\end{figure}

The largest cosmic shear survey to date, which has been published since STEP1,
comes from the {\it Canada-France-Hawaii Telescope Legacy Survey} (CFHTLS)
$i$-band data. The CFHTLS wide survey \citep{cfhtls_wide} was analysed using the
HH shear measurement method, and the CFHTLS deep survey \citep{cfhtls_deep}
using the ES1 method. These methods perform very differently on the simulated
images.

\begin{figure*} \begin{center}
\epsfig{file=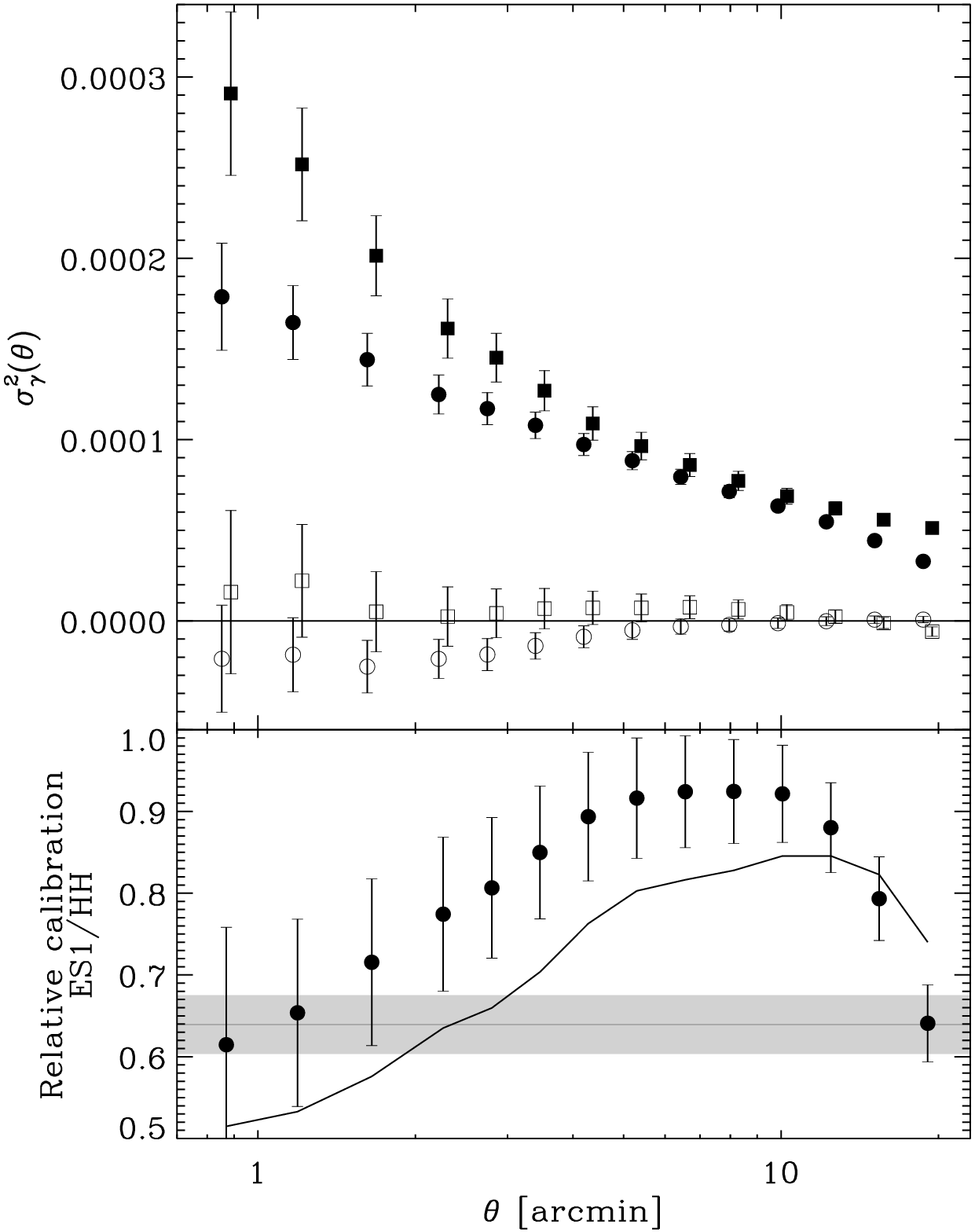,width=8.55cm,angle=0,clip=} ~
\epsfig{file=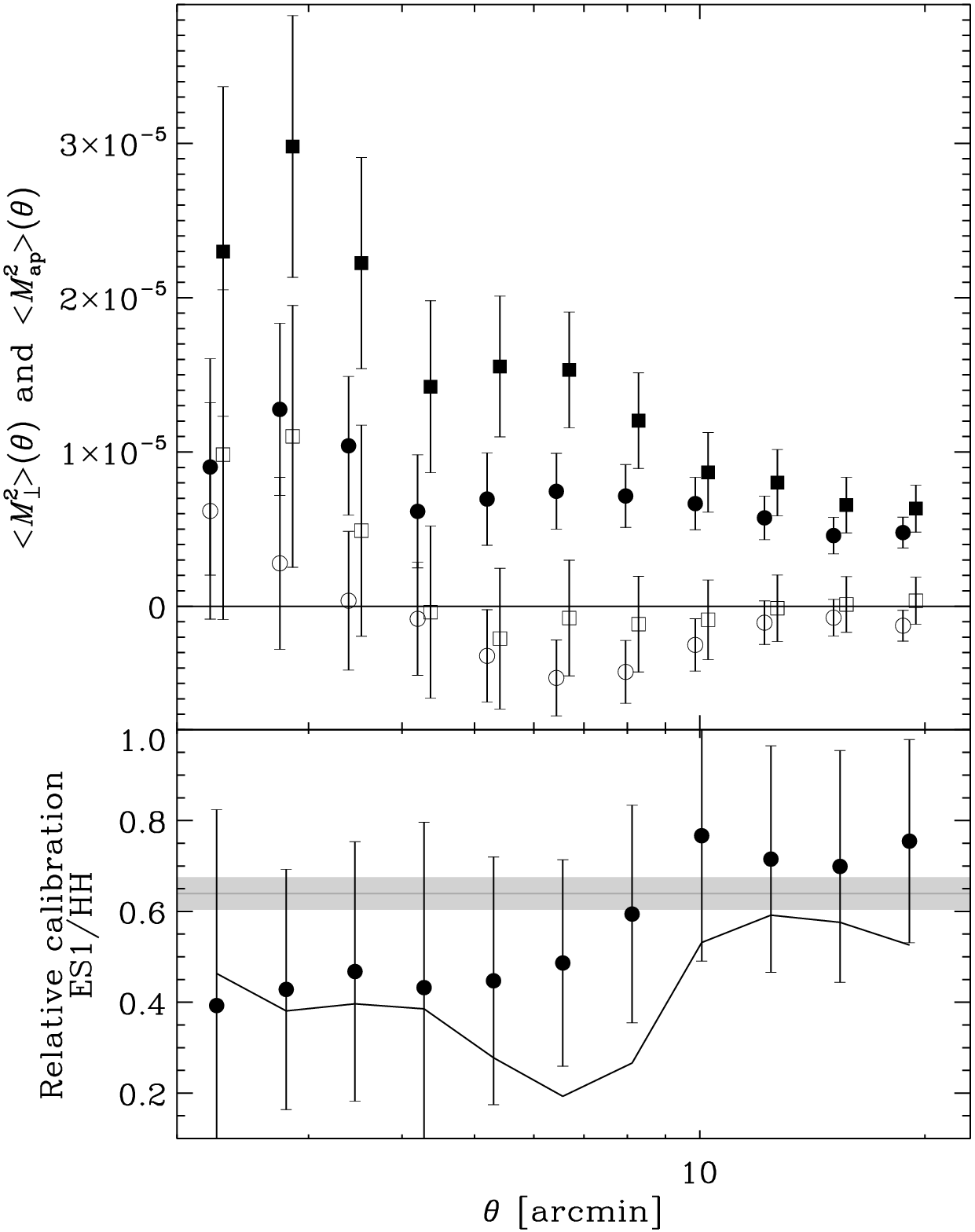,width=8.55cm,angle=0,clip=} \caption{Comparison of
shear-shear correlation functions measured from real CFHTLS deep survey data,
after HH (squares) and ES1 (circles) analyses. The
correlation functions are split into $E$- and $B$-modes in two different ways: the
variance of the shear in cells is shown on the left as a function of cell
radius, and the variance of the mass aperture statistic is shown on the right.
In both cases, the solid points show the $E$-mode, and the open points the
$B$-mode. The error bars show statistical errors only (\ie\ no account is made for
cosmic variance since the survey region is identical), 
but note that the difference between the two data sets is in
fact more significant than indicated, because the same galaxies are used in each
analysis, so noise enters only from the shape measurement process and not from
variation in intrinsic galaxy ellipticities.
In the lower panels, the points show the ratio of the $E$-modes calculated from
the two analyses, and the lines show the ratio of the $E$-modes plus $B$-modes. 
The grey bands indicate the relative calibration of the two methods in simulated
image set~C, which is the most closely matched to actual observing conditions.}
\label{fig:cfhtls_tophat}
\end{center} \end{figure*}

The HH method recovers shear in the STEP2 images with remarkable success. The
seeing in the CFHTLS data is most similar to that in image set~C, for which the
overall shear calibration is within $1\%$: well within the current error budget.
\citet{cfhtls_wide} also featured a parallel analysis using an independent KSB+
pipeline, which agreed with the HH results, and also demonstrates the potential
robustness of KSB+ at this level of precision (similar comparisons have also
been performed by \citet{Massey} and \citet{gemscs2}, and these also give
results consistent with that work). The HH method had difficulty only with the
calibration of very faint galaxies, due to its non-smooth fitting of $P^\gamma$
as a function of magnitude. If a similar bias is present in the CFHTLS analysis,
it will have lowered the effective redshift distribution of source galaxies, and
slightly diluted the overall signal. Both of these effects would have led to an
underestimation of $\sigma_8$, although only by a small amount, due to the low
weight given to faint galaxies. As discussed by \citet{vWzfudge}, a more
significant bias (which acts in the opposite sense) arises from using the Hubble
Deep Field to infer the redshift distribution of galaxies. As the survey area of
the CFHTLS grows, and the statistical error bars decrease, it may be prudent for
this analysis to conservatively use slightly fewer galaxies.

The ES1 method underestimates shear in the STEP2 images by 20\% overall, and by
as much as 30\% for the faintest galaxies. We have verified this result
retrospectively in STEP1 simulations, and also confirmed it in real images, by
comparing the results of the HH and ES1 shear measurement pipelines on the same
CFHTLS deep data. Of course, the true ``input'' shear is not known for real
data. Figure~\ref{fig:cfhtls_wvsd} shows the {\it relative} calibration of the
two methods in real data, with the dashed line indicating their relative
calibration in simulated image set~C. This should not be interpreted as a strict
prediction, since the simulation was not designed to mimic this specific survey:
the simulated and real data have very different noise properties, and the only
similarity between their PSFs is their size. Nonetheless, the agreement is
impressive. Figure~\ref{fig:cfhtls_wvsdsizemag} shows a further comparison of
the methods' relative calibration, in which galaxies have been split by size and
magnitude. Once again, overlaying the performance of ES1 from image set~C
confirms the results of the STEP simulations with remarkable success. A likely
source of the shear calibration bias is in the smoothing of $P^\gamma$ as a
function of $r_g$ and magnitude. Tests indicate that the shear susceptibility is
more stable if it is instead fitted as a smooth function of size and magnitude,
or even by using the raw values. The strong magnitude dependence is probably
related to the sudden drop at small sizes. Note
also that both pipelines started from scratch with the individual exposures,
reducing them and stacking them independently. All of the available exposures
are stacked in both versions, so the two sets of images have effectively the
same depth. The full data reduction pipeline of both groups is being tested, and
the differences could therefore have been introduced at any stage.

Figure~\ref{fig:cfhtls_tophat} shows the two-point correlation functions of the
matched shear catalogues (using the weights of the individual catalogues), which
are normally used to constrain cosmological parameters at the end of a weak
lensing analysis. Although the ES1 analysis consistently measures a lower signal
than the HH analysis, the discrepancy is not uniform on all scales. The relative
bias is most pronounced on small scales when measuring the variance of the
aperture mass statistic, and on both small and large scales for the shear
variance in cells. Such variation is not seen in the galaxy-by-galaxy
comparison of relative shear calibration. For example, the signal in
figure~\ref{fig:cfhtls_wvsdsizemag} is stable to changes in the size of the area
over which the shears are averaged.

We hypothesise that there may therefore be an {\it additional} source of bias in
the ES1 CFHTLS analysis, due to PSF anisotropy residuals. Since the PSF
anisotropy varies spatially, the residual would average out across the survey,
and not affect the overall bias. The correlation functions were calculated using
the procedure in \citet{vWb04}, which deals with an unknown constant of
integration in the calculation of $\sigma^2_\gamma(\theta)$ by forcing the
$B$-modes of to zero on large scales. This prior on the $B$-modes can add
spurious power to the $E$-modes, and could have artificially re-raised the
cosmic shear signal. Indeed, the ratio of the sum of the $E$- and $B$-modes
between analyses is flatter than that of the $E$-modes alone. Furthermore, the
star-star correlation functions \citep{cfhtls_deep} show an excess before PSF
correction, on similar scales to that observed in the left-hand panel of
figure~\ref{fig:cfhtls_tophat}.

A na\"ive correction for a 20\% shear calibration bias in the CFHTLS deep survey
\citep{cfhtls_deep} would raise the measured value of $\sigma_8$ almost
proportionally. This would remain within the estimated error budget for the
lensing analysis due to non-Gaussian cosmic variance \citep{sem_nongauss}, but
adds tension to an existing discrepancy with the three year results from WMAP
\citep{wmap3}. In practice, a more sophisticated recalibration will probably be
required. If our hypothesis of an additional systematic is correct, this would
have partially cancelled the shear calibration bias. Judging by the ratio of the
observed correlation functions, the net underestimation of $\sigma_8$ could have
been around 10--15\%. More work is needed to test this hypothesis; but it is
beyond the scope of this paper. A full reanalysis of the CFHTLS survey,
including the latest data, will therefore follow. 

The striking confirmation of the STEP results on real data demonstrates the
success of our simulation project, and highlights the vital role that artificial
images will play in the exploitation of future surveys. Ideally, they ought not
be relied upon for simple empirical recalibration, but they will be essential to
verify the performance of methods derived from first principles. The STEP images
remain publicly available to test future weak lensing analyses. Simultaneously,
the complexity of our correlation functions results also highlight the
importance of subtleties in weak shear measurement that may arise only within
the complex environment of real observational data. To fully understand such
effects, we shall pursue further development of the {\it data}STEP
project$^{\ref{url:step}}$, an ongoing comparison of the output from various
shear measurement methods on a common sample of real data.


\section{Conclusions}
\label{sec:conc}

Performance has improved since STEP1, and the STEP project continues to drive
progress and innovation in shear measurement methods. The most accurate methods,
with better than $\sim2\%$ level calibration errors for most of the tested
observing conditions, were the MJ2 implementation of BJ02, the TS and HH
implementations of KSB+, the KK and JB implementations of shapelets and the RM
implementation of Reglens. Particular advances are apparent in methods that used
the results of STEP1 to tune their algorithms, which bodes well for the future
of this project. For example, the introduction of a calibration factor to the TS
method has proved reassuringly robust with our new, more realistic simulated
images. We have also verified the STEP results on real data, finding striking
confirmation of methods' relative shear calibration in the CFHTLS deep survey. 

There is no one shear measurement method that is doing everything
best. With the increased precision possible in this analysis, we can now
distinguish all of the methods from {\it perfect} performance. Since absolute
shear calibration can not be directly ascertained from real data, this remains
the most important issue. The calibration bias in most methods leads to a slight
underestimation of shear. Both the shear calibration (multiplicative) errors and
anisotropic PSF correction (additive) errors are also found to depend upon
characteristics of the PSF. Technical advances in individual methods will
therefore still be required. Ideally, one would attempt to take the most
successful aspect of several methods and combine them. The fundamentally
different approaches to the two main tasks in shear measurement make this
difficult, but there is common ground (\eg\ object detection algorithms, the
shapelet basis functions, and galaxy weighting  schemes), so the individual
lessons learned with each method may not necessarily be irreconcilable. To this
end, we have developed a classification scheme for shear measurement methods,
and have described all existing methods in a common language so that their
similarities and differences are apparent. Development is continuing in earnest.


We have used our improved simulations to identify various aspects of shear
measurement that have been effectively solved at the current level of precision.
We have also uncovered other, specific areas that remain problematic. Studying
these may provide a route to the most rapid technological advances. Development
needs to be focussed towards:

\begin{itemize}
\item Pixellisation
\item Correlated background noise
\item PSF measurement
\item Galaxy morphology evolution.
\end{itemize}

\noindent These four points are explained below.\\

This is the first STEP project in which the input shear has been applied in
arbitrary directions relative to the pixel grid. That this direction affects the
calibration of shear measurement methods, even for images with a circular PSF
and no other preferred direction, implies that pixellisation is not fully
controlled.  Pixel effects may also explain the general tendency of methods to
underestimate shear. Since no explicit provision is made for pixellisation in
many methods, this result is not surprising. This work has quantified just how
much of an effect it has, and thereby emphasised the importance of a proper
treatment in the future. High \etal\ (in preparation) are specifically
investigating pixellisation through tailor-made image simulations with varying
pixel scales. 

Although not all data sets have background noise that is significantly
correlated between adjacent pixels, it is particularly apparent in natively
undersampled data, for which several exposures dithered by sub-pixel shifts must
be co-added. The introduction of correlated background noise to the STEP2
simulations hindered several methods: during the detection of faint objects, the
modelling of objects to a specified fidelity, and the weighting of individual
shear estimators. Now that this issue has been raised, work is underway in the
context of several of the shear measurement methods.

Various schemes have been developed to improve PSF interpolation across a field
of view \citep{Hoekstra03,JarvisPSF}, but some methods seem to be having trouble
with the initial measurements of the PSF from individual stars. The measurement
of the shape of each star affects shear estimates from many galaxies, and is
therefore of vital importance. When the PSF is highly elliptical, this work has
revealed some peculiar residual shear offsets, in the directions orthogonal (at
$45^\circ$) to that ellipticity. We have not yet found a satisfactory
explanation for this, but speculate that it might be caused by difficulties
measuring the centroid and the ellipticity of stars that have substructure,
skewness, and no single, well-defined ellipticity. Methods that model the full
PSF, and especially those that attempt PSF  deconvolution, are less affected,
but at the expense of a having smaller number density of useable galaxies for
which the complicated deconvolution algorithms currently converge. This issue
will require further investigation, and questions about the residual shears
cannot be addressed until this is resolved.

Issues of galaxy morphology evolution become particularly important for those
methods whose calibration relies on the overall distribution of galaxies'
intrinsic ellipticities. High redshift galaxies are both more elliptical and
more irregular; and evolution in the ellipticity variance directly affects the
shear calibration. For a 2D cosmic shear survey, even if the mean shear is
correctly measured, this can bias the effective redshift distribution of source
galaxies and the geometrical interpretation of the lensing signal, with all the
consequences discussed in \citet{vWzfudge}. For a 3D analysis, it can change the
apparent redshift evolution of the signal and hence the apparent cosmological
matter distribution.\\

The next STEP project will analyse a set of simulated space-based images. With
their higher spatial resolution, we expect that variation in galaxy morphology
will more profoundly affect shear measurement. We will therefore repeat the
exercise of comparing the analysis of complex shapelet galaxies with more
idealised objects, and also separate the galaxy populations by morphological
class. The cuspy space-based PSFs will provide a different (easier) r\'{e}gime
in which to test centering, and we shall explicitly avoid PSF interpolation
errors by allowing methods to assume that the PSF is constant. This should make
interpretation easier. Background noise will also be left intentionally
uncorrelated. However, variations in the pixel scale will be introduced, to
specifically test methods' robustness to pixellisation effects.

Such ongoing improvements are vital to the success of gravitational lensing as a
viable probe of cosmology. Although the measurement of weak lensing is not
limited by unknown physical processes, the technical aspect of galaxy shape
measurement at such high precision remains computationally challenging. In this
paper, we have demonstrated that simulated images can drive progress in this
field, and can provide a robust test of shear measurement on real data. Previous
cosmic shear measurements would have benefitted from access to STEP, and the
future exploitation of dedicated surveys relies upon the development of methods
that are being tested here first. Both the tools and the collective will are now
in place to meet this challenge. The STEP simulations remain publicly available,
and the weak lensing community is progressing to the next level of technical
refinement in a spirit of open cooperation. We conclude with the hope that, by
accessing the shared technical knowledge compiled by the STEP projects, all
future shear measurement methods will be able to reliably and accurately measure
weak lensing shear.


\section*{Acknowledgements}

Funding for the development of the shapelet image simulation pipeline and STEP
telecons was provided by DoE grant \#96859. Further funding for group 
communication was provided by CITA and NSERC. We thank the NASA Jet Propulsion
Laboratory for financial and administrative support of the STEP workshop. We
thank the COSMOS collaboration, particularly Anton Koekemoer and Nick Scoville,
for providing the high resolution HST images from which the population of galaxy
morphologies was drawn. We thank Matt Ferry and Mandeep Gill for continued help
in developing the shapelets image simulation pipeline. We thank CalTech ADPF
staff and particularly Patrick Shopbell for help with the computing resources
required to manufacture and to distribute the simulated images. CH is supported
by a CITA national fellowship.

\bibliographystyle{mn2e}
\bibliography{ceh_2005}

\appendix

\begin{table*}
\begin{center}
\begin{tabular}{cr@{$\pm$}lr@{$\pm$}lr@{$\pm$}lr@{$\pm$}lr@{$\pm$}lr@{$\pm$}l}
{\bf{Author}} & 
\multicolumn{2}{c}{\bf{Image set A}} &
\multicolumn{2}{c}{\bf{Image set B}} & 
\multicolumn{2}{c}{\bf{Image set C}} & 
\multicolumn{2}{c}{\bf{Image set D}} & 
\multicolumn{2}{c}{\bf{Image set E}} & 
\multicolumn{2}{c}{\bf{Image set F}} \\
\hline\hline
\multirow{2}{*}{JB} &    2.34 &   2.89 &   -1.46 &   2.46 &    5.08 &   3.06 &    7.31 &   3.08 &    3.44 &   3.02 &    1.92 &   3.14 \\
                    &    5.34 &   2.91 &   -7.78 &   2.05 &   -2.57 &   2.87 &    2.12 &   3.01 &   -6.85 &   3.77 &   -1.25 &   3.31 \\\hline
\multirow{2}{*}{C1} &   -9.33 &   1.12 &   -6.30 &   1.12 &  -15.78 &   1.27 &  -17.01 &   1.05 &  -15.60 &   1.09 &   -9.18 &   1.24 \\
                    &   -7.44 &   1.07 &   -6.06 &   0.97 &  -11.69 &   1.19 &  -18.11 &   0.97 &  -18.90 &   1.35 &   -9.22 &   1.32 \\\hline
\multirow{2}{*}{C2} &   -7.97 &   1.13 &   -4.13 &   1.14 &  -12.68 &   1.35 &   -7.39 &   1.19 &   -7.64 &   1.19 &   -5.50 &   1.28 \\
                    &   -6.05 &   1.12 &   -4.43 &   0.97 &   -8.31 &   1.24 &   -9.16 &   1.13 &  -11.99 &   1.49 &   -6.50 &   1.37 \\\hline
\multirow{2}{*}{MH} &    3.73 &   1.30 &    7.76 &   1.46 &    6.26 &   1.55 &    2.51 &   1.37 &    0.82 &   1.41 &    1.78 &   1.56 \\
                    &    3.67 &   1.24 &    7.70 &   1.23 &    3.39 &   1.52 &    4.57 &   1.39 &   -2.88 &   1.75 &    0.86 &   1.63 \\\hline
\multirow{2}{*}{HH} &   -0.05 &   0.81 &   -3.57 &   0.84 &    1.53 &   0.94 &   -1.94 &   0.82 &   -1.33 &   0.83 &   -0.30 &   0.90 \\
                    &   -1.88 &   0.79 &   -3.33 &   0.70 &    0.28 &   0.88 &   -2.50 &   0.81 &   -4.95 &   1.04 &   -1.89 &   0.94 \\\hline
\multirow{2}{*}{MJ~~}&  -1.05 &   1.18 &   -1.07 &   1.01 &   -0.35 &   1.31 &   -0.98 &   1.20 &   -1.92 &   1.21 &   -2.81 &   1.30 \\
                    &   -4.07 &   1.10 &   -2.11 &   0.84 &    1.09 &   1.21 &   -0.75 &   1.16 &   -3.18 &   1.49 &   -3.54 &   1.33 \\\hline
\multirow{2}{*}{MJ2}&   -0.74 &   0.97 &   -3.99 &   0.89 &    0.81 &   1.04 &   -0.73 &   0.94 &   -0.01 &   0.94 &   -1.77 &   0.96 \\
                    &   -3.04 &   0.90 &   -3.20 &   0.75 &    1.58 &   0.98 &   -1.49 &   0.92 &   -4.10 &   1.14 &   -1.14 &   1.01 \\\hline
\multirow{2}{*}{KK} &   -1.06 &   1.05 &   -0.74 &   1.21 &   -6.28 &   1.26 &   -3.38 &   1.15 &   -3.04 &   1.13 &   -2.58 &   1.19 \\
                    &   -2.20 &   1.01 &   -1.96 &   1.01 &   -4.34 &   1.17 &   -2.38 &   1.10 &   -4.74 &   1.36 &   -4.51 &   1.28 \\\hline
\multirow{2}{*}{RM} &   -1.88 &   0.97 &   -4.05 &   0.90 &    1.08 &   1.14 &   -1.13 &   1.04 &   -0.99 &   1.04 &   -0.39 &   1.14 \\
                    &   -3.58 &   0.94 &   -3.91 &   0.75 &   -0.65 &   1.11 &   -3.67 &   0.99 &   -6.17 &   1.26 &   -4.20 &   1.22 \\\hline
\multirow{2}{*}{RN} &   -2.28 &   1.27 &   -0.79 &   1.16 &   -4.16 &   1.57 &   -3.52 &   1.33 &   -3.90 &   1.35 &   -6.20 &   1.46 \\
                    &   -4.85 &   1.21 &   -3.04 &   0.96 &   -6.55 &   1.48 &   -5.26 &   1.28 &   -7.68 &   1.66 &   -6.18 &   1.53 \\\hline
\multirow{2}{*}{SP} &  -10.52 &   1.25 &   -7.52 &   1.40 &  -12.60 &   1.49 &  -12.67 &   1.55 &  -14.41 &   1.34 &  -12.20 &   1.44 \\
                    &   -3.96 &   1.25 &   -3.49 &   1.31 &   -6.89 &   1.55 &   -5.66 &   1.56 &   -9.62 &   1.87 &   -6.91 &   1.60 \\\hline
\multirow{2}{*}{MS1}&  -15.19 &   1.15 &  -13.40 &   1.00 &  -22.79 &   1.30 &  -11.85 &   1.22 &  -15.45 &   1.25 &  -13.93 &   1.29 \\
                    &  -15.79 &   1.11 &  -12.76 &   0.85 &  -21.68 &   1.24 &  -11.92 &   1.19 &  -19.01 &   1.45 &  -14.87 &   1.56 \\\hline
\multirow{2}{*}{MS2}&   -3.40 &   1.75 &   -8.09 &   1.30 &  -12.55 &   2.31 &   -0.70 &   2.08 &   -0.68 &   1.97 &   -1.99 &   2.10 \\
                    &   -2.94 &   1.75 &   -4.18 &   1.19 &   -6.55 &   2.21 &    5.13 &   2.07 &  -11.98 &   2.61 &   -1.70 &   2.40 \\\hline
\multirow{2}{*}{TS} &   -1.43 &   1.47 &    2.82 &   1.57 &    0.26 &   1.87 &   -2.76 &   1.55 &   -3.69 &   1.58 &   -2.04 &   1.74 \\
                    &   -0.97 &   1.38 &    1.88 &   1.30 &   -2.54 &   1.67 &   -1.11 &   1.56 &   -7.81 &   1.98 &   -2.60 &   1.79 \\\hline
\multirow{2}{*}{ES1}&  -15.51 &   1.27 &   -8.11 &   1.29 &  -19.03 &   1.34 &  -19.09 &   1.26 &  -17.31 &   1.26 &  -12.45 &   1.45 \\
                    &  -18.07 &   1.21 &   -8.02 &   1.06 &  -21.05 &   1.19 &  -19.65 &   1.17 &  -20.60 &   1.60 &  -16.80 &   1.51 \\\hline
\multirow{2}{*}{ES2}&   13.66 &   3.28 &   11.68 &   3.34 &   -1.36 &   3.47 &    3.03 &   2.97 &    1.06 &   2.85 &    3.00 &   3.47 \\
                    &    4.61 &   3.10 &   14.64 &   2.70 &   -4.93 &   3.20 &    3.10 &   2.73 &   -3.82 &   3.61 &   -7.25 &   3.74 \\\hline
\hline

\hline
\end{tabular}
\end{center}
\caption{Tabulated values of shear calibration bias ($\times10^{-2}$) from figure \ref{fig:mcresults}.
In each entry, the top line refers to the first component of shear, and the bottom line to the second.} 
\label{tab:appendixm}
\end{table*}

\begin{table*}
\begin{center}
\begin{tabular}{cr@{$\pm$}lr@{$\pm$}lr@{$\pm$}lr@{$\pm$}lr@{$\pm$}lr@{$\pm$}l}
{\bf{Author}} & 
\multicolumn{2}{c}{\bf{Image set A}} &
\multicolumn{2}{c}{\bf{Image set B}} & 
\multicolumn{2}{c}{\bf{Image set C}} & 
\multicolumn{2}{c}{\bf{Image set D}} & 
\multicolumn{2}{c}{\bf{Image set E}} & 
\multicolumn{2}{c}{\bf{Image set F}} \\
\hline\hline
\multirow{2}{*}{JB} &  -6.8 &  6.5 & -17.2 &  5.4 & -34.5 &  7.5 &  24.5 &  7.6 &  83.7 &  8.0 &  17.4 &  7.3 \\
                    &   1.3 &  6.6 & -15.0 &  5.5 &  -1.0 &  7.5 & -80.3 &  7.6 &  46.5 &  8.0 &  10.6 &  7.5 \\\hline
\multirow{2}{*}{C1} &  21.2 &  2.5 &  26.7 &  2.5 &  -5.2 &  3.2 & 124.2 &  2.6 &  64.1 &  2.9 & -11.8 &  2.9 \\
                    &  21.2 &  2.5 &  -5.4 &  2.6 &  23.2 &  3.1 & -70.0 &  2.5 & 130.2 &  2.9 &   8.5 &  2.9 \\\hline
\multirow{2}{*}{C2} &  -3.3 &  2.5 &  -1.1 &  2.5 & -21.6 &  3.3 & 259.4 &  2.9 &  29.7 &  3.2 &  -6.2 &  2.9 \\
                    &  38.3 &  2.6 &  18.8 &  2.6 &  39.8 &  3.2 & -36.6 &  2.9 & 276.6 &  3.2 &   3.6 &  3.0 \\\hline
\multirow{2}{*}{MH} &  10.2 &  3.0 &  19.8 &  3.3 &  19.6 &  3.9 & 101.2 &  3.4 &  91.6 &  3.8 &  -4.4 &  3.6 \\
                    &   5.4 &  3.0 &  21.9 &  3.3 &   6.7 &  4.0 & -84.2 &  3.6 &  99.3 &  3.8 &   6.3 &  3.6 \\\hline
\multirow{2}{*}{HH} &   1.6 &  1.8 &  -4.8 &  1.8 &  -6.1 &  2.3 &   3.7 &  2.0 &  75.2 &  2.2 &  -2.2 &  2.1 \\
                    &  -4.6 &  1.8 &  -3.1 &  1.8 &  -0.6 &  2.2 & -65.5 &  2.0 &  -5.9 &  2.1 &   9.8 &  2.1 \\\hline
\multirow{2}{*}{MJ~~}&-11.8 &  2.5 &  -9.5 &  2.2 &  -6.5 &  3.2 &  18.2 &  2.9 &  13.8 &  3.1 &  -2.2 &  2.8 \\
                    &  -0.9 &  2.6 &   6.0 &  2.1 &   1.6 &  3.1 & -12.7 &  2.8 &  16.6 &  3.0 &   1.2 &  2.8 \\\hline
\multirow{2}{*}{MJ2}& -10.3 &  1.9 &  -4.8 &  1.7 &   0.4 &  2.3 &  23.9 &  1.9 &  15.5 &  2.2 &  -0.8 &  1.9 \\
                    &   1.5 &  1.9 &   3.1 &  1.7 &   2.3 &  2.2 & -16.8 &  2.0 &  19.7 &  2.1 &   1.2 &  1.9 \\\hline
\multirow{2}{*}{KK} &  -2.1 &  2.4 &  -5.2 &  2.7 & -14.0 &  3.1 & -71.6 &  2.8 &  66.6 &  3.0 &   0.1 &  2.8 \\
                    &  -2.7 &  2.4 &  -6.6 &  2.7 &   2.1 &  3.0 & -69.5 &  2.7 & -56.9 &  2.9 &  -3.9 &  2.8 \\\hline
\multirow{2}{*}{RM} &  22.9 &  2.2 &  14.9 &  2.0 &  26.5 &  2.9 & -33.5 &  2.5 & 112.0 &  2.8 &   0.1 &  2.6 \\
                    &  -9.9 &  2.2 &  -3.1 &  1.9 &  -5.8 &  2.8 &-105.7 &  2.5 & -19.4 &  2.7 &   2.4 &  2.7 \\\hline
\multirow{2}{*}{RN} &  -5.3 &  2.8 &  -5.0 &  2.5 &  -6.3 &  3.8 & -34.9 &  3.1 &  43.1 &  3.4 &   2.5 &  3.1 \\
                    &   1.8 &  2.7 &  -0.1 &  2.5 &   8.9 &  3.7 & -33.1 &  3.2 & -26.8 &  3.3 &   4.6 &  3.2 \\\hline
\multirow{2}{*}{SP} &  -1.1 &  2.5 &  -3.4 &  2.9 &  -4.5 &  3.3 & -69.9 &  3.3 &  71.6 &  3.4 &   5.5 &  3.0 \\
                    &  -1.1 &  2.7 &  -7.6 &  3.2 &  -4.6 &  3.8 & -55.3 &  3.6 & -13.3 &  3.5 &   4.1 &  3.3 \\\hline
\multirow{2}{*}{MS1}&  -5.6 &  2.5 & -22.3 &  2.1 &   4.9 &  3.2 & 105.1 &  3.0 &  58.5 &  3.3 &  -7.6 &  3.1 \\
                    &  10.3 &  2.6 &  23.1 &  2.1 &   7.2 &  3.2 & -45.7 &  3.0 &  83.8 &  3.1 &   6.1 &  3.2 \\\hline
\multirow{2}{*}{MS2}&  -7.9 &  3.9 & -21.3 &  2.8 &   3.2 &  5.6 & 140.5 &  5.0 &  41.5 &  5.2 &  -0.4 &  4.9 \\
                    &  14.4 &  4.0 &  24.3 &  3.0 &  19.5 &  5.9 & -28.7 &  5.1 & 154.4 &  5.4 &   9.2 &  5.3 \\\hline
\multirow{2}{*}{TS} &  -2.9 &  3.3 &  -4.3 &  3.5 &   2.7 &  4.5 & -46.2 &  3.9 &  70.3 &  4.3 &  -3.5 &  4.0 \\
                    &  -3.0 &  3.2 &  -1.3 &  3.6 &   0.4 &  4.4 & -65.4 &  3.9 & -40.3 &  4.2 &  -3.1 &  4.0 \\\hline
\multirow{2}{*}{ES1}&  -9.1 &  2.8 &  -4.1 &  2.9 &   5.7 &  3.3 & 153.1 &  3.1 &  54.3 &  3.4 &  -5.5 &  3.4 \\
                    &   4.0 &  2.8 &   8.7 &  2.8 &   9.9 &  3.2 & -58.7 &  3.0 & 132.0 &  3.4 &   0.4 &  3.3 \\\hline
\multirow{2}{*}{ES2}& -11.0 &  7.4 &   8.5 &  7.4 &  15.0 &  8.2 &  95.3 &  7.1 &  96.7 &  7.8 & -10.4 &  8.2 \\
                    & -11.2 &  7.4 &  -3.3 &  7.2 &   5.7 &  8.4 & -92.9 &  7.1 &  77.9 &  7.7 &   7.7 &  8.1 \\\hline
\hline

\hline
\end{tabular}
\end{center}
\caption{Tabulated values of residual shear offset ($\times10^{-4}$) from figure \ref{fig:mcresults}.
In each entry, the top line refers to the first component of shear, and the bottom line to the second.} 
\label{tab:appendixc}
\end{table*}

\label{lastpage}

\end{document}